\documentclass[]{JFM-FLM_Au}
\usepackage{graphicx,caption,subcaption}
\usepackage{epstopdf, epsfig}
\usepackage{bm}
\usepackage{ulem}
\usepackage{newtxtext}
\usepackage{newtxmath}
\usepackage{amsmath}
\usepackage{natbib}
\usepackage{hyperref}
\hypersetup{
	colorlinks=true,
	urlcolor=blue,
    citecolor=blue
}

\lefttitle{Prateek Anand and Ganesh Subramanian}
\righttitle{Journal of Fluid Mechanics}
\title{Inertial migration of slender prolate and thin oblate spheroids in plane Poiseuille flow}
\author{Prateek Anand\aff{1} \and Ganesh Subramanian\aff{2}}
\corresau{Ganesh Subramanian, \email{sganesh@jncasr.ac.in}}
\affiliation{\aff{1}International Centre for Theoretical Sciences, Bengaluru-560089, India.
\aff{2}Engineering Mechanics Unit, Jawaharlal Nehru Centre for Advanced Scientific Research, Bengaluru-560064, India.}

\begin{document}
\maketitle

\begin{abstract}
A neutrally buoyant rigid spheroid suspended in a wall-bounded
 plane Poiseuille flow undergoes cross-stream migration, driven by fluid inertia. We examine theoretically the migration of a spheroid of aspect ratio $\kappa$ in the limit of small but finite particle Reynolds number\,($Re_p$), and for small confinement ratios\,($\lambda$), with the channel Reynolds number, $Re_c = Re_p/\lambda^2$, assumed arbitrary; here, $\lambda=L/H$ with $L$ being the semi-major axis of the spheroid and $H$ denoting the separation between channel walls. For small $\lambda$, the asymptotic separation between the rotation-cum-orbital-drift and migration time scales implies that, to begin with, inertia rapidly drives the spheroid towards the tumbling orbit~(orbit constant, $C=\infty$) with negligible migration; migration is subsequently driven by a time-averaged lift velocity, the average being over the orientations sampled in the inertially stabilized tumbling orbit. While spheroids with $\kappa\sim O(1)$ rotate with the Jeffery angular velocity to a very good approximation, deviations from Jeffery rotation, for both large and small $\kappa$, lead to a time-averaged inertial lift profile with equilibrium locations\,(zero crossings) that differ from the classical Segre-Silberberg predictions for a sphere. Beyond a threshold $Re_c$, both slender spheroids and thin disks attain a steady orientation in the neighborhood of the walls, and with increasing $Re_c$, these rotation-arrested regions grow in extent, moving in towards the channel centerline. In contrast to spheres, but consistent with experiments, onset of rotation arrest causes the aforementioned equilibrium positions to move towards the centerline; further, for thin disks alone, the equilibrium positions themselves become rotation-arrested beyond a threshold $Re_c$. The $\kappa$-dependence of the equilibrium positions can be leveraged towards developing passive shape-sorting protocols on microfluidic platforms. 
\end{abstract}

\section{Introduction}
A spherical particle is known to undergo cross-stream migration in plane Poiseuille flow under the influence of fluid inertia. This phenomenon of inertial migration was first observed experimentally by~\cite{segre1961} for neutrally buoyant spheres in pipe Poiseuille flow, and has been subsequently investigated in both the pipe and channel\,(plane Poiseuille flow) geometries by a number of theoretical~(
\cite{coxbrenner1968}, \cite{holeal1974},\cite{vasseur1976}, \cite{schonberghinch1989}, \cite{asmolov1999}, \cite{matas2009}), numerical~(\cite{shao2008inertial}, \cite{seki2017}, \cite{nakayama2019}, \cite{glowinski2021}) and
experimental~(\cite{tachibana1973}, \cite{aoki1979}, \cite{matas2004}, \cite{masaeli2012}, \cite{nakayama2019}) studies, carried out over the past several decades. The aforesaid body of work has shown that, for small pipe or channel Reynolds numbers\,($Re_c \ll 1$) and with confinement effects being sufficiently weak\,($\lambda \ll 1$), a sphere in the said shearing flows migrates to an equilibrium position(s) intermediate between the wall(s) and the centerline, where the inertial lift force equals zero; here, $\lambda$ denotes the ratio of the sphere radius to the confining length scale, either the separation between channel walls or the pipe diameter. The equilibrium position moves towards the walls with increasing $Re_c$. In contrast, a rigid spheroid in plane Poiseuille flow has a lateral\,(cross-stream) velocity component, even in the Stokes limit, due to hydrodynamic interactions with the channel walls; the component being $O(\lambda^2)$ in the limit $\lambda \ll 1$. While reversibility of the Stokes equations does not preclude an instantaneous lift force on a spheroid, a net cross-stream migration is nevertheless prohibited, and accordingly, a neutrally buoyant spheroid in the inertialess limit exhibits an oscillatory motion, with an amplitude of $O(\lambda^2)$, about any streamline of the ambient plane Poiseuille flow. A time-averaged cross-stream migration of a spheroid in a Newtonian fluid still requires inertia, and in this paper, we examine the nature of this time-averaged migration, with an emphasis on how it differs from the migration of spheres described above.


The ability of inertial lift forces to differentially order suspended particles continuously flowing at a high rate through micro-channels, in the absence of external fields, is expected to have a broad range of applications in bioparticle separation, high-throughput cytometry, and large-scale filtration systems~\citep{dicarlo2007}. This ability is contingent on the lift-induced equilibria being dependent on intrinsic particle characteristics such as shape, size, deformability and so on. A few examples emphasizing the importance of particle shape are worthy of mention. Shape remains one of the most important factors in the specific identification of a bioparticle and has several applications in biotechnology. For instance, drug delivery systems benefit from using non-spherical particles based on their influence on cellular internalization and vascular dynamics~\citep{decuzzi2009intravascular,mitragotri2009}. Synthesized barcoded anisotropic particles can be used in microchannels to achieve high-throughput screening for genetic analysis, combinatorial chemistry and clinical diagnostics~\citep{pregibon2007}. The discoid shape of red blood cells is known to affect the magnitude of the inertial lift, as well as the stable focusing positions in a blood filtration device~\citep{dicarlo2010}.

\begin{figure}
	\centering
	\includegraphics[width=\textwidth]{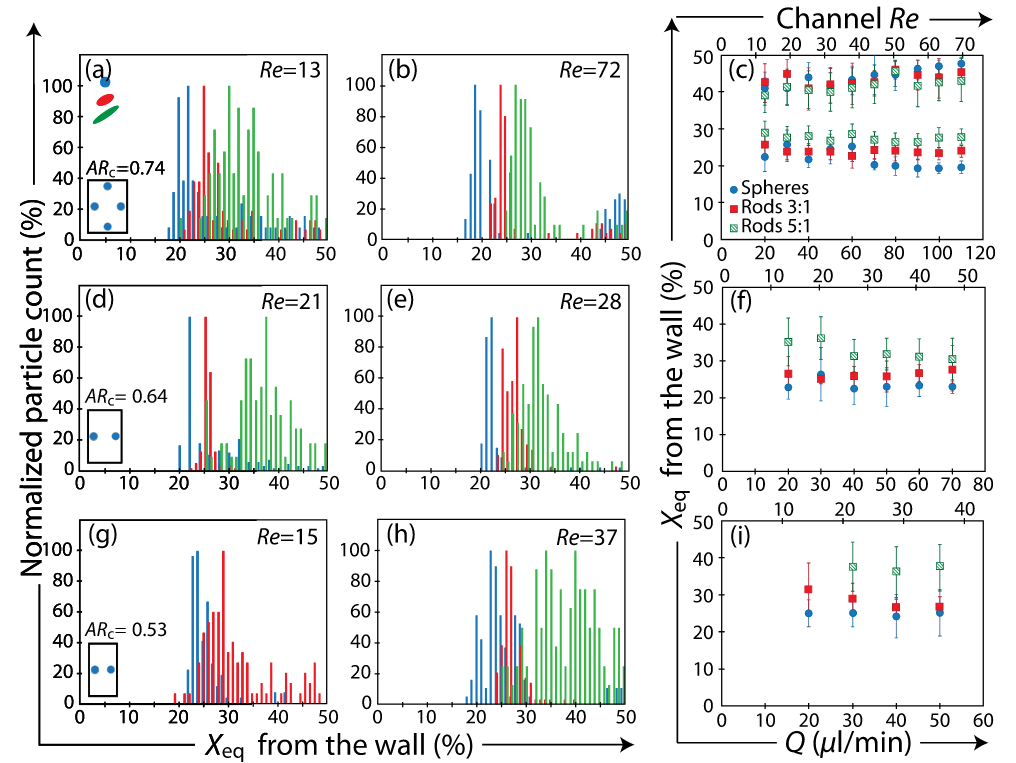}
	\caption{Experimental results, reproduced from Figure 3 of \cite{masaeli2012}, that support the existence of shape-sensitive equilibria. (a), (b), (d), (e), (g) and (h) are histograms of the equilibria for spheres, and spheroids with $\kappa=3$ and $5$, for different $H/W$ and for different flow rates $Q$ (corresponding to different $Re_c$; denoted as $Re$ in the plots). For $H/W=0.74$, the particles migrate to four intermediate equilibria, with two each located symmetrically in the horizontal and vertical mid-planes. For $H/W=0.53$ and $0.64$ (closer to the idealized plane Poiseuille flow configuration with $H/W \to 0$), particles migrate to only the two equilibria in the horizontal midplane. As evident from (c), (f) and (i), the equilibrium locations of spheroids with $\kappa = 5$ are closer to the centerline at any $Re_c$, and continue to move towards it with increasing $Re_c$.}
	\label{fig:Masaeliequilibria}
\end{figure}

Specific information pertaining to shear-induced migration of anisotropic particles, with evidence of shape-sensitive equilibria, is available from the experiments of \citet{hur2011} and \citet{masaeli2012}. \citet{hur2011} investigated the motion of non-spherical particles of various shapes in a straight rectangular micro-channel with $H/W=0.65$; $H$ and $W$ being the height and width of the cross-section, respectively. Experiments were done for duct Reynolds numbers\,(defined using $H$ and the maximum velocity) upto $160$, and for $Re_p\gtrsim O(1)$, with confinement ratios ranging over the interval $0.15\leq\lambda\leq0.35$. The authors observed cylinders and disks to exhibit tumbling and spinning motions, respectively, consistent with theory\,(see below). Rather surprisingly, they found the equilibrium positions for spheroids to not depend on their aspect ratios, but instead on their cross-sectional diameters, with spheres and spheroids having identical cross-sectional diameters migrating to the same transverse location. \cite{masaeli2012} performed both experiments and numerical simulations on spheres and spheroids of various aspect ratios, suspended in pressure-driven flow through a rectangular duct with $0.53\leq H/W\leq 0.74$. Experiments were done for duct Reynolds numbers\,(defined in the same manner as above) upto $80$, and for $Re_p\sim O(1)$, with the confinement ratio ranging over the intervals $0.04<\lambda<0.12$ for spheres, $0.09<\lambda<0.25$ for prolate spheroids with $\kappa=3$, and $0.125<\lambda<0.35$ for $\kappa=5$. Prolate spheroids were mainly observed to tumble, this being consistent with both predictions of the small-$Re_p$ analyses of \cite{navaneeth2016} and \cite{einarsson2015}, and the results of finite $Re_p$ computations~\citep{huang2012,rosen2014effect,rosen2015dynamical}. Further, the higher aspect ratio spheroids were found to tumble at a rate slower than the Stokesian prediction, with the rotation time period significantly exceeding the Jeffery value\,($=2\pi(2\,U_\text{avg}/H)^{-1}(\kappa+\kappa^{-1})$, $U_\text{avg}$ being the average velocity in the duct), this again being consistent with theory\,\citep{subkoch2005,navaneeth2017}. 
With regard to shear-induced migration behaviour, the authors observed that, as $Re_c$ increases, spheres migrate to equilibrium locations closer to the walls compared to spheroids. This may be seen from the peaks of the histograms shown in Figure~\ref{fig:Masaeliequilibria}\,(corresponding to Figure~3 in \citet{masaeli2012}). For all duct cross-sectional ratios, with $Re_c$ fixed, the equilibria for spheroids with $\kappa = 5$ are located closest to the centerline\,(largest $X_{eq}$), those for spheres are closest to the walls\,(smallest $X_{eq}$), and those for spheroids with $\kappa=3$ are at intermediate locations. Further, for a given duct with increasing $Re_c$, the peak for spheres moves towards the wall, while that for spheroids with $\kappa=5$ moves towards the centerline. Note also that, for the smaller cross-sectional aspect ratios $H/W=0.53$ and $0.64$, which are closer to the idealized plane Poiseuille configuration\,($H/W \rightarrow 0$), particles migrated to two equilibrium positions lying in the plane midway between the longer edges of the cross-section. In contrast, for $H/W=0.74$, the particles focused onto four equilibrium positions, one in the neighborhood of each cross-sectional edge. The equilibrium positions above compared well with hybrid Lattice-Boltzmann simulations performed by the same authors. Many of these aspects have been reviewed in \citet{dicarlo2014}. It will be seen later, in $\S$\ref{sec:results}, that the aforementioned features of the \citet{masaeli2012} experiments are consistent with predictions of our small-$Re_p$ theory. In particular, an inertia-induced increase in the rotation period, culminating in rotation arrest for sufficiently large $\kappa$, does indeed lead to migration of slender spheroids towards the centerline. 

Motivated by the aforementioned applications, and the underlying fundamental question of shape-sensitive equilibria, we 
examine the inertial migration of a neutrally buoyant spheroid of an arbitrary aspect ratio, in plane Poiseuille flow, for small but finite $Re_p$. $Re_p$ here is the micro-scale Reynolds number based on the spheroid semi-major axis\,($L$), being related to $Re_c$ as $Re_p = \lambda^2 Re_c$; following \citet{schonberghinch1989}, $\lambda$ is also assumed to be small, with $Re_c$ being arbitrary. 
In a recent study, \cite{anand2023Jeff} characterized the inertial migration of spheroids in the above limit, using a point-particle formulation similar to that used for spherical particles in the pioneering efforts of  \citet{holeal1974}, \citet{vasseur1976} and \citet{schonberghinch1989}. While \citet{anand2023Jeff} accounted for an $O(Re_p)$ drift of the spheroid orientation across Jeffery orbits, the rotation within the inertially stabilized orbit\,(the tumbling mode for prolate spheroids with $\kappa > 1$; the spinning mode for oblate spheroids with $0.14 < \kappa < 1$, and either of tumbling or spinning modes, depending on initial orientation, for oblate spheroids with $\kappa < 0.14$; see \citet{navaneeth2016,navaneethRapids}) was still assumed to occur with the Stokesian\,(Jeffery) angular velocity. Thus, \citet{anand2023Jeff} calculated the time-averaged lift based on the Jeffery angular velocity, and the resulting lift profiles for spheroids were found to have the same equilibrium locations as those for a sphere; although, the magnitude of the lift was a function of $\kappa$. The $\kappa$-independence of the equilibria within a Jeffery-averaged approximation implies that both spheres and spheroids would end up converging to the same positions in a sufficiently long channel, under the action of inertial lift forces, albeit at different rates. Therefore, any shape-separation protocol must rely on a precise control of the length of the microfluidic channel used for separation. In a later effort, \cite{anand2024FiniteSize} have shown that the inclusion of finite-size effects\,(that is, small but finite $\lambda$) for spherical particles leads to the emergence of a pair of new equilibrium locations, closer to the centerline, beyond a threshold $Re_c$. This effort helped characterize small-$Re_p$ migration of spheres on the $\lambda-Re_c$ plane, thereby harmonizing the findings of multiple experiments~\citep{hur2011,nakayama2019} and computations~\citep{shao2008inertial,glowinski2021}, dating back to the original effort of \citet{matas2004}. Importantly, it was argued by the authors that the locations of the new equilibria, and the threshold $Re_c$ for their emergence, are both expected to depend on $\kappa$; although, an explicit calculation remains to be done for non-spherical particles. 

In the present effort, we highlight the possibility of $\kappa$-dependent equilibria, even within the point-particle framework used in \cite{anand2023Jeff}, provided one goes beyond the Jeffery-averaged approximation. 
The need for such a correction is suggested by the experiments of~\cite{masaeli2012}  above, where a correlation was found between particle rotation and inertial migration - specifically, inertial slowdown of the rotation of larger aspect ratio spheroids appeared to retard wallward migration. By including the correction to the leading order Jeffery angular velocity in the tumbling mode, in our analysis of inertial migration, we show that the resulting predictions are consistent with the above observation. For spheroids of order unity aspect ratios, which were the focus in \cite{anand2023Jeff}, the aforementioned slowdown is small, being $O(Re_p^{3/2})$\citep{lin1970,navaneeth2017}, and thereby, smaller than the $O(Re_p)$ orbital drift; the smallness of the inertial angular velocity correction, for $Re_p \lesssim 1$, is also suggested by very recent experiments~\citep{guazzelli2025}. This leads to an extremely weak dependence of the equilibrium positions on $\kappa$. For sufficiently large or small $\kappa$, however, the inertial angular velocity correction becomes comparable to the Jeffery contribution for near flow\,(gradient)-aligned prolate\,(oblate) spheroids. This in turn leads to a substantial slowing down of spheroid rotation, eventually leading to rotation arrest beyond a threshold $Re_p$. Such an arrest was first predicted for a slender prolate spheroid by \citet{subkoch2005}, in which case the inertial angular velocity correction is $O(Re_p/\ln \kappa)$, and the threshold $Re_p$ comes out to be $O(\ln \kappa/\kappa)$; the threshold for a thin oblate spheroid is $O(\kappa)$\citep{navaneeth2017}. 
 For both slender prolate and thin oblate spheroids, inertia-induced slowdown and eventual rotation arrest lead to equilibrium positions that are $\kappa$-dependent, and that transition from an initial wallward movement to an eventual movement towards the centerline with increasing $Re_c$. Further, for sufficiently large $Re_c$, thin oblate spheroids adopt a stationary orientation even at their equilibrium positions.

The paper is organized as follows. In $\S$\ref{sec:form}, we write down the governing equations and boundary conditions for a freely rotating spheroid in plane Poiseuille flow, after which we discuss the nature of the inertia-induced orbital drift for both prolate and oblate spheroids based on earlier work by \citet{einarsson2015} and \citet{navaneeth2016}. We then write down the expression for the angular velocity of a spheroid rotating in the inertially stabilized tumbling orbit, which allows for a discussion of inertial slowdown of rotation, leading to arrest. In $\S$\ref{sec:Reclarge}, exploiting the separation between the rotation and migration time scales, we first time-average the governing equations for a tumbling spheroid, and then outline the procedure to calculate the time-averaged lift velocity for $Re_c \gtrsim O(1)$. The result is valid when the inertial screening length\,($L Re_p^{-\frac{1}{2}}$) is of order the channel width or smaller, and involves application of a shooting method to the partially Fourier-transformed system of governing equations. This method was first used by \citet{schonberghinch1989} for spheres, and has been discussed in detail in \cite{anand2023Jeff}. 
In $\S$\ref{sec:results}, we present the lift velocity profiles calculated using the shooting method, for a wide range $Re_c$'s of order unity or greater, with an emphasis on slender prolate and thin oblate spheroids. We show, in particular, that when rotation arrested zones emerge within the channel, owing to $Re_p$ exceeding the local rotation-arrest threshold, the lift-induced equilibrium location moves towards the centerline with increasing $Re_c$, and at a rate that depends on $\kappa$; this movement being stronger for thin oblate spheroids. 
In $\S$\ref{sec:conclusion}, we first summarize the main results of this paper, provide a sketch of the mechanism underlying the $\kappa$-dependent migration, and then return to the implications of $\kappa$-dependent equilibrium locations for passive shape sorting. At the end of this section, we discuss extensions of the present effort to include finite-size contributions to inertial migration that have recently been shown to qualitatively affect migration of spheres at large $Re_c$\,\citep{anand2024FiniteSize}. In Appendix~\ref{App:LiftSmallRec}, the equations in $\S$\ref{sec:Reclarge} are simplified for $Re_c\ll1$, and the $\kappa$-dependence of the equilibrium positions is established by calculating the lift velocity 
 semi-analytically using a regular perturbation expansion. This semi-analytical effort forms the basis for the mechanistic discussion in section $\S$\ref{sec:conclusion}.

\section{Problem formulation and governing equations}\label{sec:form}

\begin{figure}
	\centering
    \begin{subfigure}[b]{0.59\textwidth}
		\includegraphics[width=\textwidth]{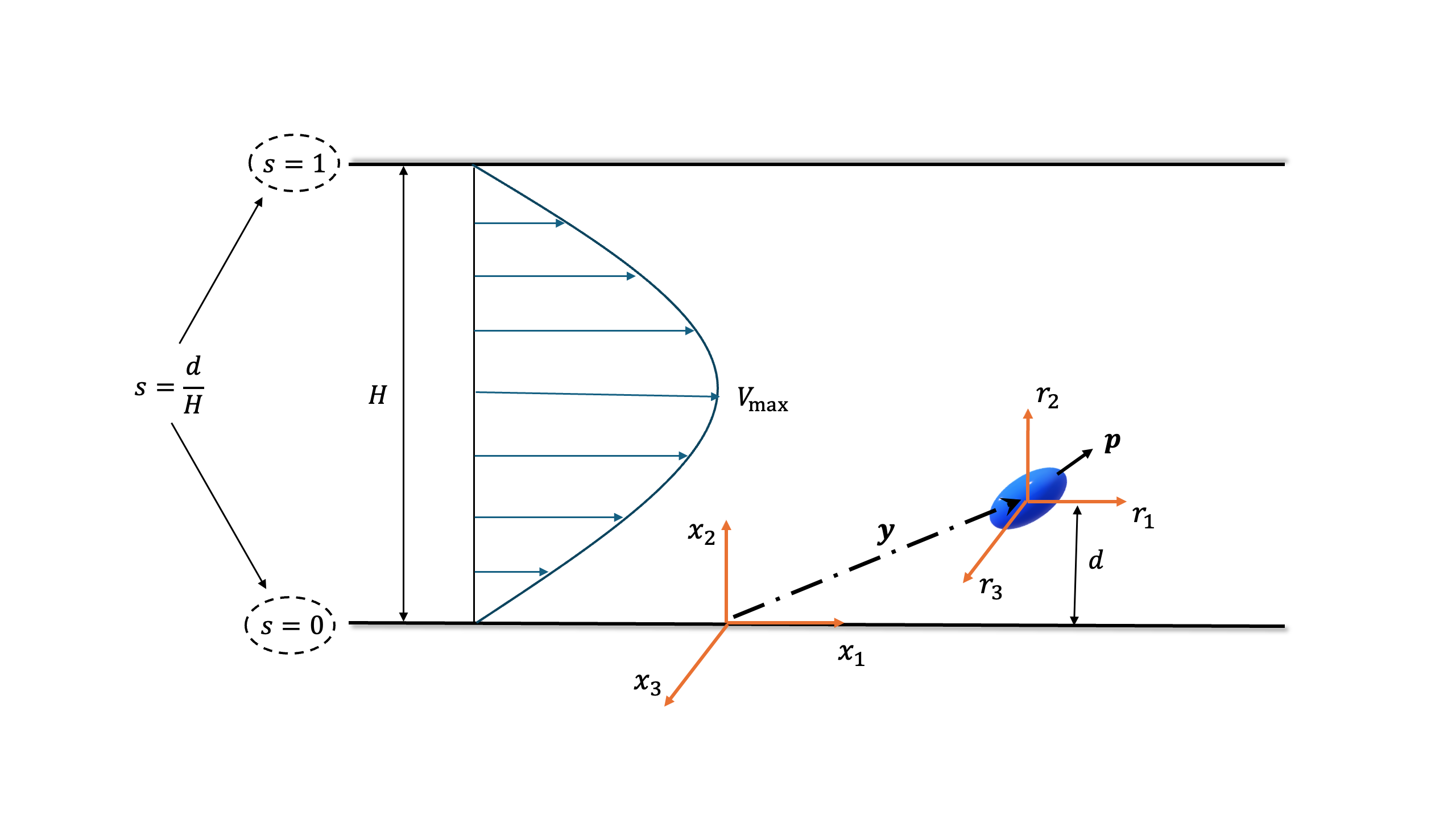}
        \caption{}
    \end{subfigure}
    \hfill
    \begin{subfigure}[b]{0.39\textwidth}
		\includegraphics[width=\textwidth]{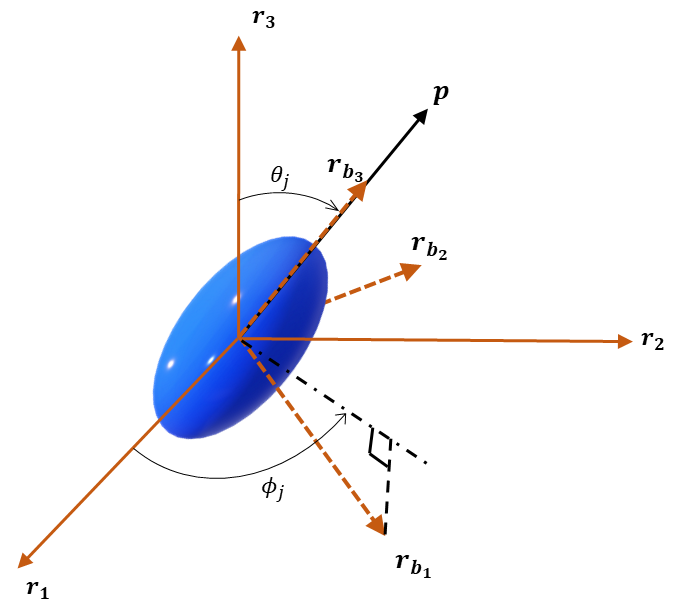}
        \caption{}
    \end{subfigure}
	\caption{(a) A neutrally buoyant spheroid with symmetry axis $\bm{p}$ in plane Poiseuille flow. The position of the spheroid relative to the lab frame ($x_1,x_2,x_3$) is denoted by $\bm{y}$; $(r_1,r_2,r_3)$ represents the Cartesian frame with origin at the spheroid center, and translating with it. (b) shows the body-fixed coordinate system aligned with $\bm{p}$, along with the polar\,($\theta_j$) and azimuthal\,($\phi_j$) angles that define the spheroid orientation. The dot-dashed line is the projection of the $\bm{r}_{b_1}$-axis on the flow-gradient plane.}
	\label{fig:channelGeometry}
\end{figure}

Figure~\ref{fig:channelGeometry}a shows a neutrally buoyant spheroid of aspect ratio $\kappa=L/b$\,($L$ and $b$ are the semi-major and minor axes) freely suspended in a wall-bounded plane Poiseuille flow at a distance $d$ from the lower wall; $\kappa<1$ and $>1$ for oblate and prolate spheroids, respectively.
The spheroid motion is governed by the following equations:
\begin{subequations}
	\begin{align}
	\nabla^2 \bm{u}-\bm{\nabla} p =Re_p\Bigg(\frac{\partial\bm{u}}{\partial t}+\frac{d\bm{U}_p}{dt}&+\bm{u}\cdot\bm{\nabla u}+\bm{u}\cdot\bm{\nabla u}^\infty+\bm{u}^\infty\cdot\bm{\nabla u}\Bigg),\\
	\bm{\nabla}\cdot \bm{u}&=0,
	\end{align} \label{eq:NS1}
\end{subequations}
for the disturbance velocity field, in a reference frame with its origin at the spheroid center, and translating with the spheroid velocity ${\bm U}_p$. Here, $\bm{u}$ satisfies:
\begin{subequations}
	\begin{align} \bm{u}&=\bm{\Omega}_p\wedge \bm{r}-\bm{u}^\infty \text{ for } \bm{r}\in S_p,\\
	\bm{u}&\rightarrow 0 \text{ for } r_1,r_3\rightarrow \infty\,(r_2\hspace*{0.05in}\text{fixed}),\\
	\bm{u}&=0 \text{ at } r_2=-s\lambda^{-1}, (1-s)\lambda^{-1},
	\end{align} \label{eq:BC1} 
\end{subequations}
where $S_p$ denotes the spheroid surface. $\bm{U}_p$ and $\bm{\Omega}_p$ in (\ref{eq:NS1}a) and (\ref{eq:BC1}a) are the spheroid translational and angular velocities. In (\ref{eq:NS1}a-\ref{eq:BC1}c), all variables are non-dimensionalized using $L$ and the velocity scale $V_c=V_\text{max}L/H$, this being the order of the velocity change across the ends of the spheroid. The particle Reynolds number is therefore given by $Re_p=V_c L/\nu = V_\text{max}L^2/H\nu$, $\nu$ being the kinematic viscosity of the fluid. 
The ambient plane Poiseuille flow, $\bm{u}^\infty$ in (\ref{eq:NS1}a) and (\ref{eq:BC1}a), is given by,
\begin{align}
\bm{u}^\infty=(\alpha+\beta r_2 +\gamma \lambda {r_2}^2)\bm{1}_1-\bm{U}_p,
\label{eq:uamb}
\end{align}
where $\alpha \bm{1}_1 - \bm{U}_p$, with $\alpha=4 \lambda^{-1} s(1-s)$, is the slip velocity at the spheroid center, $\beta=4(1-2s)$  is the local shear rate that varies linearly across the channel, and $\gamma=-4$ is the constant curvature of the plane Poiseuille profile. $s = d/H$ denotes the non-dimensional spheroid location, with $s=1/2$ corresponding to the channel centerline. $\lambda=L/H$ is the confinement ratio, with $Re_p$ and $\lambda$ being assumed small, while allowing for the channel Reynolds number,  $Re_c=Re_p/\lambda^2$, to be arbitrary. The analysis below also assumes $s, 1-s \gg \lambda$, and is therefore restricted to spheroid motion at distances greater than $O(L)$ from either wall.

For small $Re_p$, $\bm{U}_p$ and $\bm{\Omega}_p$ may be determined using the Faxen's laws for translation and rotation of a force-free and torque-free spheroid \citep{brennerbook,anand2023Jeff}. This, at leading order in $\lambda$, yields: 
\begin{align}
&\bm{U}_{p}= \left[\alpha+\frac{\gamma\lambda}{3\kappa^2}[\cos^2\phi_j+(\cos^2\theta_j+\kappa^2\sin^2\theta_j)\sin^2\phi_j] \right]\bm{1}_1, \label{eq:UpFaxen}\\
&\bm{\Omega}_{p}=-\frac{\beta(\kappa^2-1)}{4(\kappa^2+1)}\cos\phi_j\sin 2\theta_j\bm{1}_1+\frac{\beta(\kappa^2-1)}{4(\kappa^2+1)}\sin\phi_j\sin 2\theta_j\bm{1}_2\nonumber\\
&+\frac{\beta}{2}\big[-1+\frac{(\kappa^2-1)}{(\kappa^2+1)}\cos 2\phi_j\sin^2 \theta_j\big]\bm{1}_3, \label{eq:OmegapFaxen}
\end{align}
where the Cartesian unit vectors correspond to directions shown in Figure~\ref{fig:channelGeometry}a.

The local linear flow approximation of the plane Poiseuille profile is a simple shear flow, and the inertial screening length for a particle in simple shear is $L Re_p^{-1/2}$ or $H Re_c^{-1/2}$. The analysis for shear-induced migration depends on $Re_c$. When $Re_c\ll 1$, the inertial screening length is much larger than the channel width, implying that the inner region, characterized by scales of $O(H)$, gives the dominant contribution to the lift velocity; inertia in this case is a regular perturbation. When $Re_c\gtrsim O(1)$, the inertial screening length is of order the channel width or smaller, and the contributions from scales of $O(H)$ and $O(H Re_c^{-1/2})$ must be accounted for via a matched asymptotic expansions approach\citep{schonberghinch1989,anand2023Jeff}. Scaling arguments towards the end of this section show that, for the rotation-arrested scenario which turns out to be most relevant to inertial migration, the assumption of a small $Re_c$ imposes a severe restriction on spheroid aspect ratios, and we therefore focus on the case $Re \gtrsim O(1)$ below.

The Stokes velocity disturbance due to a freely rotating spheroid only decays as $1/r^2$ at distances large compared to $L$, with the algebraic decay implying  that distinct inner\,($r \sim O(L)$) and outer-region expansions\,($r \sim O(HRe_c^{-1/2})$) for the velocity and pressure fields are necessary 
to account for inertial effects for $Re_c \gtrsim O(1)$; the leading terms in the latter expansion satisfy the linearized Navier-Stokes equations. To capture the dominant outer-region contribution, one transforms to outer coordinates using $\bm{r}=Re_p^{-1/2}\bm{R}$, 
with the Stokesian rates of decay in the inner region implying  the scalings $\bm{u}=Re_p\bm{U}$ and $p=Re_p^{3/2} P$ for the leading order terms. When used in (\ref{eq:NS1})-(\ref{eq:BC1}), $\bm{U}$ and $P$ satisfy:  
\begin{subequations} 
	\begin{align} 
	\frac{\partial^2 U_i}{\partial R_j^2}-\frac{\partial P}{\partial R_i}-\frac{\partial U_i}{\partial t}-\frac{1}{Re_p}\frac{dU_{p,i}}{dt}-&U_2(\beta+2\gamma R_2 Re_c^{-1/2})\delta_{i1}-(\beta R_2+\gamma R_2^2 Re_c^{-1/2})\frac{\partial U_i}{\partial R_1}\nonumber \\
 &=\beta S_{ij}\frac{\partial\delta(\bm{R})}{\partial R_j},\\
	\frac{\partial{U}_i}{\partial x_i}&=0,
	\end{align} \label{eq:NSOuterEqn}
\end{subequations}

where $U_i$ satisfies:
\begin{subequations}
\begin{align}
    U_i &\sim \frac{3\beta R_i S_{jm}R_j R_m}{4\pi R^5} \text{ for } \bm{R}\rightarrow 0,\\
	U_i &\rightarrow 0 \text{ for } R_1,R_3\rightarrow \infty\,(R_2\hspace*{0.05in}\text{fixed}),\\
	U_i&=0 \text{ at } R_2=-s Re_c^{1/2}, (1-s)Re_c^{1/2}.
\end{align} \label{eq:NSOuterBC} 
\end{subequations}
The Faxen correction contributes at a higher order in $\lambda$\citep{anand2023Jeff} and therefore, $u^\infty_i\approx Re_p^{-1/2}(\beta R_2+\gamma\,Re_c^{-1/2} R_2^2) \delta_{i1}$ in (\ref{eq:NSOuterEqn}a). Note also that the no slip condition (\ref{eq:BC1}a) has been replaced by (\ref{eq:NSOuterBC}a), the requirement of matching to the farfield limit of the inner stresslet field, the tensorial stresslet amplitude being given by \citep{navaneeth2017}:
\begin{align}
\bm{S}(\bm{p})&= A_1 \frac{3}{2}(\bm{E}:\bm{pp})\left(\bm{pp}-\frac{\bm{I}}{3}\right) + A_2 ((\bm{I}-\bm{pp})\cdot\bm{E}\cdot\bm{pp}+\bm{pp}\cdot\bm{E}\cdot(\bm{I}-\bm{pp}))\nonumber\\& +A_3\left((\bm{I}-\bm{pp})\cdot\bm{E}\cdot(\bm{I}-\bm{pp})+(\bm{I}-\bm{pp})\frac{\bm{E}:\bm{pp}}{2}\right),
\label{eq:StressletProlate}
\end{align}
in terms of the unit vector $\bm p$ denoting the spheroid orientation; here,  $E_{ij}=\frac{\beta}{2}(\delta_{i1}\delta_{j2}+\delta_{i2}\delta_{j1})$ is the rate of strain tensor associated with the local simple shear. The above expression arises from resolving simple shear flow into an axisymmetric extension aligned with $\bm{p}$, longitudinal planar extensions in a pair of orthogonal planes containing $\bm{p}$, and a pair of transverse planar extensions in the plane perpendicular to $\bm{p}$~\citep{subkoch2006,navaneeth2016}, with $A_i$'s in \eqref{eq:StressletProlate} being the $\kappa$-dependent scalar amplitudes corresponding to these component flows~\citep{kimkarrila,navaneeth2017}. 
For $\kappa>1$, these amplitudes are given by:
\begin{align}
A_1&=-\frac{16\pi(\kappa^2-1)^{5/2}}{9\kappa^3[-3(\kappa^2-1)^{1/2}\kappa+2\kappa^2 \cosh^{-1}(\kappa)+\cosh ^{-1}(\kappa)]},\\
A_2&=-\frac{16\pi(\kappa^2-1)^3}{3\kappa^2(\kappa^2+1)(\kappa^4+\kappa^2-3(\kappa^2-1)^{1/2}\kappa\cosh^{-1}(\kappa)-2)},\\
A_3&=-\frac{32\pi(\kappa^2-1)^3}{3\kappa^3(2\kappa^5-7\kappa^3+3(\kappa^2-1)^{1/2}\cosh^{-1}(\kappa)+5\kappa)}.
\end{align}
The corresponding expressions for an oblate spheroid ($\kappa<1$) may be obtained by first substituting $\kappa=\xi_0/(\xi_0^2-1)^{1/2}$ in terms of the coordinate label\,($\xi_0>1$) for the spheroid, and then using the transformation $d\to-\iota d, \xi_0\to\iota(\xi_0^2-1)^{1/2}$ in the dimensional stresslet, obtained from multiplying $\bm{S}$ above by $\mu L^3 V_\text{max}/H$ with $L= d\xi_0$~\citep{navaneeth2017}.

The stresslet given by (\ref{eq:StressletProlate}) is a function of time owing to $\bm p$ in general being time dependent. The nature of this time dependence for small $Re_p$ depends on the Jeffery orbit in which the spheroid rotates. For $Re_p = 0$, a spheroid can rotate in any of an infinite number of these orbits parameterized by an orbit constant $C \in [0,\infty)$, with the limiting values corresponding to spinning\,($C=0$) and tumbling\,($C= \infty$) modes. Rotation along a Jeffery orbit generates an unsteady disturbance field for any $C \neq 0$, and the associated fluid inertial forces in a region of $O(L^3)$ around the rotating spheroid, lead to an $O(Re_p)$ drift across orbits. This orbital drift stabilizes the tumbling mode for a prolate spheroid, and either the spinning or tumbling mode for oblate spheroids with $\kappa \lesssim 0.14$~\citep{einarsson2015,navaneeth2016}. The Jeffery time period remains unchanged to $O(Re_p)$ for spheroids with $\kappa \sim O(1)$, with the first correction only occurring at $O(Re_p^{3/2})$~\citep{navaneeth2017}. The angular velocity, to $O(Re_p)$, of a spheroid, tumbling stably in the flow-gradient plane\,($\theta_j = \frac{\pi}{2}$), can thus be written in the form~\citep{navaneeth2016}:
\begin{align}
\dot{\phi}_j=\dot{\phi}_j^{(0)}+ Re_s \dot{\phi}_j^{(1)} + O(Re_p^{3/2}),
\label{eq:Fullangularvelocity}
\end{align}
where $Re_s=|\beta| Re_p$ is the particle Reynolds number based on the magnitude of the local shear rate. The Jeffery angular velocity $\dot{\phi}_j^{(0)}$, and the first inertial correction $\dot{\phi}_j^{(1)}$, are given by:
\begin{align}
\dot{\phi}_j^{(0)} &=\beta \left(-\frac{1}{2}+\frac{\kappa^2-1}{2 (\kappa^2+1)}\cos 2\phi_j\right)\label{eq:Jeffangvel},\\
\dot{\phi}_j^{(1)}&=\beta\sin \phi_j\cos\phi_j\![G_1(\kappa)\!+\!G_2(\kappa)\cos 2\theta_j\!+\!G_3(\kappa)\cos 2\phi_j\!+\!G_4(\kappa)\cos 2\theta_j \cos 2\phi_j], \label{inertcorr_Re}
\end{align}
where the expression for $\dot{\phi}_j^{(0)}$ may be inferred from \eqref{eq:OmegapFaxen}. In (\ref{eq:Jeffangvel}) and (\ref{inertcorr_Re}), $\theta_j$ and $\phi_j$ are the polar and azimuthal angles depicted in Figure~\ref{fig:channelGeometry}, and that define the spheroid orientation; $\bm p = \sin\theta_j \cos \phi_j {\bm 1}_1 + \sin\theta_j \sin \phi_j {\bm 1}_2 + \cos \theta_j {\bm 1}_3$, with the orbit constant $C$ mentioned above being given by $C=\kappa^{-1}\tan\theta_j(\kappa^2\sin^2\phi_j+\cos^2\phi_j)^{1/2}$. The expressions for $G_1-G_4$ in (\ref{inertcorr_Re}) are given in Appendix~\ref{App:AngVel}. 

\begin{figure}
	\centering
    \begin{subfigure}{0.49\textwidth}
		\includegraphics[width=\textwidth]{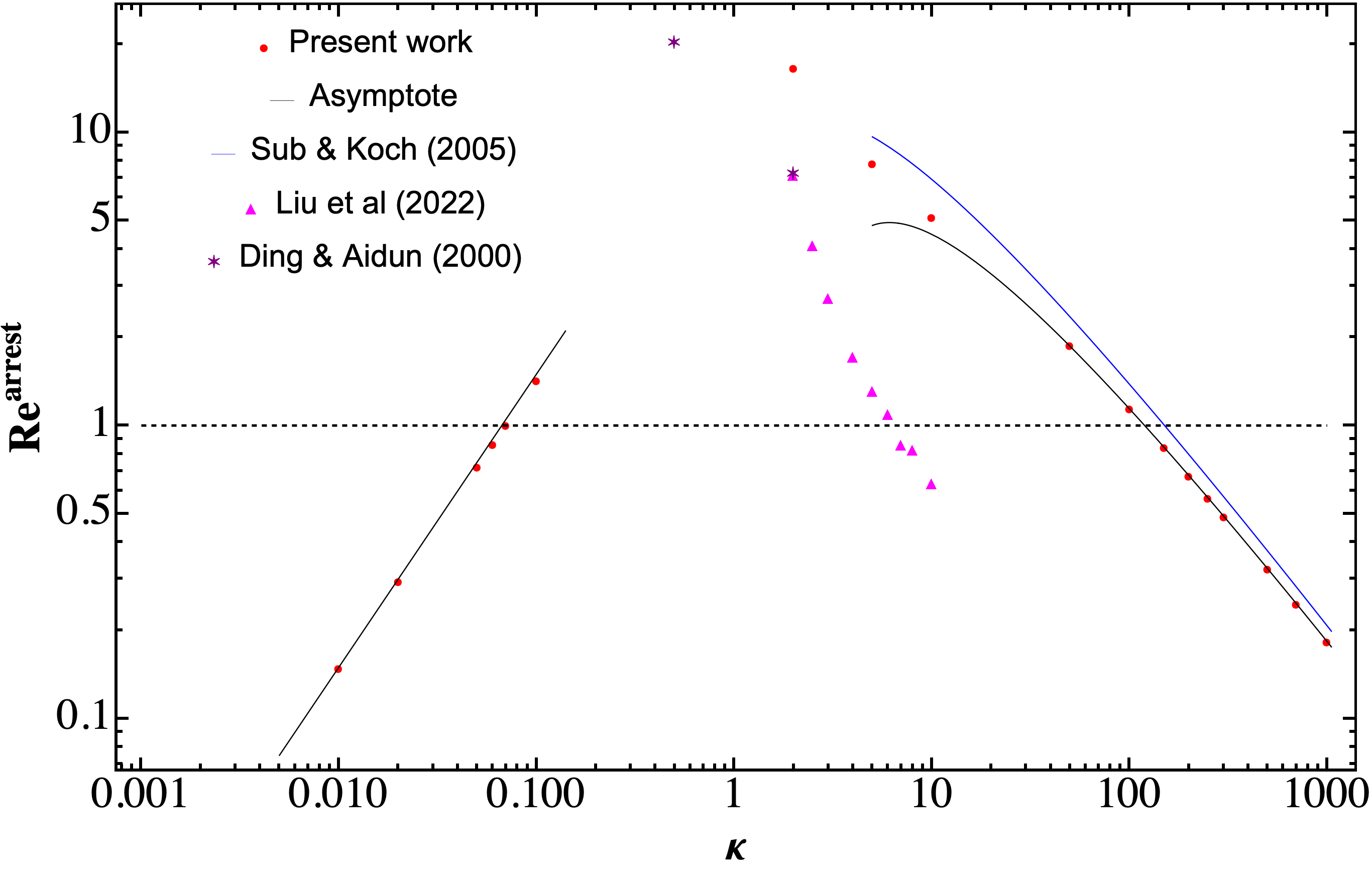}
		\caption{}
	\end{subfigure}
    \begin{subfigure}{0.49\textwidth}
		\includegraphics[width=\textwidth]{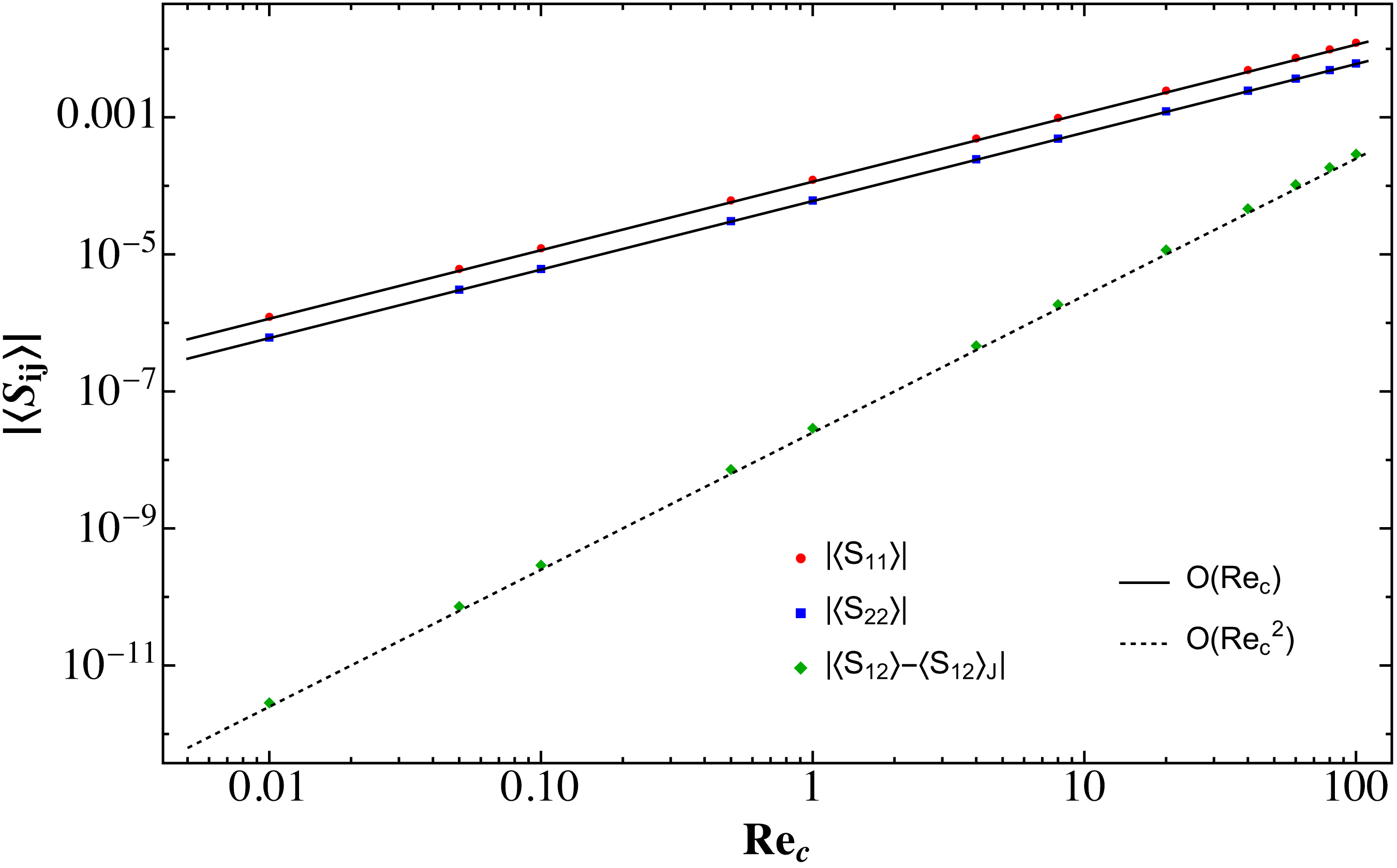}
		\caption{}
	\end{subfigure}
    \begin{subfigure}{0.49\textwidth}
		\includegraphics[width=\textwidth]{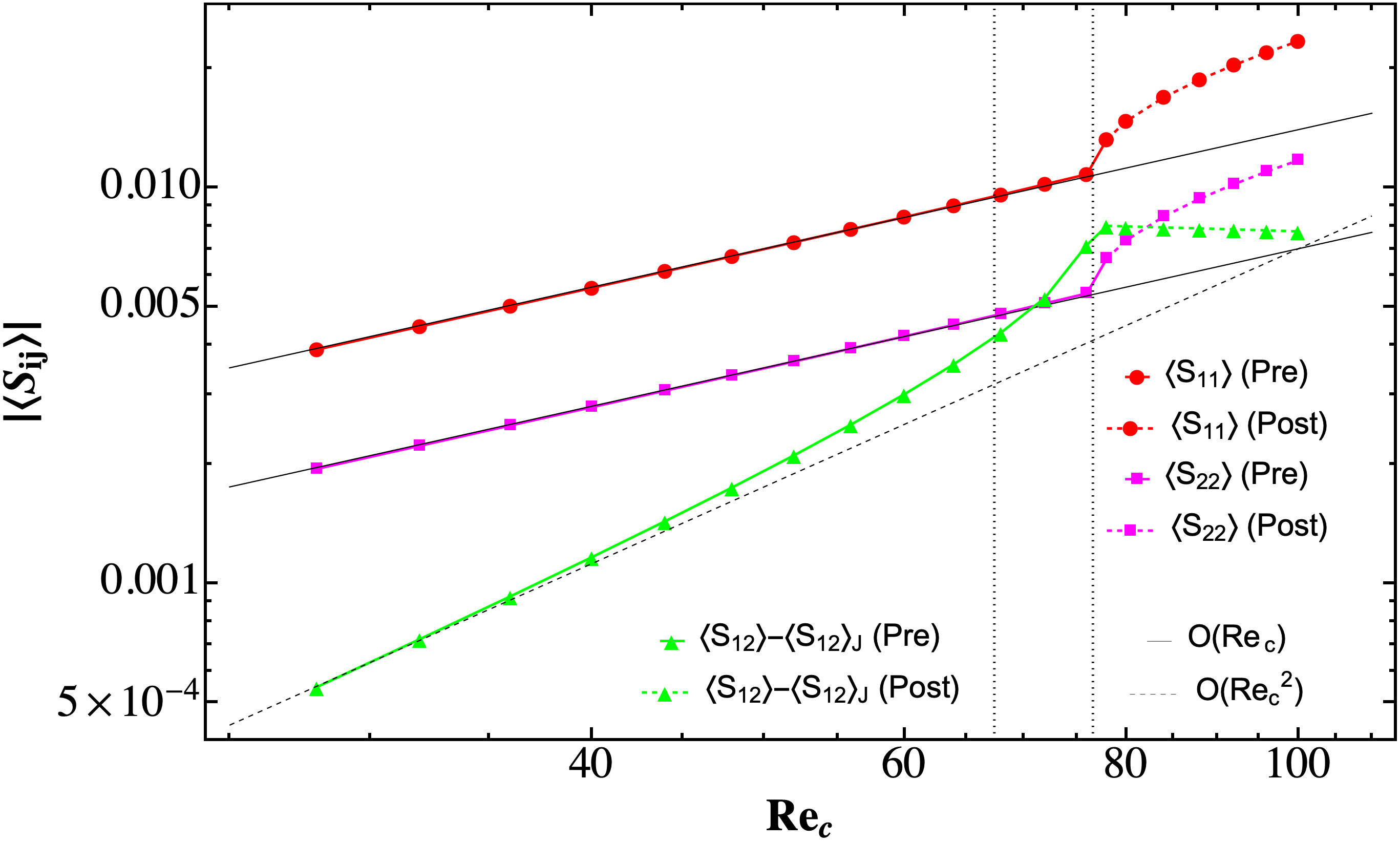}
		\caption{}
	\end{subfigure}
    \begin{subfigure}{0.49\textwidth}
		\includegraphics[width=\textwidth]{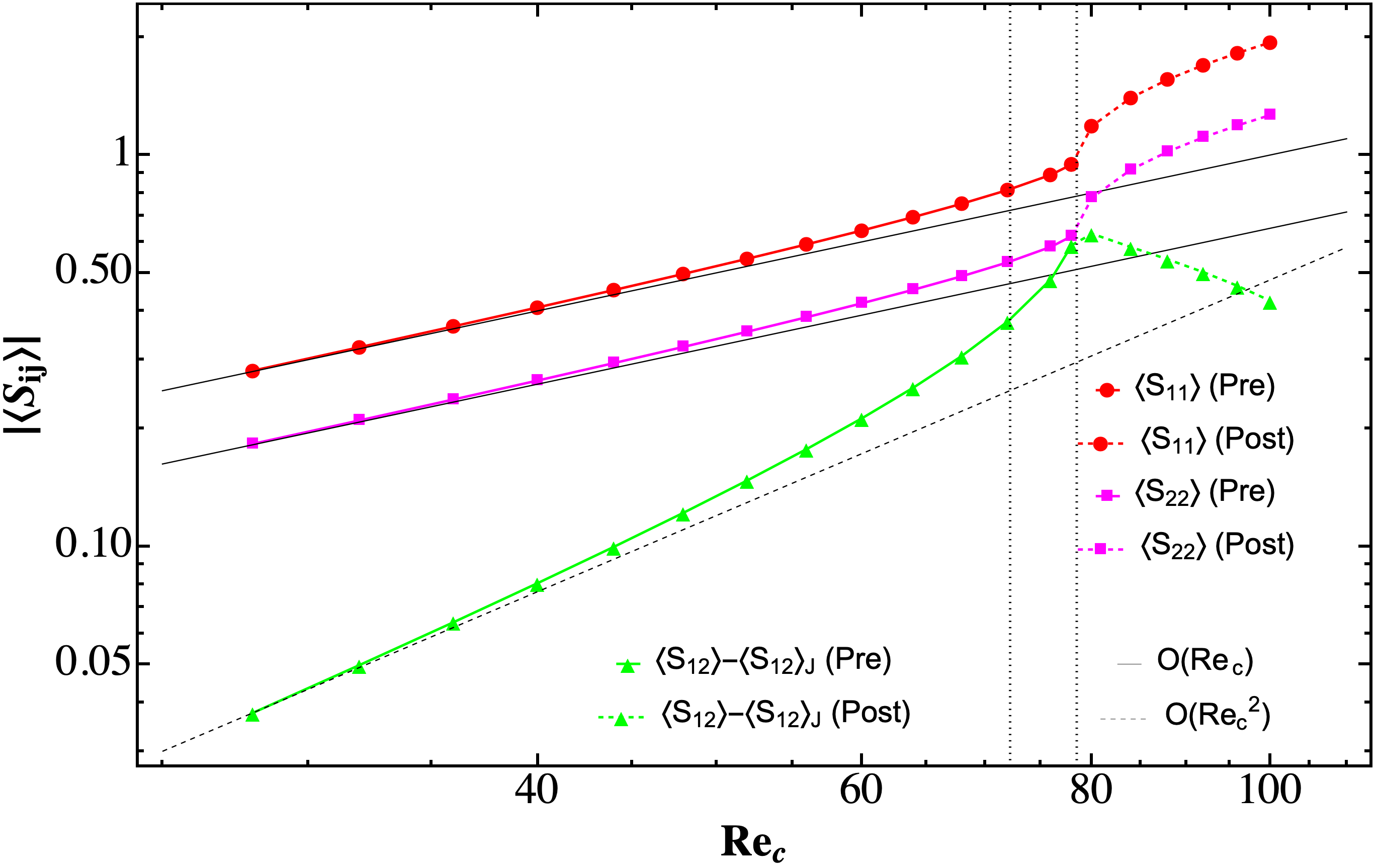}
		\caption{}
	\end{subfigure}
	\caption{(a) 
    The analytical predictions for $Re^\text{arrest}$ plotted as a function of $\kappa$; these include
    (1) $(30\ln(2\kappa)-45)/\kappa$, and its leading logarithmic order approximation $30\ln \kappa/\kappa$, for $\kappa\gg1$; and (2) $15\kappa$ for $\kappa\ll1$. 
    The region below the horizontal dashed line, $Re^\text{arrest}=1$, is a rough indication of the range of validity of a small-$Re_p$ theory. Also included are results for $Re^\text{arrest}$, obtained from the lattice-Boltzmann simulations of \citet{aidun2000} and \citet{liu2022}, and shown as purple stars and pink triangles, respectively; the lone result for $\kappa < 1$ is for an oblate spheroid with $\kappa = 0.5$, while the remaining ones pertain to 2D simulations involving elliptic cylinders in plane Couette flow. The various components of the inertially averaged stresslet, $\langle S_{ij}\rangle$, plotted as a function of $Re_c$, for a fixed $s$: (b) $\kappa=10$, $s = 0.3$ - in this case, the diagonal components $\langle S_{11} \rangle$ and $\langle S_{22} \rangle$ exhibit a linear scaling, while the inertial correction to the Jeffery-averaged stresslet, $\langle S_{12} \rangle_J$ exhibits a quadratic scaling over the entire range of $Re_c$, (c) $\kappa=200$, $s = 0.06$ - this location becomes rotation arrested for $Re_c\gtrsim 78$, and (d) $\kappa=0.05$, $s = 0.04$ - which becomes rotation arrested for $Re_c\gtrsim 80$. For both (c) and (d), there is a departure from the aforementioned $O(Re_c)$ and $O(Re_c^2)$ scalings when $Re_c$ approaches $Re_c^\text{arrest}$.}
	\label{fig:Rearrest&stresslet}
\end{figure}

While $\dot{\phi}_j^{(0)} \sim O(1)\,\forall\,\phi_j$ for spheroids with $\kappa \sim O(1)$, this is no longer true for spheroids of extreme-aspect-ratios\,($\kappa,\kappa^{-1}\gg1$), where one may differentiate between aligned and non-aligned phases of rotation. $\dot{\phi}_j^{(0)}$ remains order unity only during the latter phase, but becomes asymptotically small during periods of near-alignment - it is $O(\kappa^{-2})$\,($O(\kappa^2))$ for a prolate\,(oblate) spheroid nearly aligned with the flow\,(gradient) direction. To see this and the resulting amplification of inertial effects, consider the limiting forms of \eqref{eq:Fullangularvelocity} for small $\phi_j$ ($\pi/2-\phi_j$) for a prolate\,(oblate) spheroid:
\begin{subequations}
\begin{align}
\dot{\phi}_j\beta^{-1}&= -\left(\phi_j^2+\frac{1}{\kappa^2}\right)+ \frac{2\phi_j Re_s}{30\ln (2\kappa)-45}+ O\left(\frac{\phi_j^2}{\kappa^2}\right)\,\,\,\text{for}\,\,\,\, \kappa\gg 1,\\
\dot{\phi}_j\beta^{-1}&=-\left[\left(\frac{\pi}{2}-\phi_j\right)^2+\kappa^2\right]-\frac{2 Re_s}{15}\left(\frac{\pi}{2}-\phi_j\right)+O\left(\left(\frac{\pi}{2}-\phi_j\right)^2\kappa^2\right)\,\,\,\text{for}\,\,\,\, \kappa^{-1}\gg 1,
\end{align} \label{eq:AngVelNearArrest}
\end{subequations}
where the limiting forms of the aspect-ratio functions in (\ref{eq:Jeffangvel}) and (\ref{inertcorr_Re}) for large and small $\kappa$ have been used in (\ref{eq:AngVelNearArrest}a) and (\ref{eq:AngVelNearArrest}b), respectively.
The Jeffery contribution, corresponding to the first bracketed term in the above equations, is seen to be asymptotically small for extreme aspect ratios. It is $O(\kappa^{-2})$ for $\phi_j \sim O(\kappa^{-1})$ in (\ref{eq:AngVelNearArrest}a), and $O(\kappa^2)$ for $\frac{\pi}{2}-\phi_j \sim O(\kappa)$ in (\ref{eq:AngVelNearArrest}b), leading to a prolate\,(oblate) spheroid spending a time of $O(\kappa H/V_{max})$\,($O(\kappa^{-1}H/V_{max}$)) in the near-aligned phase, which then controls the Jeffery time period. The inertial terms are $O(Re_s/\kappa\ln \kappa)$ and $O(Re_s\kappa)$ for the prolate and oblate cases, and therefore, become comparable to the leading Jeffery contributions during the aligned phase, leading to a threshold $Re_s$ value beyond which the spheroid adopts a stationary orientation. The rotation arrest occurs at $Re_s=Re^\text{arrest}=(30\ln(2\kappa)-45)/\kappa$ for slender prolate spheroids\,(this threshold, to leading logarithmic order, was predicted by \cite{subkoch2005}). An analogous arrest occurs for flat disks at $Re_s=Re^\text{arrest}=15\kappa$~\citep{navaneeth2017}. Note that both thresholds fall within the purview of a small-$Re_p$ theory in the respective limits.
 Figure~\ref{fig:Rearrest&stresslet} plots all three asymptotes mentioned above as a function of $\kappa$
; the region below the horizontal dashed line\,(corresponding to $Re_s = 1$) may be regarded as the regime of validity of the theory.

There is numerical support for inertia-induced rotation arrest, although not in the specific case of three dimensions and with $Re_p$ small, which precludes a quantitative comparison. 
\cite{aidun2000} performed lattice Boltzmann simulations for vorticity-aligned circular and elliptic cylinders, the latter with a cross-sectional aspect ratio of $2$, and an oblate spheroid with $\kappa=0.5$, all of which were suspended freely in plane Couette flow with $0.05\leq\lambda\leq0.25$; $Re_s=Re_p$ for plane Couette flow. While the rotation rate of the circular cylinder approached zero only for $Re_s \rightarrow \infty$, rotation of the elliptic cylinder was arrested for $Re^\text{arrest}\simeq 7.25$~(defined using the semi-major axis), with the period of rotation diverging as $(Re^\text{arrest}-Re_s)^{-1/2}$ for $Re_s \rightarrow Re^\text{arrest}$. The tumbling oblate spheroid, which appears to be the only 3D case examined in literature, attained a steady orientation at $Re^\text{arrest}\simeq20.25$. More recently, \cite{liu2022} studied the motion of elliptic cylinders with cross-sectional aspect ratios in the interval $[2,10]$, using the immersed boundary-lattice Boltzmann method, again in plane Couette flow, and found $Re^\text{arrest}$ to decrease from $7.2$ down to $0.64$ over the aspect-ratio interval above; this inverse dependence is qualitatively similar to that for a slender fibre, albeit without the additional logarithmic factor that comes from slender body theory in three dimensions. The $Re_s^\text{arrest}$ values of \cite{liu2022}, and those for the oblate spheroid in \cite{aidun2000}, are plotted alongside the analytical predictions in Figure~\ref{fig:Rearrest&stresslet}a, with the comparison highlighting the amplified effect of inertia in 2D. It is worth reiterating that the increase in time period for $\kappa \sim O(1)$ is controlled by the $O(Re_s^{3/2})$ correction in (\ref{eq:Fullangularvelocity})\citep{navaneeth2017}. For such $\kappa$'s, one may expand the time period integral\,($\textstyle\int_{2\pi}^0 d\phi_j \dot{\phi}_j^{-1}$) for small $Re_s$, and the $O(Re_s)$ correction involving $\dot{\phi}_j^{(1)}$ causes equal amounts of slowing down and speeding up in the extensional and compressional quadrants, respectively, leading to no net change in the tumble period. This is no longer true for extreme aspect ratios where, as seen above, $\dot{\phi}_j^{(1)}$ becomes comparable to $\dot{\phi}_j^{(0)}$, and causes a divergence of the time period\,(rotation arrest). In our emphasis on extreme aspect ratios, we have neglected the $O(Re_s^{3/2})$ contribution since it remains smaller than $\dot{\phi}_j^{(0)}$ regardless of $\kappa$. 

We now analyze the implications of rotation arrest from the perspective of calculating the inertial lift velocity. This is important since, as will be seen in section \ref{sec:results} below, onset of rotation arrest heralds a deviation of the actual lift profile from the Jeffery-averaged approximation obtained in \citet{anand2023Jeff}. As indicated earlier in this section, the nature of the migration calculation depends on $Re_c$, with a semi-analytical approach based on a reciprocal theorem formulation being possible when $Re_c \ll 1$. In contrast, one resorts to a numerical implementation of a matched asymptotic expansions approach for $Re \gtrsim 1$. The discussion above shows that arrest for slender spheroids occurs at $Re^\text{arrest} \approx 30\ln \kappa/\kappa$. Using the maximum value, $Re_s = 4Re_p$, corresponding to the channel walls\,($s=0,1$), and $Re_p = \lambda^2 Re_c$, the arrest criterion above corresponds to $Re_c \approx 15\ln \kappa/(2\lambda^2\kappa)$. The assumption of small $Re_c$ would therefore imply $(15\ln\kappa/2\kappa)^{1/2}\ll\lambda\ll1$; for thin oblate spheroids, one obtains $(15\kappa/4)^{1/2} \ll \lambda \ll 1$. Assuming $\lambda = 0.1$, $(15\ln\kappa/2\kappa)^{1/2}$ dips below $0.1$ only for $\kappa \gtrsim 5000$! The oblate-side criterion yields $\kappa \lesssim 0.002$. Clearly, the assumption of a small $Re_c$ is very restrictive, and most migration scenarios involving spheroids of extreme aspect ratios correspond to $Re_c \gtrsim O(1)$. In light of this, rather than having separate sections devoted to the analyses for small and order unity $Re_c$, as in \citet{anand2023Jeff}, we relegate the details of the small-$Re_c$ analysis to Appendix \ref{App:LiftSmallRec}, and only focus on $Re_c \gtrsim O(1)$ in the next section.

\section{Inertial lift velocity for $Re_c\gtrsim O(1)$}\label{sec:Reclarge}

As discussed in detail in \cite{anand2023Jeff}, for small $\lambda$, the time scale over which the spheroid migrates to its equilibrium position, under the influence of inertia, is asymptotically larger than that required for it to complete a single tumble. This implies that the leading order migration is the result of an averaged lift velocity, the average being over the orientations sampled during a tumble period. Therefore, defining the time averaging operator 
as:
\begin{align}
\langle f(\phi_j)\rangle=\frac{1}{T}\int_{2\pi}^{0}f(\phi_j) \dfrac{d\phi_j}{\dot{\phi}_j},
\label{eq:NonJeffTimeavg}
\end{align}
with $\dot{\phi}_j$ given by (\ref{eq:Fullangularvelocity}), and with $T=\int_{2\pi}^{0}\frac{d\phi_j}{\dot{\phi}_j}$ being the tumble time period, 
we average equations (\ref{eq:NSOuterEqn}a,b) and the boundary conditions (\ref{eq:NSOuterBC}a-c), for a fixed ${\bm R}$. This gives:
\begin{subequations} 
	\begin{align} 
	\frac{\partial^2 \langle U_i\rangle}{\partial R_j^2}-\frac{\partial \langle P\rangle}{\partial R_i}-\langle U_2\rangle(\beta+&2\gamma R_2 Re_c^{-1/2})\delta_{i1}-(\beta R_2+\gamma R_2^2 Re_c^{-1/2})\frac{\partial \langle U_i\rangle}{\partial R_1} 
 =\beta \langle S_{ij}\rangle\frac{\partial\delta(\bm{R})}{\partial R_j},\\
	\frac{\partial{\langle U}_i\rangle}{\partial x_i}&=0,
	\end{align} \label{eq:NSOuterEqnTA}
\end{subequations}
where $\langle U_i\rangle$ satisfies:
\begin{subequations}
\begin{align}
    \langle U_i\rangle &\sim \frac{3\beta R_i \langle S_{jm}\rangle R_j R_m}{4\pi R^5} \text{ for } \bm{R}\rightarrow 0,\\
	\langle U_i\rangle &\rightarrow 0 \text{ for } R_1,R_3\rightarrow \infty\,(R_2\hspace*{0.05in}\text{fixed}),\\
	\langle U_i\rangle &=0 \text{ at } R_2=-s Re_c^{1/2}, (1-s)Re_c^{1/2},
\end{align} \label{eq:NSOuterBCTA} 
\end{subequations}
with the translational\,($d\bm{U}_p/dt$) and unsteady\,($\frac{\partial U_i}{\partial t}$) acceleration terms in (\ref{eq:NSOuterEqn}a) integrating to zero over a single period.

Approximating $\dot{\phi}_j$ by $\dot{\phi}_j^{(0)}$ in the time-averaging operation leads to the Jeffery-averaged analysis in \cite{anand2023Jeff}, with $T=T_\text{jeff}=\int_{2\pi}^{0}d\phi_j/\dot{\phi}_j^{(0)}= 2\pi\beta^{-1}(\kappa+\kappa^{-1})$ being the non-dimensional Jeffery time period. Note that the 
Jeffery time-averaging operator is independent of the ambient flow owing to $\dot{\phi}_j^{(0)}$ being only a function of $\kappa$, and the Jeffery-averaged stresslet must therefore be linear in $\bm E$, as for a sphere. This implies that, for the plane Poiseuille profile, the forcing in (\ref{eq:NSOuterEqnTA}a)  must only involve $\langle S_{12} \rangle_J$, and is of the form $\beta\langle S_{12} \rangle_J \delta(R_3)[\delta(R_1)\delta'(R_2){\bm 1}_1 + \delta'(R_1)\delta(R_2){\bm 1}_2]$
\,(the subscript $J$ denoting a Jeffery-averaged quantity), with $\langle S_{12}\rangle_J$ defined by:
\begin{align}
\langle S_{12}\rangle_J &= \frac{ (3 A_1+A_3)\kappa+2 A_2 (\kappa ^2+1)}{4(\kappa +1)^2}. \label{eq:S12J}
\end{align}
The above led \citet{anand2023Jeff} to conclude that the Jeffery-averaged lift velocity of a tumbling spheroid, which is a linear functional of $\langle U_2\rangle$\,(see (\ref{eq:VpHinch}) below), can be written as:
\begin{align}
\langle V_p \rangle_J(s;Re_c,\kappa) = \frac{\langle S_{12}\rangle_J(\kappa)}{(-10\pi/3)} V_p^\text{sphere}(s;Re_c),
\label{eq:VpJeffAvg}
\end{align}
and differs from that for a sphere only by a multiplicative function of $\kappa$; the full expression is given in Appendix \ref{App:LiftSmallRec} - see \eqref{eq:VpJeffAvgd} therein. The relation (\ref{eq:VpJeffAvg}) implies that the Jeffery-averaged equilibria, corresponding to $\langle V_p \rangle_J = 0$, are independent of $\kappa$, being identical to those for a sphere~\citep{schonberghinch1989} at the same $Re_c$. Note that the proportionality relation in (\ref{eq:VpJeffAvg}) is always true for a spinning spheroid within the small-$Re_p$ framework used here. This is due to the steady velocity disturbance, even though the rate of spin is altered by inertia\,(at $O(Re_p^{3/2}$)). Since inertia stabilizes the spinning mode for oblate spheroids with $\kappa \gtrsim 0.14$, our focus is on oblate spheroids with aspect ratios less than this threshold value.

Intuitively speaking, one expects a Jeffery-averaged analysis to give the inertial lift for sufficiently small $Re_c$, for spheroids of order unity aspect ratios, since the inertial correction to the spheroid angular velocity is asymptotically small in this limit. The Jeffery-averaged lift is $O(Re_p)$ for $Re_c \ll 1$, owing to $V_p^\text{sphere}$ in (\ref{eq:VpJeffAvg}) being of this order. It turns out, however, that there is a second contribution at the same order, and that crucially relies  on the alteration of spheroid rotation by weak inertia. This latter $O(Re_p)$ contribution was missed out in \citet{anand2023Jeff}, owing to the authors writing down the lift velocity in terms of a Jeffery-averaged volume integral involving the $O(Re_p)$ inertial acceleration. The correction to the Jeffery-averaged acceleration, arising from $\dot{\phi}_j^{(1)}$ in the time-averaging operator defined in (\ref{eq:NonJeffTimeavg}), would only contribute to the lift velocity at $O(Re_p^2)$, and this smaller contribution was accordingly neglected in the said analysis. While the neglect is justified for the inertial acceleration, the second $O(Re_p)$ contribution to the inertial lift comes from a different source. As discussed in more detail in Appendix \ref{App:LiftSmallRec}, the origin of this contribution is the $O(Re_p)$ term in the stresslet forcing in (\ref{eq:NSOuterEqnTA}a) that arises on using an inertial, rather than Jeffery, orientation average. The explicit form of this forcing is given below, and involves both the diagonal components of the time-averaged stresslet and an inertial correction to $\langle S_{12} \rangle_J$ that appears in (\ref{eq:VpJeffAvg}) - see discussion in the next paragraph. Despite being formally of the same order, this second contribution turns out to involve an extremely small numerical pre-factor for $\kappa\sim O(1)$, as a result of which the Jeffery-averaged lift obtained in \citet{anand2023Jeff} remains a very accurate approximation of the inertial lift, for all $Re_c$, for $\kappa \sim O(1)$; for instance, see Figure~\ref{fig:JeffvsInertprofiles}a in section \ref{sec:results} below. Notwithstanding its smallness, the calculation in Appendix \ref{App:LiftSmallRec} does show that a $\kappa$-dependence of the equilibrium position arises even for $\kappa \sim O(1)$ and $Re_c \ll 1$, with the equilibrium positions of spheroids being marginally closer to the centerline than those of spheres. The direction of shift\,(towards the centerline) remains consistent with that for spheroids of extreme aspect ratios analyzed in section \ref{sec:results}. Further, and importantly, the analysis shows that the essential mechanism underlying this shift is that of a repulsive interaction with an image stresslet induced by the wall boundary conditions, and this is considered in more detail in section \ref{sec:conclusion}.




We now move beyond the Jeffery-averaged analysis, considering $Re_c \gtrsim O(1)$. With the inertial angular velocity included in the time-averaging operator, the forcing function in (\ref{eq:NSOuterEqnTA}a) also involves the diagonal elements in addition to $\langle S_{12}\rangle$, being of the form:
\begin{align}
\langle S_{ij}\rangle\frac{\partial\delta(\bm{R})}{\partial R_j} = &\langle S_{12}\rangle\delta(R_3)[\delta(R_1)\delta'(R_2)\bm{1}_1+\delta'(R_1)\delta(R_2)\bm{1}_2]+\langle S_{11}\rangle\delta'(R_1)\delta(R_2)\delta(R_3)\bm{1}_1 \nonumber \\
&+\langle S_{22}\rangle\delta(R_1)\delta'(R_2)\delta(R_3)\bm{1}_2+\langle S_{33}\rangle\delta(R_1)\delta(R_2)\delta'(R_3)\bm{1}_3. \label{eq:inertialstressletforcing}
\end{align}
On account of the $Re_p$-dependent orientation dynamics, the averaged stresslets in the above expression are functions of $Re_c$\,(for $\lambda$ fixed), and the $Re_c$-dependence for the different components is shown in Figs.\ref{fig:Rearrest&stresslet}b, c and d for different $\kappa$. For $\kappa = 10$, one is far from rotation arrest, and the quadratic scaling of $\langle S_{12}\rangle - \langle S_{12}\rangle_J$ and the linear scaling of $\langle S_{11}\rangle$ and $\langle S_{22}\rangle$, over the entire range of $Re_c$, merely reflect the corresponding small-$Re_p$ scalings for $\lambda$ fixed. Since $\langle S_{12} \rangle$ contributes to the suspension viscosity~\citep{batchelor1970}, the $O(Re_c^2)$ scaling of $\langle S_{12}\rangle - \langle S_{12}\rangle_J$ in Figure~\ref{fig:Rearrest&stresslet} implies a regular $O(Re_p^2)$ correction to the shear viscosity of a suspension of spheroids, arising from an inertial alteration of the orientation distribution. Earlier calculations for a non-interacting inertial suspension of spherical particles\,(drops) have shown that the leading order correction to $S_{12}$ is $O(Re_p^{3/2})$, and originates in the outer region~\citep{lin1970,sub2011}; this singular outer-region correction is not relevant for small $Re_c$ owing to the channel wall being less than the inertial screening length.

Having clarified the functional form of the stresslet forcing, we proceed to solve (\ref{eq:NSOuterEqnTA}a-\ref{eq:NSOuterBCTA}c) for the disturbance fields, which involves Fourier transforming along the spatially homogeneous flow and vorticity directions\,\citep{schonberghinch1989,anand2023Jeff}. Defining the partial Fourier transform:
\begin{align}
\hat{f}(k_1,R_2,k_3)=\int_{-\infty}^\infty\int_{-\infty}^\infty dR_1 dR_3 \,f(R_1,R_2,R_3)\,e^{\iota (k_1 R_1+k_3 R_3)},
\end{align}
one obtains, after some manipulations, the following coupled ordinary differential equations for $\langle\hat{T}\rangle$ and $\langle\hat{U}_2\rangle$:
\begin{subequations}
	\begin{align}
	\frac{d^2 \hat{\langle T\rangle}}{dR_2^2}-k_\perp^2\hat{\langle T\rangle}&=2\iota k_1 \langle\hat{U}_2\rangle(\beta+2\gamma R_2 Re_c^{-1/2}),\\
	\frac{d^2\langle\hat{U}_2\rangle}{dR_2^2}-k_\perp^2\langle\hat{U}_2\rangle&=\frac{d\hat{\langle T\rangle}}{dR_2}-\iota k_1\langle\hat{U}_2\rangle(\beta R_2+\gamma R_2^2 Re_c^{-1/2}),
	\end{align} \label{eq:FTODEs}
\end{subequations}
in the intervals $-sRe_c^{\frac{1}{2}} < R_2 < 0$ and $0< R_2 < (1-s)Re_c^{\frac{1}{2}}$. Here, $\langle \hat{T}\rangle =\langle \hat{P}\rangle + \beta\langle S_{22}\rangle\delta(R_2)$. is a modified (transformed)\,pressure field. (\ref{eq:FTODEs}ab) are subject to the following conditions:
\begin{subequations}
\begin{align} 
\langle\hat{U}_2\rangle\sim &\frac{\iota\beta\langle S_{12}\rangle e^{-k_\perp R_2} R_2\,\,k_1}{2}+\frac{\beta\,\,R_2\,\, e^{-k_\perp R_2} \left((k_1^2+2 k_3^2)\langle S_{22}\rangle -(k_1^2-k_3^2)\langle S_{11}\rangle\right)}{4 k_\perp}\nonumber\\
&\text{ for } k_1,k_3\to\infty \text{ and } R_2\rightarrow 0,\\
\langle\hat{U}_2\rangle&=\frac{d\langle\hat{U}_2\rangle}{dR_2}= 0 \text{ at } R_2=-s\,Re_c^{1/2},(1-s)\,Re_c^{1/2},
\end{align} \label{eq:FTBC}
\end{subequations}
where $k_\perp^2=k_1^2+k_3^2$. (\ref{eq:FTBC}a) is the matching requirement to the transformed stresslet field in the inner region, that now accounts for the modified stresslet tensor resulting from the inertial orientation dynamics, while (\ref{eq:FTBC}b) is the no-slip boundary condition at the channel walls. The stresslet forcing in (\ref{eq:NSOuterEqnTA}a) enters the problem in two ways. The second term in $\hat{T}$ above serves to absorb the localized contribution induced by the most singular component\,($\propto \langle S_{22} \rangle \delta''(R_2)$) of the aforementioned stresslet forcing that enters (\ref{eq:FTODEs}a). The remaining part of the stresslet forcing enters via jump conditions at $R_2 = 0$, that serve to connect the $\hat{\langle T\rangle}$ and $\hat{\langle U_2 \rangle}$-fields for positive and negative $R_2$. The detailed calculation in Appendix \ref{App:LiftSmallRec} leads to the following jump conditions:
\begin{subequations}
	\begin{align}
	\hat{\langle T\rangle}^+ (k_1,0,k_3) -\hat{\langle T\rangle}^-(k_1,0,k_3) &=2\iota k_1 \beta\langle S_{12}\rangle,\\
	\frac{d\hat{\langle T\rangle}^+}{dR_2}(k_1,0,k_3)-\frac{d\hat{\langle T\rangle}^-}{dR_2}(k_1,0,k_3) &= \beta \big[(k_1^2-k_3^2)\langle S_{11}\rangle-(k_1^2+2 k_3^2)\langle S_{22}\rangle\big],\\
	\langle\hat{U}_2\rangle^+(k_1,0,k_3)&=\langle\hat{U}_2\rangle^- (k_1,0,k_3), \\
	\frac{d\langle\hat{U}_2\rangle^+}{dR_2}(k_1,0,k_3)-\frac{d\langle\hat{U}_2\rangle^-}{dR_2}(k_1,0,k_3) &=\iota k_1 \beta \langle S_{12}\rangle,
	\end{align} \label{eq:FTJumpCondns}
\end{subequations}
where the superscripts `+' and `-' denote the transformed fields for $R_2>0$ and $R_2<0$, respectively. One solves the system (\ref{eq:FTODEs}ab), along with the boundary conditions (\ref{eq:FTBC}a-b) and jump conditions (\ref{eq:FTJumpCondns}a-d), using a shooting method originally used by \cite{schonberghinch1989} for a spherical particle, and described in more detail by \cite{anand2023Jeff} as part of their Jeffery-averaged analysis for spheroids. 

The limiting form of $\langle U_2 \rangle$ in the matching region $\bm{R}\ll 1$, obtained from the above procedure, is the sum of the singular stresslet contribution given in (\ref{eq:FTBC}a), and a uniform flow along the gradient\,(cross-stream) direction that is a consequence of fluid inertia. The neutrally buoyant spheroid being force-free is convected by the latter uniform flow which therefore equals the inertial lift velocity. It may be determined as the limit of the inverse transform for $\bm{R}\to0$:
\begin{align}
\langle V_p\rangle&= \frac{Re_p}{4\pi^2}\,\,\Re\left\{\int_{-\infty}^\infty\int_{-\infty}^\infty \,\langle\hat{U}_2\rangle^\pm(k_1,0,k_3)\,dk_1\, dk_3\right\},
 \label{eq:VpHinch}
\end{align}  
Here, $\Re\{.\}$ denotes taking the real part, which eliminates the singular, albeit purely imaginary, stresslet contribution\,(although the stresslet as a whole vanishes anyway in the limit $R_2 \rightarrow 0$; see (\ref{eq:FTBC}a)). Note that one may use either $\langle\hat{U}_2\rangle^+$ or $\langle\hat{U}_2\rangle^-$ in (\ref{eq:VpHinch}) on account of the continuity across $R_2=0$ (see (\ref{eq:FTJumpCondns}c)). 



\section{Results}\label{sec:results}

\begin{figure}
	\centering
	\begin{subfigure}{0.49\textwidth}
		\includegraphics[width=\textwidth]{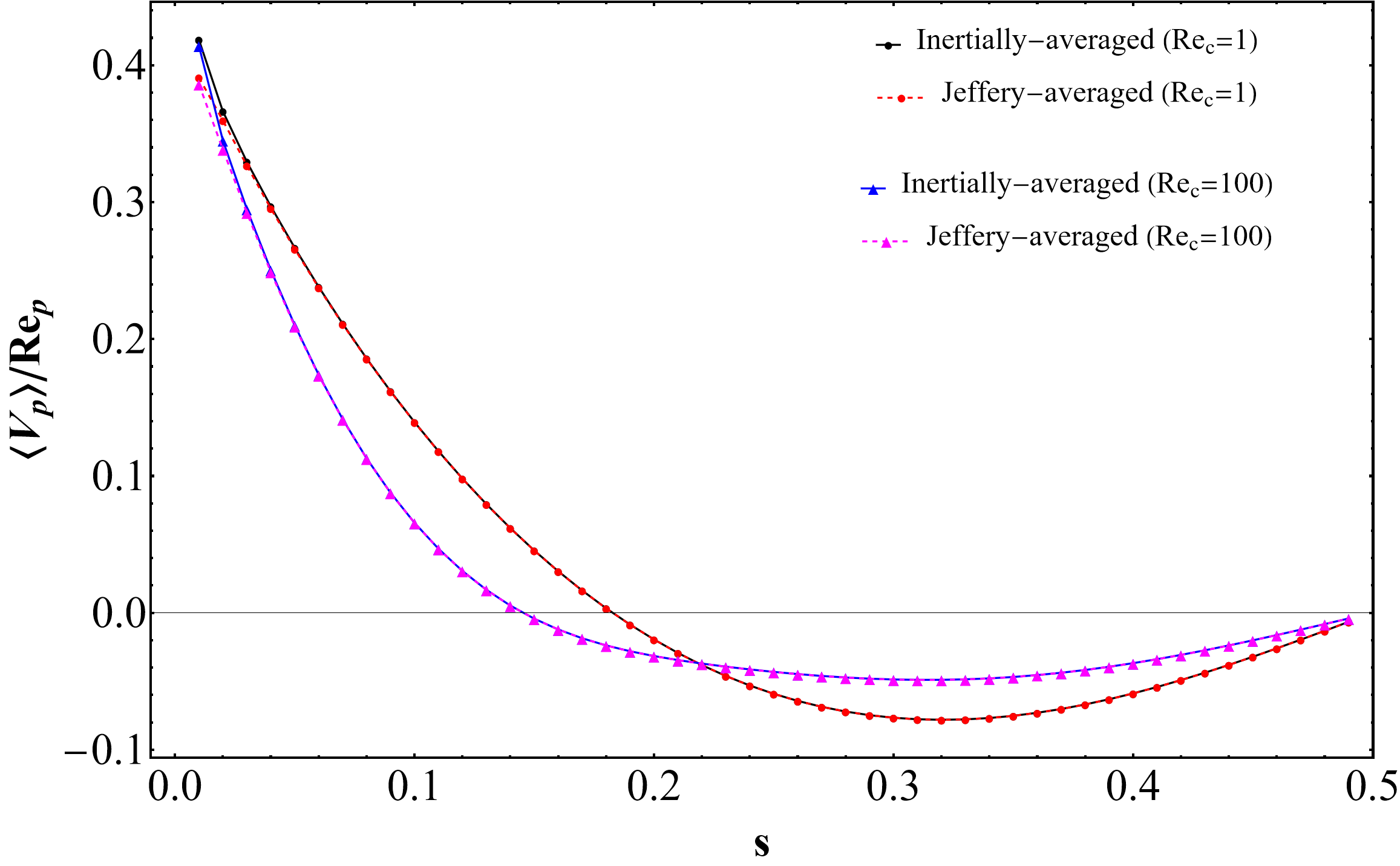}
		\caption{}
	\end{subfigure}
	\begin{subfigure}{0.49\textwidth}
		 \includegraphics[width=\textwidth]{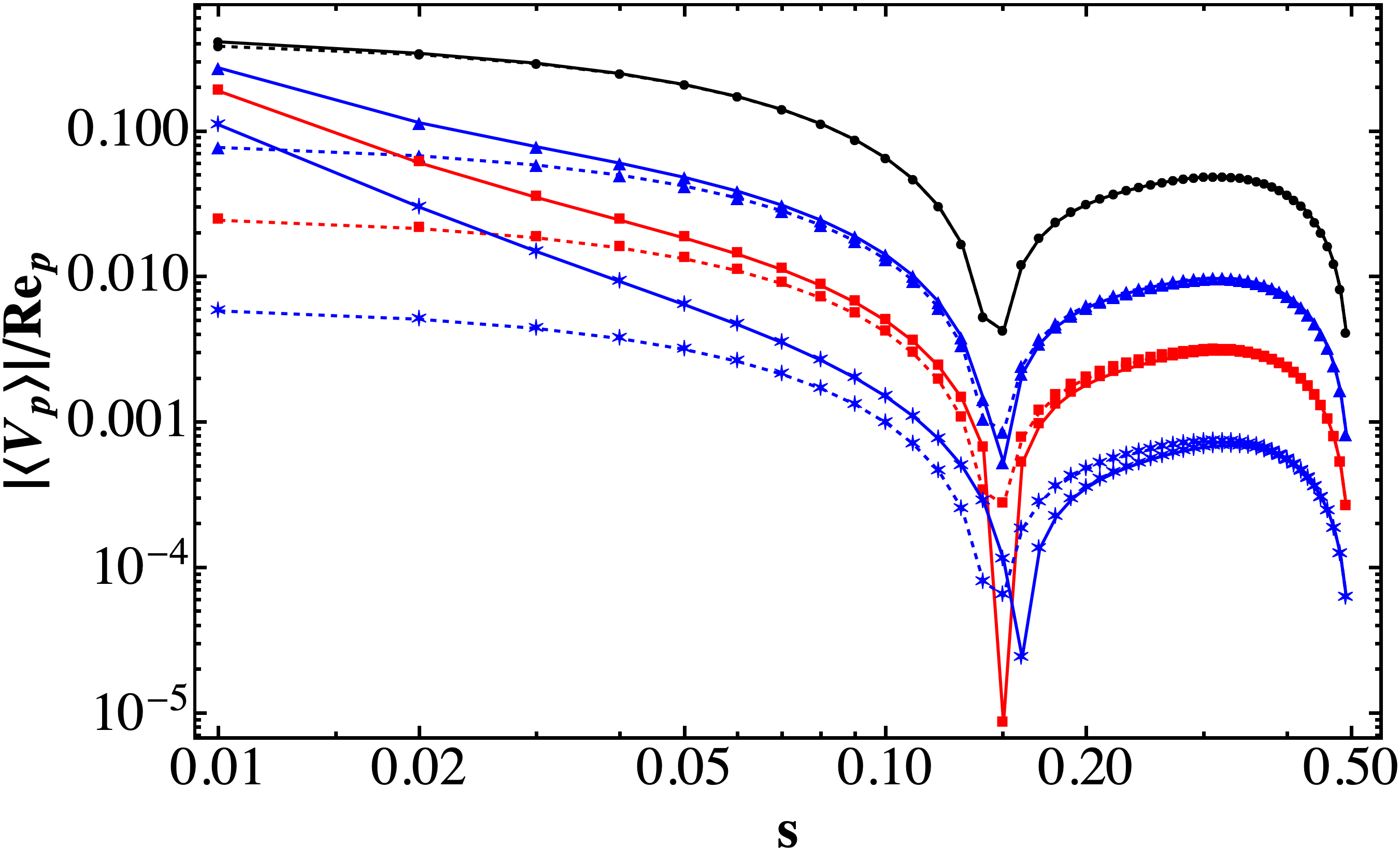}
		\caption{}
	\end{subfigure}
		\caption{(a) Comparison of lift velocity profiles, obtained using the inertial and Jeffery orientation distributions, for tumbling prolate spheroids with $\kappa=2$; $\lambda=0.05$ and $Re_c = 1$ \& $100$. The two  profiles exhibit excellent agreement over the entire range of $s$ examined. (b) A comparison of inertially-averaged~(solid curves) and Jeffery-averaged~(dashed curves) lift velocity profiles for $\lambda = 0.05, Re_c=100$, for prolate spheroids of different aspect ratios: $\kappa=2$~(circles), $5$~(triangles),  $10$~(squares) and $25$~(stars). With increasing $\kappa$, the inertially-averaged profiles start to deviate from the Jeffery-averaged ones; the deviation begins in the near-wall region, and for $\kappa = 25$, leads to a small shift of the equilibrium position towards the centerline.
        }
		\label{fig:JeffvsInertprofiles}
\end{figure}

We begin with Figure~\ref{fig:JeffvsInertprofiles} which compares inertial lift profiles obtained using the shooting method framework described in the previous section, with those obtained using the Jeffery-averaged analysis of \cite{anand2023Jeff}, for a tumbling prolate spheroid with $\kappa = 2$. Note that the said profiles, and those that follow, are only plotted in the half-channel domain\,($s \in (0,1/2)$) on account of the profile anti-symmetry about the channel centerline. Figure~\ref{fig:JeffvsInertprofiles}a shows the two sets of results to be in excellent agreement for both $Re_c = 1$ and $100$, for $\lambda = 0.05$. This confirms the smallness of the non-Jeffery contribution to the lift for $s \gtrsim 0.02$, for spheroids with order unity aspect ratios, provided $Re_p$ remains small. Figure~\ref{fig:JeffvsInertprofiles}b shows the inertial lift profiles for $Re_c = 100$, along with the corresponding Jeffery-averaged ones, for prolate spheroids of different aspect ratios. The plot is on a logarithmic scale, so the equilibrium positions\,(zero crossings) manifest as sharp dips\,(to negative infinity) in the lift profile. With increasing $\kappa$, the actual lift profile begins to deviate from the Jeffery-averaged approximation. This deviation is seen clearly in the near-wall region, with the small-$s$ interval corresponding to this region being magnified by the logarithmic scale. The actual lift profile exhibits a divergence for $s \rightarrow 0$, in contrast to the plateauing behavior of the Jeffery-averaged profile; this divergence is explained in Appendix \ref{App:LiftSmallRec}. More importantly, the equilibrium position also begins to deviate away from that of the Jeffery-averaged profile, whose equilibrium location is the same as that for a sphere, at the chosen $Re_c$~\citep{anand2023Jeff}. The deviation is evident for the highest aspect ratio in Figure~\ref{fig:JeffvsInertprofiles}b, $\kappa =25$, where the actual equilibrium is a little closer to the centerline than the Jeffery-averaged one.


\begin{figure}
	\centering
	\begin{subfigure}{0.49\textwidth}
		\includegraphics[width=\textwidth]{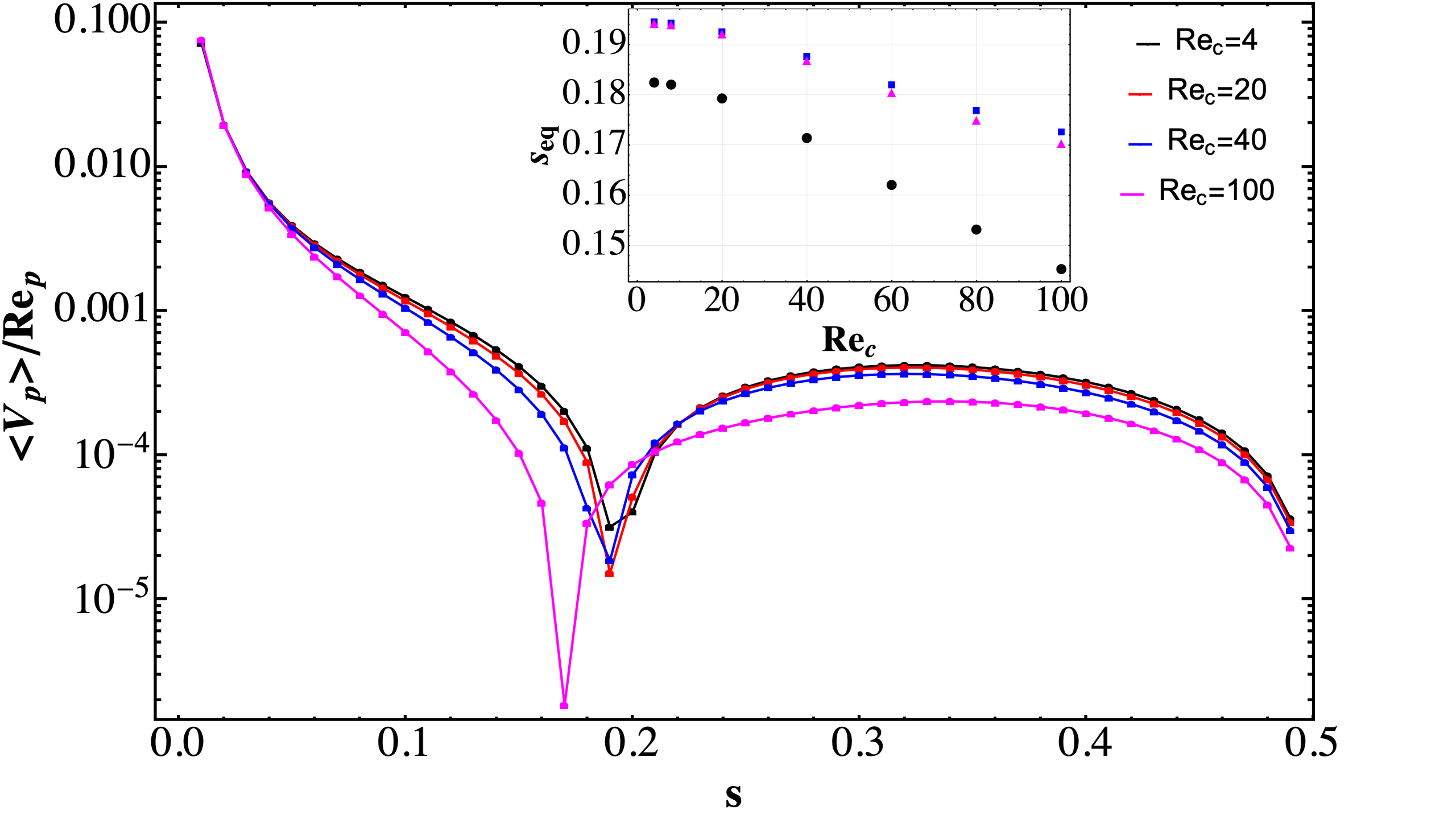}
		\caption{$\kappa=50$}
	\end{subfigure}
	\begin{subfigure}{0.49\textwidth}
		\includegraphics[width=\textwidth]{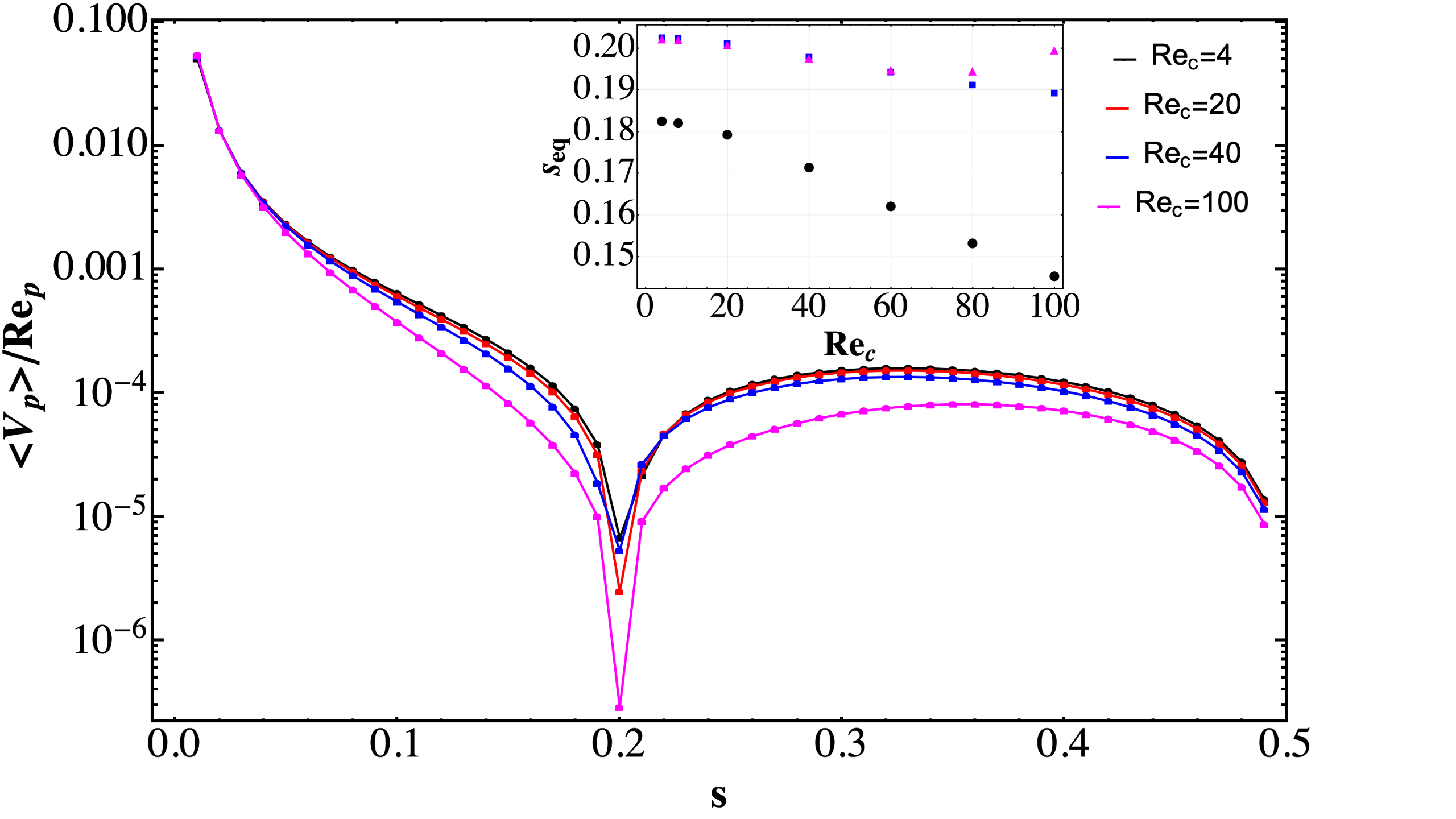}
		\caption{$\kappa=100$}
	\end{subfigure}
		\caption{Lift velocity profiles for tumbling prolate spheroids with $\kappa=50$ and $100$, for $\lambda=0.05$, for different $Re_c$. The insets compare the locus of the equilibrium locations for the actual profiles\,(pink triangles) to both the Jeffery-averaged predictions\,(black circles), and to these predictions corrected by a small-$Re_c$ contribution\,(blue squares) that accounts for the inertial perturbation to the orientation dynamics. This latter correction is calculated in Appendix \ref{App:LiftSmallRec}; see (\ref{eq:Vp2ndInert}) and (\ref{eq:N_defn}) therein.}
		\label{fig:InertialAvgdLiftkappagt1}
\end{figure}

Next, in Figures~\ref{fig:InertialAvgdLiftkappagt1}a and b, we present the inertial lift profiles for a pair of relatively slender tumbling spheroids with $\kappa=50$ and $100$, for different $Re_c$, for $\lambda = 0.05$. The plots are on a semi-log scale, so the zero crossings again appear as dips to negative infinity.  In the inset of Figure \ref{fig:InertialAvgdLiftkappagt1}a, for $\kappa=50$, the equilibrium location\,($s_{eq}$) is seen to move towards the channel wall\,($s=0$) with increasing $Re_c$, for both the Jeffery-averaged\,(black circles) and actual profiles\,(pink triangles); although, the wallward movement of the actual equilibrium is slower than the Jeffery-averaged one. Both the main plot and the inset in Figure \ref{fig:InertialAvgdLiftkappagt1}b show that, for $\kappa=100$, the actual equilibrium position remains virtually at the same location with increasing $Re_c$. In contrast, the Jeffery-averaged equilibrium shown in the inset, for comparison purposes, is independent of $\kappa$, and therefore, moves towards the wall with increasing $Re_c$, in the exact same manner as for $\kappa = 50$. Thus, for $\kappa = 100$, the wallward movement induced by channel-scale inertial forces\,(that is, by increasing $Re_c$), appears to be almost exactly compensated by the movement towards the centerline 
induced by inertial alteration of the spheroid orientation dynamics. On the whole, Figures~\ref{fig:InertialAvgdLiftkappagt1}a and b hint at the $\kappa$-dependence of inertia-lift-induced equilibria that is absent in the Jeffery-averaged profiles, and that we examine in greater detail below. 

\begin{figure}
	\centering
	\begin{subfigure}{0.49\textwidth}
		\includegraphics[width=\textwidth]{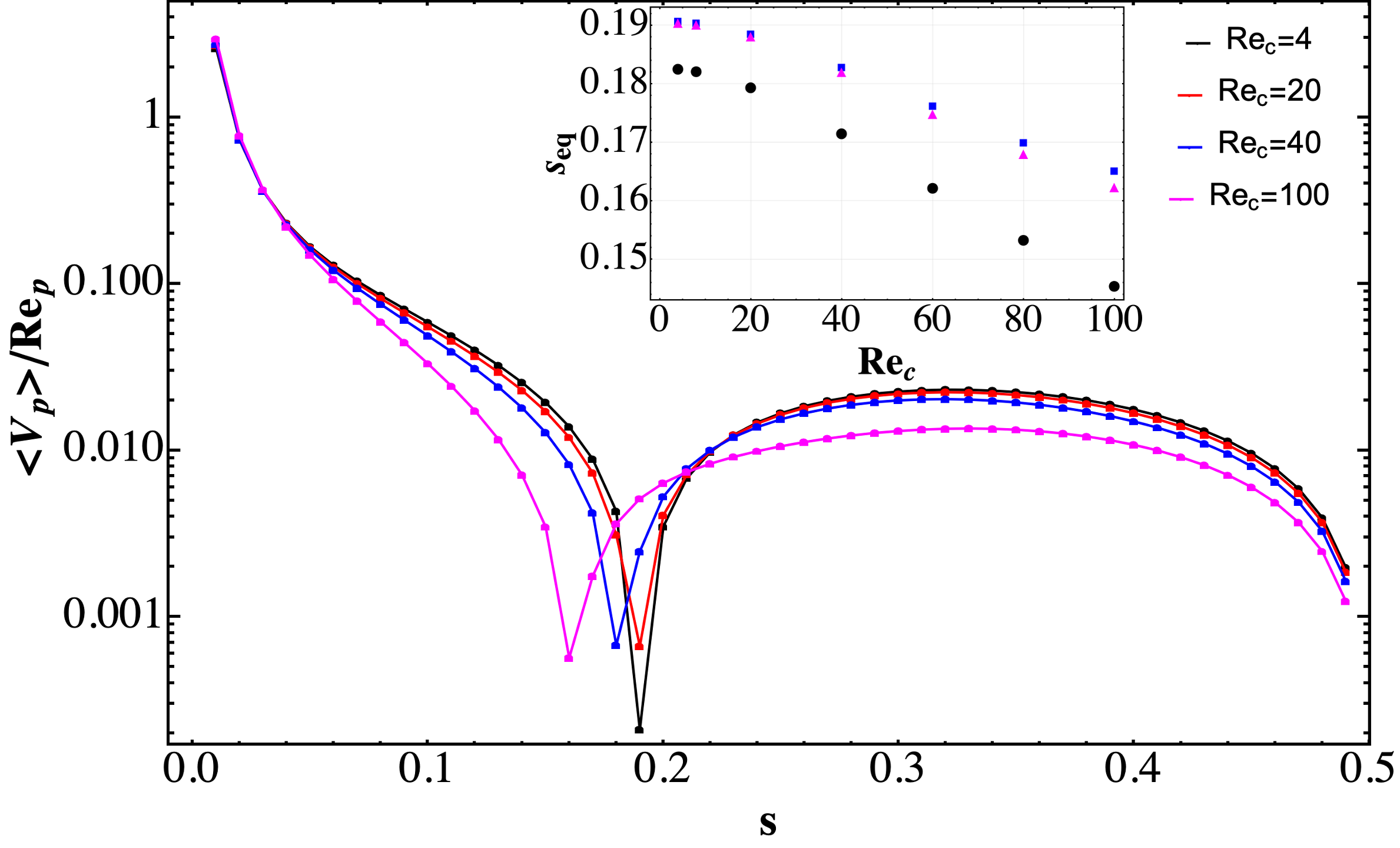}
		\caption{$\kappa=0.1$}
	\end{subfigure}
	\begin{subfigure}{0.49\textwidth}
		\includegraphics[width=\textwidth]{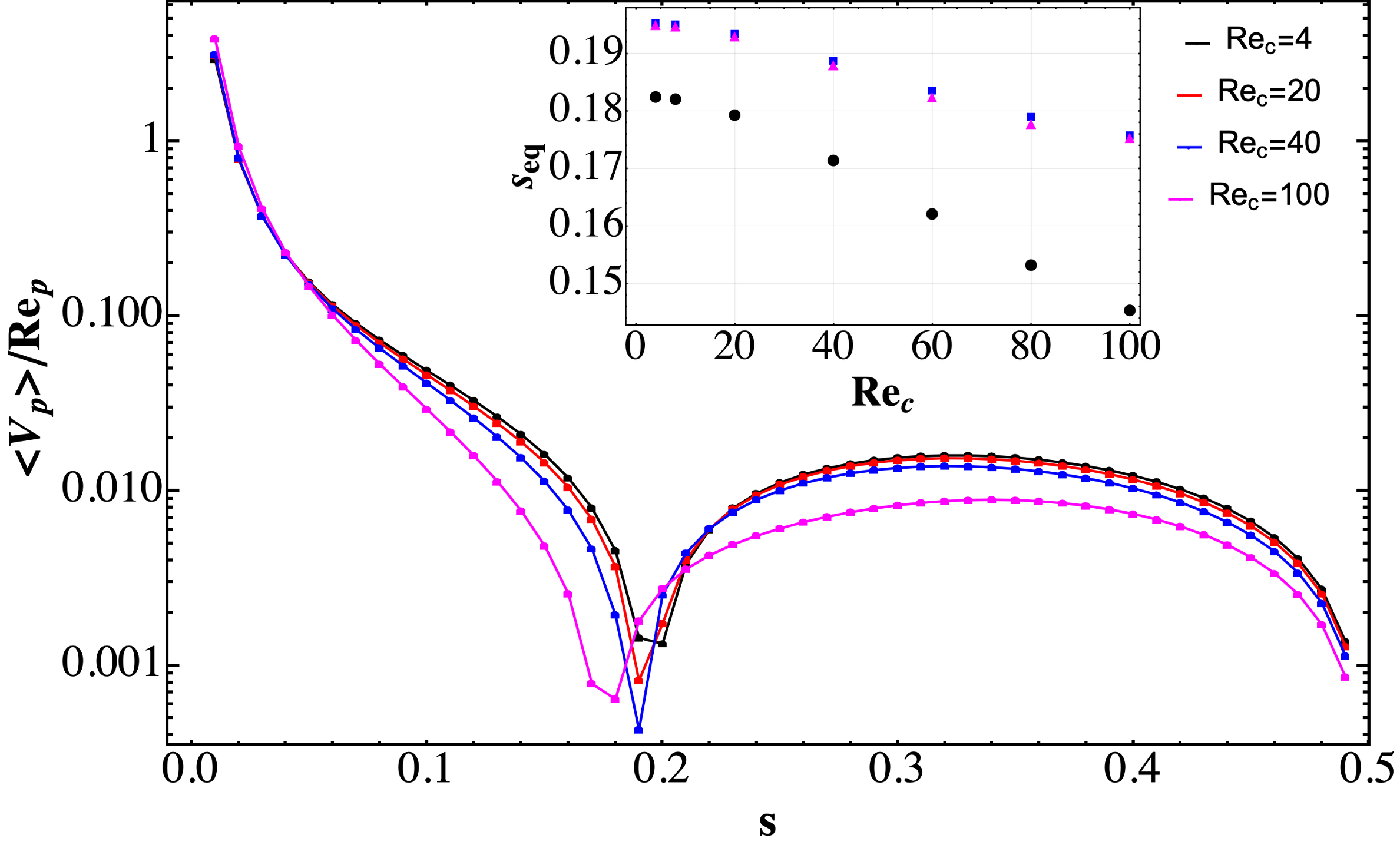}
		\caption{$\kappa=0.07$}
	\end{subfigure}
		\caption{Lift velocity profiles for tumbling oblate spheroids with $\kappa=0.1$ and $0.07$, for $\lambda=0.05$, for different $Re_c$. The insets compare the locus of the equilibrium locations for the actual profiles\,(pink triangles) to both the Jeffery-averaged equilibra\,(black circles), and to these equilibria corrected by a small-$Re_c$ contribution\,(blue squares), induced by inertial orientation dynamics, calculated in Appendix \ref{App:LiftSmallRec}; see (\ref{eq:Vp2ndInert}) and (\ref{eq:N_defn}) therein.}
		\label{fig:InertialAvgdLiftkappalt1}
\end{figure}

Figures~\ref{fig:InertialAvgdLiftkappalt1}a and b show the inertial lift profiles, for $\lambda = 0.05$, for a pair of thin oblate spheroids with $\kappa = 0.1$ and $\kappa = 0.07$, that again tumble stably in the flow-gradient plane. Similar to the prolate case, for both aspect ratios shown, the wallward movement of the actual equilibrium is slower than the Jeffery-averaged one, with the degree of slowing down being greater for the smaller aspect ratio\,($\kappa = 0.07$). Note that, in all of the insets in Figures~\ref{fig:InertialAvgdLiftkappagt1}a-\ref{fig:InertialAvgdLiftkappalt1}b, there remains a small but finite difference between the actual and Jeffery-averaged equilibrium locations even as $Re_c \rightarrow 0$. This is due to a contribution derived in Appendix \ref{App:LiftSmallRec}, that accounts for the deviation from Jeffery orientation dynamics even for small but finite $Re_c$, and that was discussed in $\S$\ref{sec:form}; see (\ref{eq:Vp2ndInert}) and (\ref{eq:N_defn}). The third sequence of symbols\,(blue squares), that appears in the insets, corrects the Jeffery-averaged equilibrium by including this second small-$Re_c$ contribution, which then leads to better agreement with the actual equilibrium locus. The agreement is considerably improved for $\kappa = 50$, $0.1$ and $0.07$, extending up to $Re_c \approx 100$. For prolate spheroids with $\kappa = 100$, however, agreement is restricted to $Re_c \lesssim 60$ in Figure~\ref{fig:InertialAvgdLiftkappagt1}b, mainly because the actual equilibrium locus exhibits an upturn, reflecting a transition from a wallward movement to a movement towards the centerline, for $Re_c > 60$. This disagreement will be amplified for the more extreme aspect ratios examined below.

The spheroids in both Figures~\ref{fig:InertialAvgdLiftkappagt1}a-b and Figures~\ref{fig:InertialAvgdLiftkappalt1}a-b tumble, regardless of their location within the channel. Tumbling ceases when the local shear Reynolds number $Re_s$ exceeds the relevant arrest threshold. The small $Re_p$ assumption implies that $Re_s$ must also be less than unity. Since $Re_s=|\beta|Re_p = 4|1-2s|Re_p$, with its maximum value, $Re_s=4Re_p$, occurring at the walls, this implies $Re_p \lesssim 1/4$, in turn implying $Re_c \lesssim Re_c^\text{max} = 1/4\lambda^2$. Thus, for a given $\lambda$, the validity of the small-$Re_p$ theory restricts the channel Reynolds number to be less than the aforesaid maximum. For the aspect ratios chosen in Figures~\ref{fig:InertialAvgdLiftkappagt1}a-\ref{fig:InertialAvgdLiftkappalt1}b, even at $Re_c^\text{max} = 1/4\lambda^2$, $Re_s$ remains smaller than the local arrest threshold for all $s$. Thus, in these cases, increasing inertia leads to a slowing down of rotation, but as mentioned above, the spheroids continue to tumble at all locations. 
Arrest first occurs when the maximum $Re_s$, corresponding to the wall locations, equals the threshold. This occurs at a threshold channel Reynolds number, $Re_c^\text{arrest}=Re^\text{arrest}/(4\lambda^2)$, with $Re^\text{arrest}=4Re_p^\text{arrest}$; the latter equals $(30\ln(2\kappa)-45)/\kappa$ and $15\kappa$ for the prolate and oblate cases, respectively, these predictions having been obtained from the ones given in the paragraph below (\ref{eq:AngVelNearArrest}). As $Re_c$ increases beyond the thresholds for either the prolate or oblate case mentioned above, a pair of arrest locations moves into the channel from the two walls. They are symmetrically disposed about the centerline, being given by $s_{\pm}^\text{arrest}=\frac{1}{2} (1 \pm \frac{Re^\text{arrest}}{4 Re_p})=\frac{1}{2}(1 \pm \frac{Re_c^\text{arrest}}{Re_c})$. 
For any $Re_c>Re_c^\text{arrest}$, one has stationary prolate\,(oblate) spheroids nearly aligned with the flow\,(gradient) directions, for $0<s<s_-^\text{arrest}$ and $s_+^\text{arrest}<s<1$; and tumbling spheroids in the central region corresponding to $s_-^\text{arrest}<s<s_+^\text{arrest}$.

\begin{figure}
	\centering
	\begin{subfigure}{0.54\textwidth}
		\includegraphics[width=\textwidth]{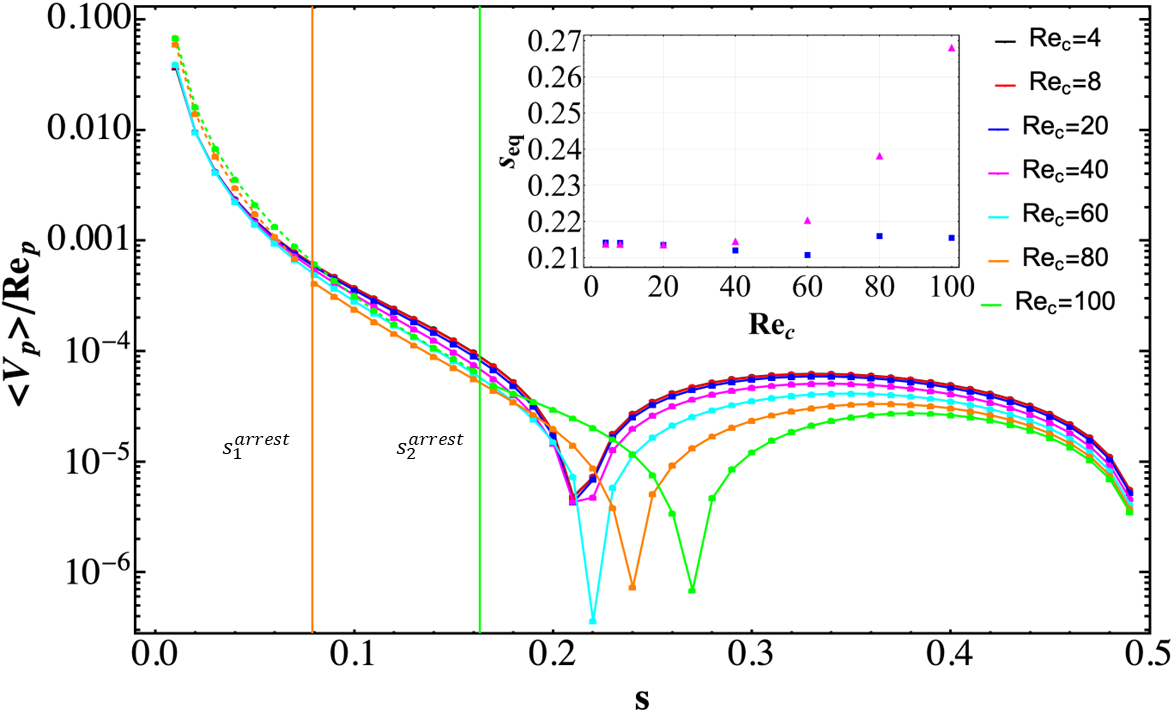}
		\caption{}
	\end{subfigure}
	\begin{subfigure}{0.45\textwidth}
		\includegraphics[width=\textwidth]{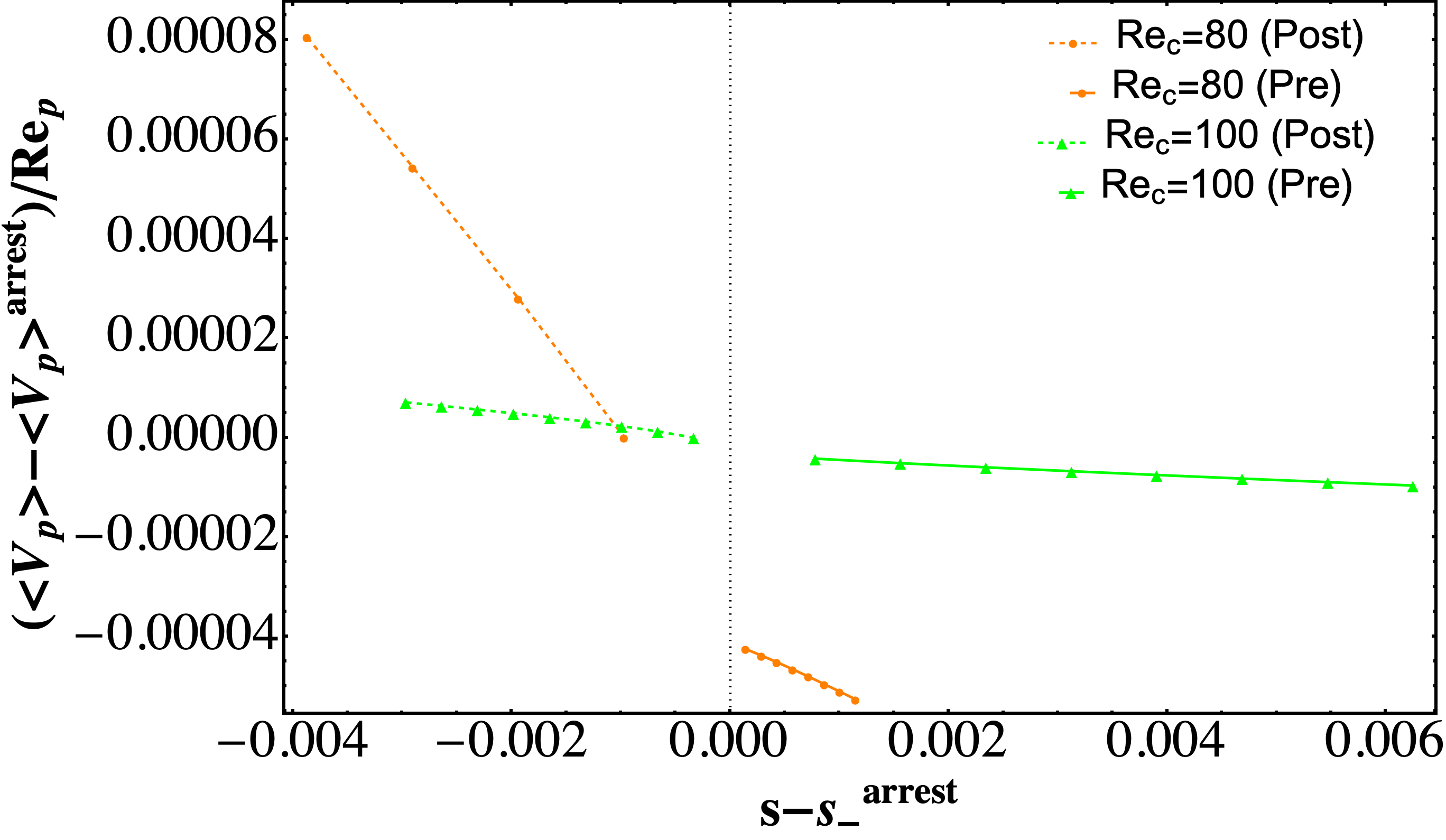}
		\caption{}
	\end{subfigure}
	\caption{(a) Lift velocity profiles for a prolate spheroid with $\kappa=200$, for $\lambda=0.05$, for different $Re_c$; $Re_c^\text{arrest}\approx67.4$. Vertical lines, corresponding to $s^\text{arrest}_-\approx0.079$ and $0.16$, demarcate regions of stationary\,($0 <  s < s^\text{arrest}_-$) and tumbling\,($s^\text{arrest}_- < s < 1/2$) spheroids for $Re_c=80$ and $Re_c=100$, respectively; the rotation-arrested portions of these lift profiles have been depicted using dashed curves. The inset compares the equilibrium locus with its small-$Re_c$ approximation. 
    (b) Magnified views of the lift velocity profiles for $Re_c=80$ and $100$ in the neighbourhood of their arrest locations $s_-^\text{arrest}$; the portions corresponding to tumbling~(solid) and stationary~(dashed) spheroids approach $s = s_-^\text{arrest}$ with differing slopes.}
	\label{fig:Liftkappa200}
\end{figure}

In Figure~\ref{fig:Liftkappa200}a, we plot the time-averaged lift profiles for $\kappa=200$ and $\lambda=0.05$, with $Re_c$ ranging from $4$ to $Re_c^\text{max}\,(=100)$. For the chosen $\kappa$ and $\lambda$, the threshold for the rotation-arrested regions to move into the channel domain is $Re_c^\text{arrest} =67.4$, which falls in the aforementioned $Re_c$-interval. This means that, for $Re_c>67.4$, there will emerge regions of stationary spheroid orientations next to both  walls. At $Re_c=80$, the locations demarcating the tumbling and rotation-arrested regions correspond to  $(s_-^\text{arrest},s_+^\text{arrest}) \equiv (0.079, 0.16)$; at $Re_c=100$, these locations move further away from the walls with $(s_-^\text{arrest},s_+^\text{arrest}) \equiv (0.16,0.84)$. The arrest locations in the lower half of the channel\,($s_-^\text{arrest}$), for the aforementioned pair of $Re_c$'s, are shown as vertical lines in Figure~\ref{fig:Liftkappa200}a. Interestingly, and in contrast to $\kappa = 50$ and $100$ in Figures~\ref{fig:InertialAvgdLiftkappagt1}a and b, the equilibrium locations for these $Re_c$'s exhibit a perceptible shift towards the channel centerline.  Figure~\ref{fig:Liftkappa200}b shows magnified views of the lift profiles for $Re_c = 80$ and $100$, in the immediate neighborhood of the respective arrest locations\,($s_-^\text{arrest} = 0.079$ and $0.16$), highlighting a jump in the profile slope across $s = s_-^\text{arrest}$. It will be seen below that this jump in slope arises from a jump in slope of the stresslet forcing, as one transitions from the tumbling to the rotation-arrested region. Similar to Figures~\ref{fig:InertialAvgdLiftkappagt1} and \ref{fig:InertialAvgdLiftkappalt1}, the inset in Figure~\ref{fig:Liftkappa200}a compares the actual equilibrium locus with the approximate one\,(the Jeffery-averaged estimate plus the small-$Re_c$ correction obtained in Appendix \ref{App:LiftSmallRec}). The two loci deviate starting at $Re_c \approx 40$, with the deviation becoming pronounced once the actual locus turns towards the centerline.

\begin{figure}
	\centering
	\includegraphics[width=0.8\textwidth]{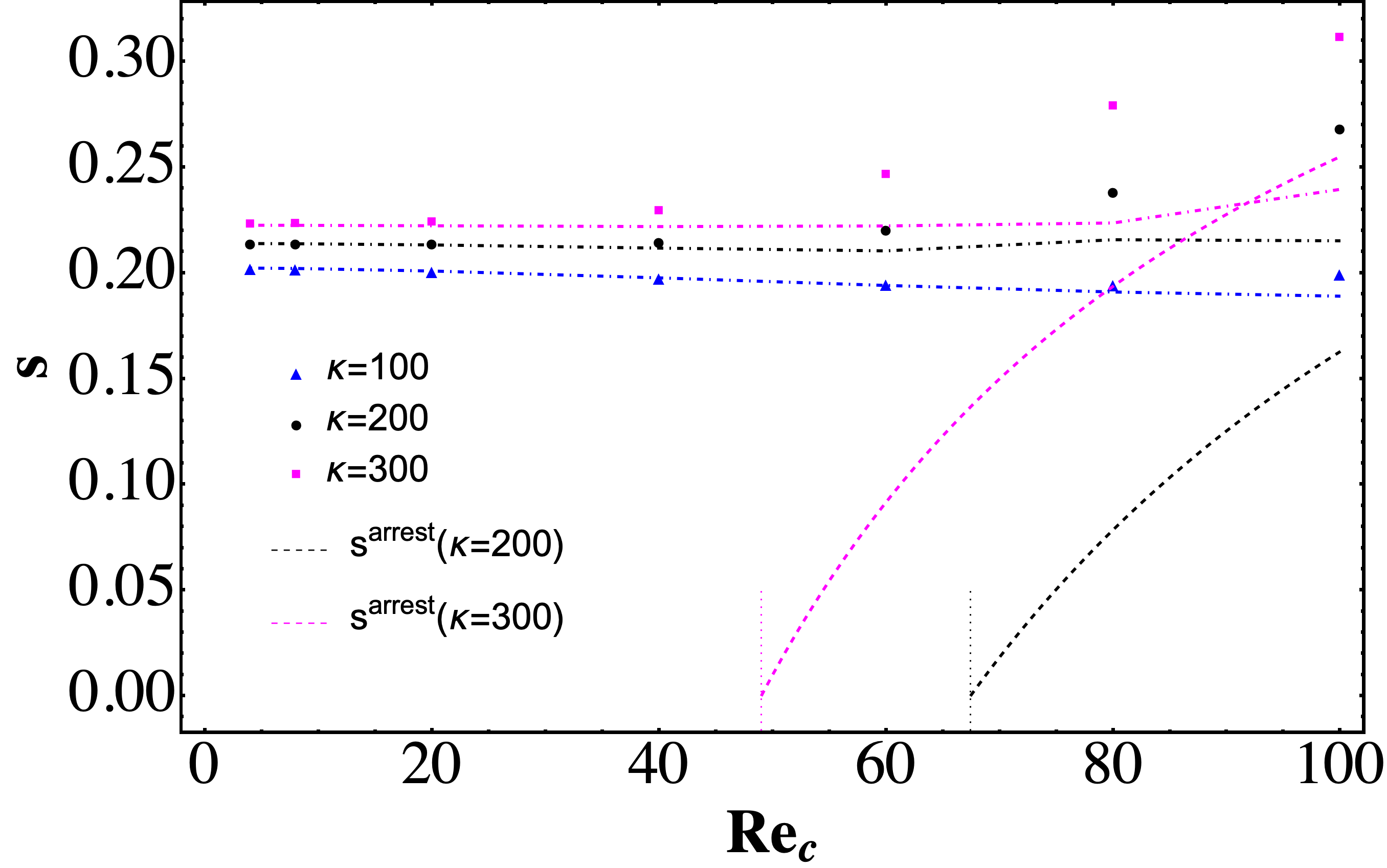}
	\caption{The symbols denote the equilibrium loci for prolate spheroids with $\kappa = 100$~(triangles), $200$~(circles) and $300$~(squares), and for $\lambda=0.05$; dash-dotted curves denote the corresponding small-$Re_c$ approximations. Rotation-arrest loci for $\kappa = 200$ and $300$ appear as dashed curves that start from $s=0$\,(the $Re_c$-axis) at $Re_c^\text{arrest}$, and asymptote to $s=1/2$ for $Re_c \rightarrow \infty$; $Re_c^\text{arrest}\approx67.4$ for $\kappa=200$ and $\approx 49$ for $\kappa=300$.
    }
	\label{fig:LociArrestProlate}
\end{figure}

The movement of the equilibrium position towards the centerline with increasing $Re_c$, for slender spheroids, is more clearly seen in Figure~\ref{fig:LociArrestProlate}, which plots the equilibrium location closer to the lower wall\,($s=0$), as a function of $Re_c$, for $\kappa=100$, $200$ and $300$, and for $\lambda=0.05$; note that the lift profiles for $\kappa = 300$ have not been included for brevity. The dash-dotted curves correspond to small-$Re_c$ approximations calculated from combining the Jeffery-averaged approximation with the additional inertial contribution given by (\ref{eq:Vp2ndInert}) and (\ref{eq:N_defn}) in Appendix \ref{App:LiftSmallRec}. For the large aspect ratios in question, the latter contribution becomes significant, leading to a perceptible dependence on $\kappa$ even for $Re_c \rightarrow 0$. Figure~\ref{fig:LociArrestProlate} also includes the rotation-arrest loci for the two larger aspect ratios. These appear as continuous curves starting at $s_-^\text{arrest} = 0$ at $Re_c = Re_c^\text{arrest}$, and asymptote to $s_-^\text{arrest} = 1/2$ for $Re_c \rightarrow \infty$. $Re_c^\text{arrest}$ equals $67.4$ for $\kappa = 200$ as mentioned in the previous paragraph, and is approximately $49$ for $\kappa = 300$; the limit $Re_c\rightarrow \infty$ is, of course, not relevant since the migration analysis pertains to the laminar regime, and one expects a transition to turbulence at $Re_c \approx 2000$. The equilibrium locus for $\kappa = 100$, in Figure~\ref{fig:LociArrestProlate}, is nearly horizontal, as is expected from the lift profiles already seen in Figure~\ref{fig:InertialAvgdLiftkappagt1}b. In contrast, the emergence of rotation-arrested regions for $\kappa = 200$ and $300$, leads to a perceptible upward movement of the locus. For $\kappa = 200$, the equilibrium position is virtually unchanged until $Re_c \approx 60$, similar to $\kappa=100$, but starts moving towards the centerline thereafter. For $\kappa = 300$, the equilibrium position starts off on a higher plateau, and begins moving towards the centerline at a smaller $Re_c$, owing to the earlier onset of rotation arrest; the subsequent trajectory being almost parallel to the rotation-arrest locus. At least, within the framework of our small-$Re_p$ analysis, the equilibrium locus for slender prolate spheroids appears to never cross the rotation-arrest locus, implying that a slender spheroid always tumbles  at its equilibrium position, regardless of $Re_c$ and $\kappa$.

\begin{figure}
	\centering
    \begin{subfigure}{0.57\textwidth}
	\includegraphics[width=\textwidth]{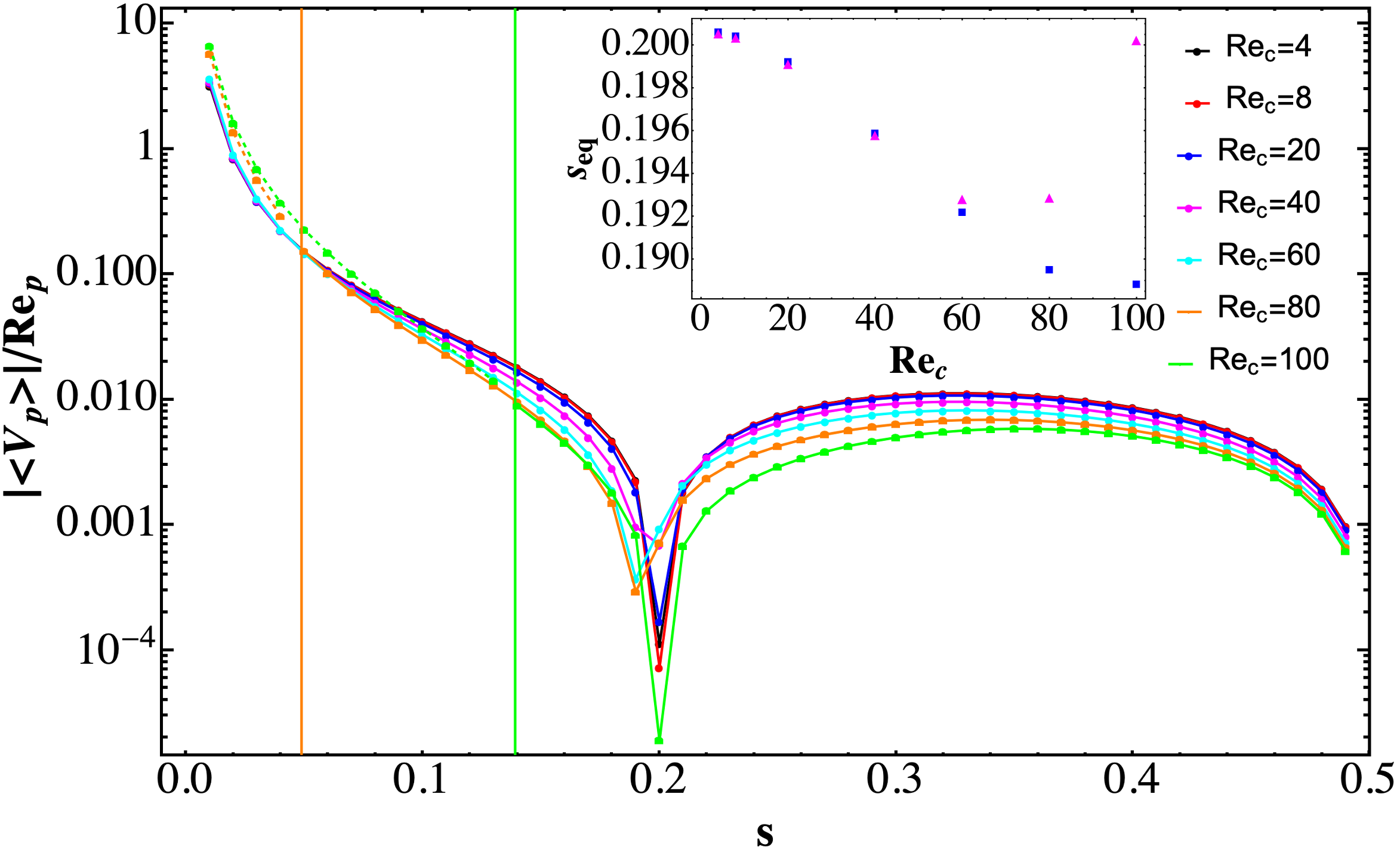}
    \caption{}
    \end{subfigure}
    \begin{subfigure}{0.42\textwidth}
    \includegraphics[width=\textwidth]{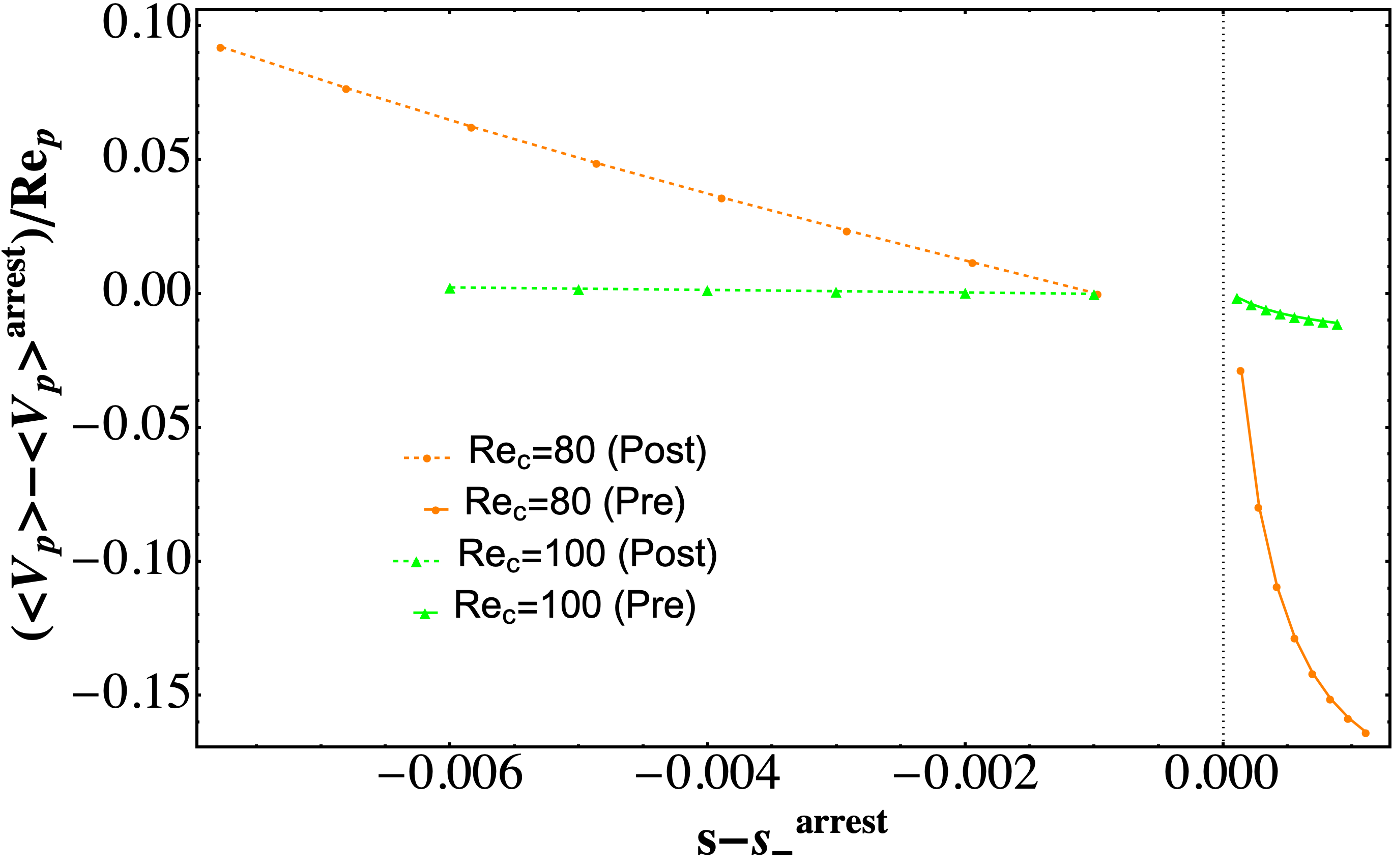}
    \caption{}
    \end{subfigure}
	\caption{(a) Lift velocity profiles for an oblate spheroid with $\kappa=0.05$, for $\lambda=0.05$, for different $Re_c$; $Re_c^\text{arrest}\approx72.2$. The pair of vertical lines, corresponding to $s^\text{arrest}_-\approx0.048$ and $s^\text{arrest}_-\approx0.14$, demarcate regions of stationary\,($0 < s < s^\text{arrest}_-$) and tumbling\,($s^\text{arrest}_- < s < 1/2$) spheroids for $Re_c=80$ and $Re_c=100$, respectively; the rotation-arrested portions of these lift profiles have been depicted using dashed curves. The inset compares the equilibrium locus with its small-$Re_c$ approximation. (b) Magnified views of the lift profiles for $Re_c = 80$ and $100$ highlight the jump in slope across $s =s^\text{arrest}_-$.}
	\label{fig:InertialAvgdLiftkappap05}
\end{figure}

\begin{figure}
        \centering
        \begin{subfigure}{0.54\columnwidth}
		\includegraphics[width=\textwidth]{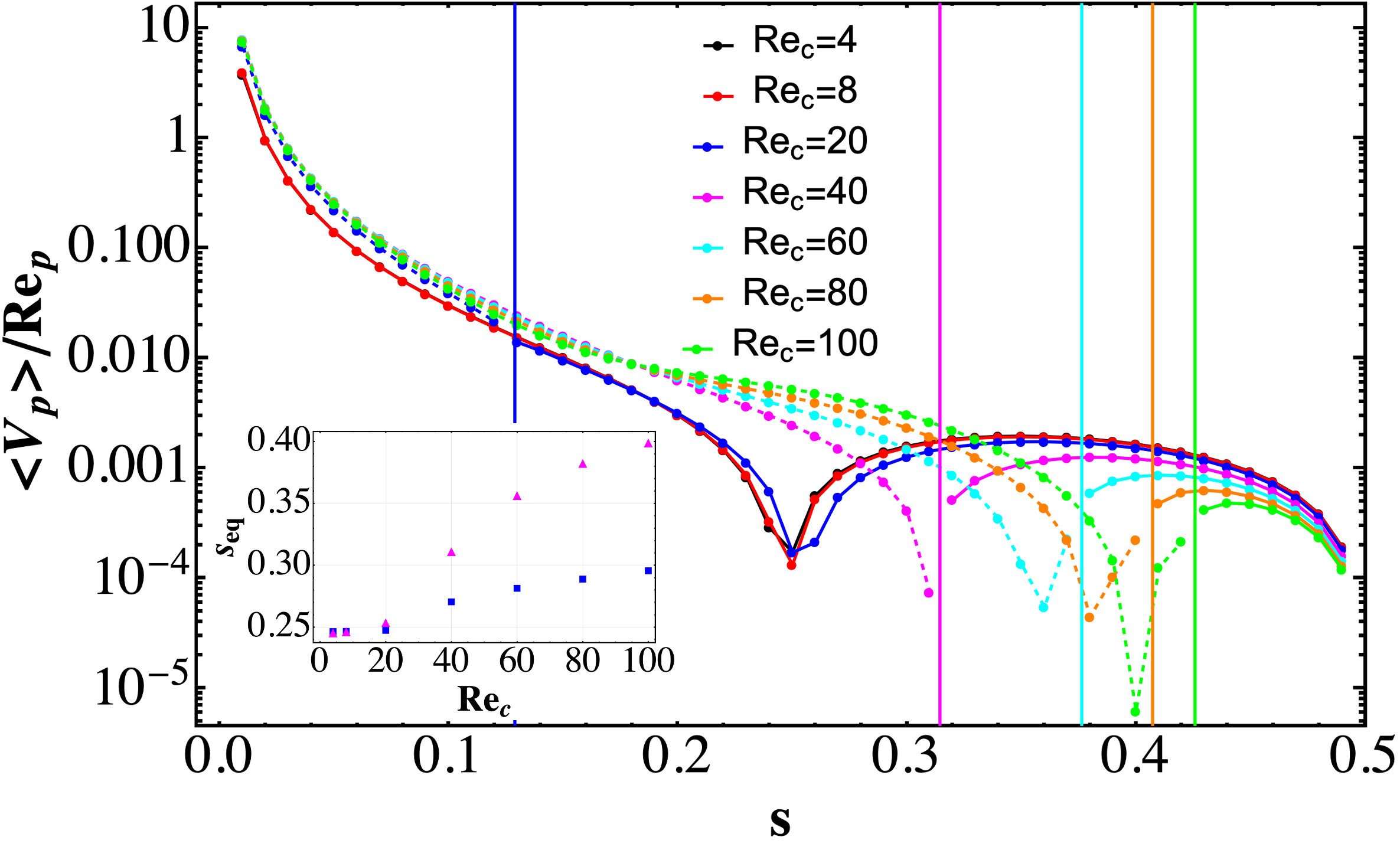}
        \caption{}
        \end{subfigure}
        \begin{subfigure}{0.45\columnwidth}
		\includegraphics[width=\textwidth]{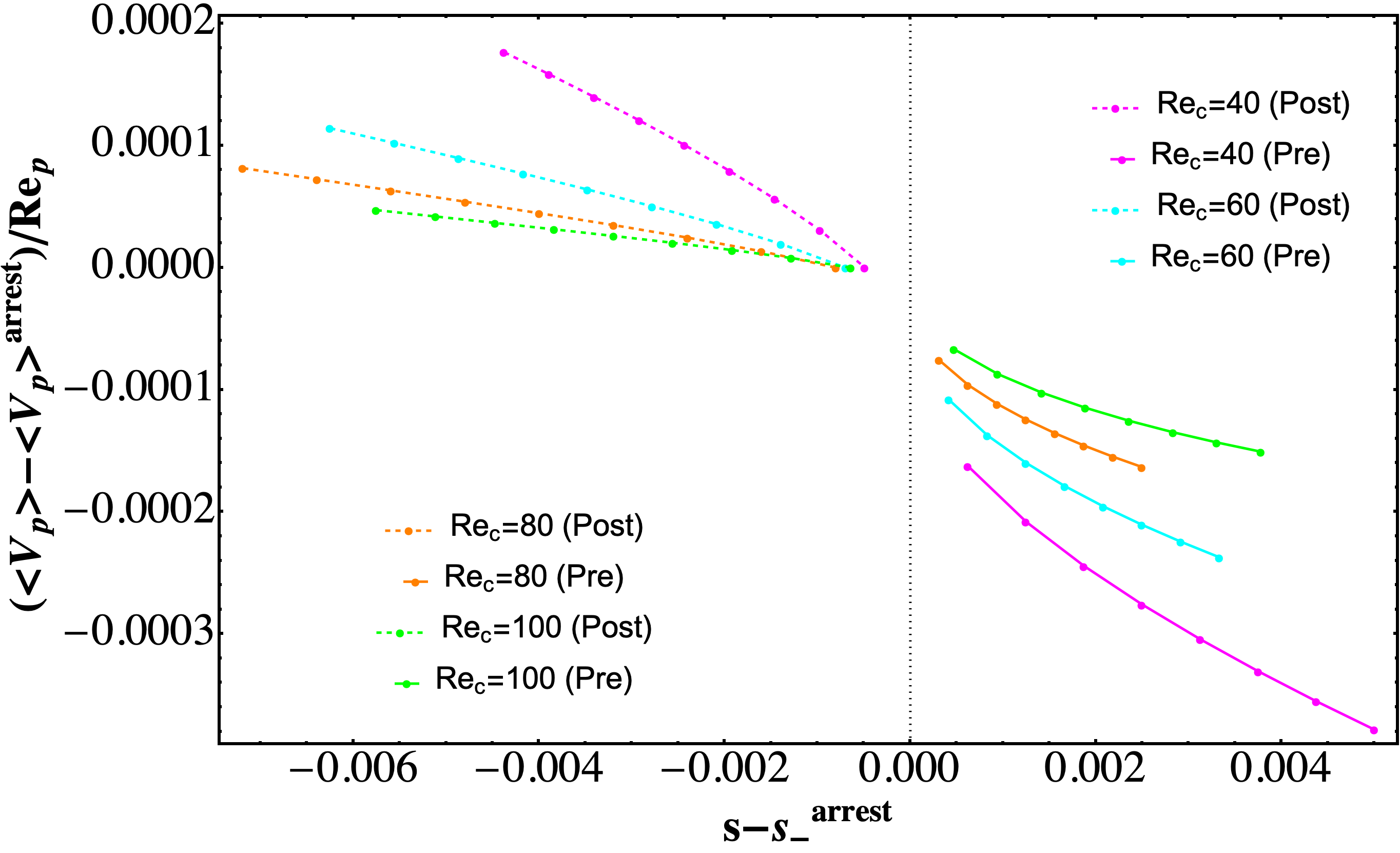}
        \caption{}
        \end{subfigure}
		\caption{(a)~Lift velocity profiles for an oblate spheroid with $\kappa=0.01$, for $\lambda=0.05$, for different $Re_c$; $Re_c^\text{arrest}\approx14.85$. Vertical lines, corresponding to $s^\text{arrest}_-\approx0.128$, $s^\text{arrest}_-\approx0.31$ and $s^\text{arrest}_-\approx0.376$, $s^\text{arrest}_-\approx0.41$ and $s^\text{arrest}_-\approx0.42$, demarcate regions of stationary\,($0 <<s < s^\text{arrest}_-$) and tumbling\,($s^\text{arrest}_- < s < 1/2$) spheroids for $Re_c=20$, $Re_c=40$, $Re_c=60$, $Re_c=80$ and $Re_c=100$, respectively; the rotation-arrested portions of these profiles being depicted using dashed curves. The inset compares the equilibrium locus with its small-$Re_c$ approximation. (b) Magnified views of the lift profiles for $Re_c > 20$ highlight the jump in slope across $s =s^\text{arrest}_-$.}
		\label{fig:InertialAvgdLiftkappap01}
\end{figure}

Figures \ref{fig:InertialAvgdLiftkappap05}a and \ref{fig:InertialAvgdLiftkappap01}a show the time-averaged lift velocity profiles for oblate spheroids with $\kappa=0.05$ and $0.01$, respectively, for $Re_c$ ranging from $4$ to $100$; $\lambda =0.05$. The differences relative to the corresponding profiles based on the Jeffery-averaged approximation\,(not shown) can be seen in both figures. Figure \ref{fig:InertialAvgdLiftkappap05}a is similar to Figure \ref{fig:InertialAvgdLiftkappagt1}b for $\kappa = 100$, in that the equilibrium location remains virtually invariant to changing $Re_c$. There is still a difference from the prolate case in that, for $\kappa = 0.05$, rotation-arrested regions emerge and move into the channel for $Re_c \gtrsim 72.2$. Thus, the lift velocity profiles for $Re_c = 80$ and $100$ exhibit both tumbling and rotation-arrested regions, demarcated by the pair of vertical lines in Figure \ref{fig:InertialAvgdLiftkappap05}a; Figure \ref{fig:InertialAvgdLiftkappap05}b highlights the jump in profile-slope for the aforesaid $Re_c$'s, as one crosses over from the tumbling to the rotation-arrested region. It is also worth noting that the rotation-arrest location in Figure \ref{fig:InertialAvgdLiftkappap05}a exhibits a significant shift towards the centerline as $Re_c$ increases from $80$ to $100$, even as the equilibrium location remains nearly unchanged. In Figure \ref{fig:InertialAvgdLiftkappap01}a, for the thinner oblate spheroid, rotation-arrested regions emerge for $Re_c \gtrsim 14.85$. The more rapid movement of the rotation-arrest location towards the centerline in Figure \ref{fig:InertialAvgdLiftkappap01}a now leads to it overtaking the equilibrium location, as a result of which the latter becomes rotation-arrested for the three largest $Re_c$'s\,($60$, $80$ and $100$); magnified views in Figure~\ref{fig:InertialAvgdLiftkappap01}b again highlight the jumps in profile slope for all $Re_c$'s that lead to rotation arrested regions within the channel. Finally, the insets in Figures~\ref{fig:InertialAvgdLiftkappap05}a and \ref{fig:InertialAvgdLiftkappap01}a show that, similar to $\kappa = 200$, the pronounced movement of the equilibrium locus towards the centerline is no longer captured by the small-$Re_c$ approximation; the failure of the latter approximation extends to most of the $Re_c$-interval examined in Figure~\ref{fig:InertialAvgdLiftkappap01}a. 

\begin{figure}
	\centering
	\includegraphics[width=0.8\textwidth]{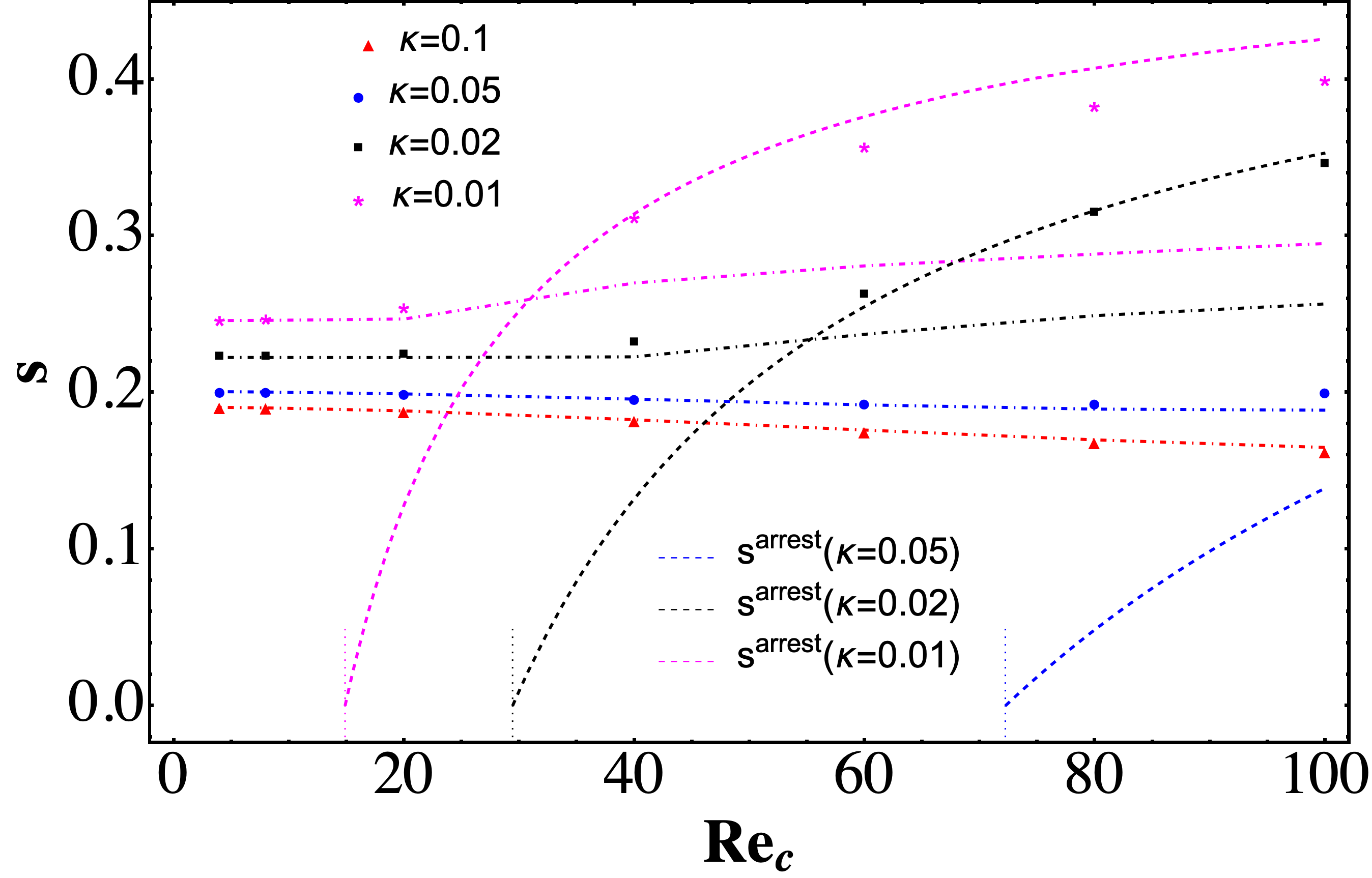}
	\caption{The symbols denote the equilibrium loci for oblate spheroids with $0.1$~(triangles), $0.05$~(circles), $0.02$~(squares) and $0.01$~(stars); $\lambda=0.05$; dash-dotted curves denote the corresponding small-$Re_c$ approximations Rotation-arrest loci for $\kappa = 0.05,\,0.02$ and $0.01$ appear as (dashed)~continuous curves that start from the $s=0$\,(the $Re_c$-axis) at $Re_c^\text{arrest}$, and asymptote to $1/2$ for $Re_c \rightarrow \infty$; $Re_c^\text{arrest}\approx72.2$ for $\kappa=0.05$, $\approx 29.4$ for $\kappa=0.02$ and $\approx 14.9$ for $\kappa=0.01$.
    }
	\label{fig:LociArrestoblate}
\end{figure}

Following along the lines of the prolate case, both the equilibrium and rotation-arrest loci, for different oblate spheroids with $\kappa < 0.14$, are plotted in Figure~\ref{fig:LociArrestoblate}; the small-$Re_c$ approximations of the equilibrium loci again appear as dash-dotted curves. The equilibrium locus for $\kappa = 0.1$ exhibits a slight decrease, corresponding to a weak wallward movement, while that for $\kappa = 0.05$ is nearly horizontal\,(consistent with the lift profiles in Figure~\ref{fig:InertialAvgdLiftkappap05}). The latter locus might enter the rotation-arrested region\,(that begins at $Re_c^\text{arrest} \approx 72.2$), and then curve upward for $Re_c > O(100)$; although, verifying this is beyond the range of validity of small-$Re_p$ theory. Thus, for the range of $Re_c$ examined, the equilibrium locus first crosses into the rotation-arrested region for $\kappa \approx 0.02$. For $\kappa = 0.01$, the equilibrium locus becomes rotation arrested for $Re_c \gtrsim 40$. The loci for both $\kappa =0.02$ and $0.01$ exhibit a strong upward movement\,(towards the centerline) following rotation arrest. On the whole, a comparison of Figures~\ref{fig:LociArrestProlate} and \ref{fig:LociArrestoblate} shows that the stronger disturbance velocity field induced by oblate spheroids leads to smaller rotation-arrest thresholds, and stronger movement of the equilibria towards the centerline. This is  also evident from the expressions for the respective rotation-arrest thresholds given earlier, where the logarithmically weak velocity field induced by a slender spheroid leads to the threshold $Re$ being larger by a factor $\ln \kappa$.

Having established the significance of rotation arrest towards inertial migration, we now examine the inertially averaged stresslet tensor, components of which appear in (\ref{eq:inertialstressletforcing}), in the neighborhood of arrest. Recall that the angular velocity given by \eqref{eq:Fullangularvelocity}, \eqref{eq:Jeffangvel} and (\ref{inertcorr_Re}), was expanded in (\ref{eq:AngVelNearArrest}a,b) for spheroids with $\kappa,\kappa^{-1}\gg1$, for orientations close to flow and gradient-alignment, where rotation is first arrested with increasing $Re_s$. When these asymptotic forms are used in the time period integral, $T = \textstyle\int_0^{2\pi} \frac{d\phi_j}{\dot{\phi}_j}$, one finds $T \approx \frac{\pi\kappa}{\beta[2(1-Re_s/Re^\text{arrest})]^{1/2}}\,(\approx \frac{\pi\kappa^{-1}}{\beta[2(1-Re_s/Re^\text{arrest})]^{1/2}})$ for $\kappa \gg 1\,(\ll 1)$, for $Re_s \rightarrow Re^\text{arrest}$. The inverse square-root divergence in the vicinity of arrest was first established by \citet{aidun2000} for rotation of a vorticity-aligned elliptic cylinder, and is characteristic of a saddle-node bifurcation. Thus, the nature of the divergence is independent of $\kappa$, with extreme $\kappa$ values allowing for a divergence even in the small $Re_s$ limit\,\citep{subkoch2005,PavanElasticDrift_JFM2025}. Note that the said bifurcation leads to a pair of stationary orientations in the flow-gradient plane for $Re_s > Re^\text{arrest}$, one being a stable node\,($\phi_j^+ \approx \kappa^{-1}(1+[2(Re_s/Re^\text{arrest}-1)]^{1/2}$ for $\kappa\gg1$ and $\approx\pi/2+\kappa(1+[2(Re_s/Re^\text{arrest}-1)]^{1/2}$ for $\kappa\ll1$) and the other being an unstable node\,($\phi_j^- \approx \kappa^{-1}(1-[2(Re_s/Re^\text{arrest}-1)]^{1/2}$ for $\kappa\gg1$ and $\approx\pi/2+\kappa(1-[2(Re_s/Re^\text{arrest}-1)]^{1/2}$ for $\kappa\ll1$), with the spheroid adopting the stable orientation for $Re_s > Re^\text{arrest}$. 

\begin{figure}
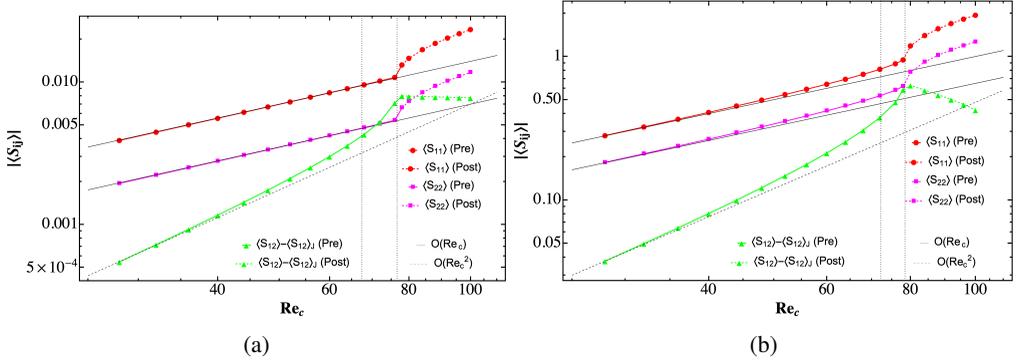

    \centering
    \begin{subfigure}{0.49\textwidth}
		\includegraphics[width=\textwidth]{Figs/Stresslet_vs_Rec_kappa_200.png}
		\caption{}
	\end{subfigure}
    \begin{subfigure}{0.49\textwidth}
		\includegraphics[width=\textwidth]{Figs/Stresslet_vs_Rec_kappa_p05.png}
		\caption{}
	\end{subfigure}
	\caption{Components of the inertially averaged stresslet, $\langle S_{ij}\rangle$, plotted as a function of $Re_c$, for fixed $\kappa$ and $s$; note that $\langle S_{33} \rangle = -(\langle S_{11} \rangle + \langle S_{22} \rangle$, and $\langle S_{13} \rangle = \langle S_{23} \rangle = 0$ due to symmetry. (a) $\kappa = 200$, $s=0.06$ and (b) $\kappa = 0.05$, $s =0.04$. For both aspect ratios, rotation arrest occurs within the $Re_c$ interval examined, and this coincides with a jump in slope for all stresslet components. The first vertical line in each plot corresponds to $Re_c^\text{arrest}$, when the arrest location just enters the channel; $Re_c^\text{arrest} \approx 67.4$ in (a), and $\approx 72.2$ in (b). The second vertical line denotes the higher $Re_c$ at which the arrest location coincides with the chosen $s$.}
	\label{fig:inertialstresslet_arrest}
\end{figure}

One may also evaluate the averaged stresslet close to arrest. The latter is defined by (\ref{eq:NonJeffTimeavg}), with $f(\phi_j)$ replaced by $\bm S(\phi_j)$, and with ${\bm S}(\phi_j)$ in turn defined by (\ref{eq:StressletProlate}) with $\theta_j = \frac{\pi}{2}$ in $\bm p$\,(owing to inertial stabilization of the tumbling mode). 
On using the relevant asymptotic forms, one finds:
\begin{subequations}
\begin{align}
\langle S_{ij}\rangle&\approx \frac{4\pi}{9\kappa\ln\kappa}(-2\delta_{i1}\delta_{j1}+\delta_{i2}\delta_{j2}+\delta_{i3}\delta_{j3})-\frac{8\pi}{3\kappa^2}(\delta_{i1}\delta_{j2}+\delta_{i2}\delta_{j1})\,\,\,\,\text{for}\,\,\,\,\kappa\gg1,\\
\langle S_{ij}\rangle&\approx \frac{16\kappa}{9}(-3\delta_{i1}\delta_{j1}+2\delta_{i2}\delta_{j2}+\delta_{i3}\delta_{j3})-\frac{4\pi\kappa}{3}(\delta_{i1}\delta_{j2}+\delta_{i2}\delta_{j1})\,\,\text{for}\,\,\kappa^{-1}\gg1.
\end{align} \label{eq:SijNearArrestTimeAvgd}
\end{subequations}
The above expressions also correspond to the stresslet associated with the post-arrest stable orientations in the two limiting cases, implying that the stresslet is continuous across arrest. Since averaging brackets are redundant after arrest, the continuity means $\displaystyle \lim_{Re_s \rightarrow {Re^\text{arrest}}^-} \langle \bm S \rangle = \lim_{Re_s \rightarrow {Re^\text{arrest}}^+} \bm S$. Further, it may also be shown that $d\langle \bm S \rangle/ds$ undergoes a jump across $Re_s = Re^\text{arrest}$. Numerical evidence for this latter behavior is shown in Figures~\ref{fig:inertialstresslet_arrest}a and b which, unlike Figure~\ref{fig:Rearrest&stresslet}b, plot the stresslet components as a function of $Re_c$ for sufficiently extreme aspect ratios, as to lead to rotation arrest within the $Re_c$-interval examined. For $Re_c$ values much less than the arrest threshold, the stresslet components exhibit the scaling behavior already seen in Fig.\ref{fig:Rearrest&stresslet}b; that is $\langle S_{11} \rangle, \langle S_{22} \rangle \sim O(Re_c), \langle S_{12} \rangle - \langle S_{12} \rangle_J \sim O(Re_c^2)$. There is a deviation from this simple scaling behavior for larger $Re_c$, with a jump in slope occurring at the critical value at which the arrest location coincides with the chosen location\,($s$) within the channel; note that this critical value equals $Re_c^\text{arrest}$ for $s = 0$ or $1$. 
It is this jump in the slope of the stresslet components that, via the forcing in (\ref{eq:NSOuterEqnTA}a), leads to a jump in the slope, across $s^\text{arrest}_\pm$, of the lift velocity profiles. The forcing determines the magnitude of the lift velocity, and a jump in its rate of change with $Re_c$ leads to the amplitude of the rotation-arrested portion of the lift profile changing at a different rate, with changing $Re_c$, compared to the tumbling portion. This implies a jump in slope at the location\,($s^\text{arrest}_\pm$) marking the transition between the two parts - see Figures~\ref{fig:Liftkappa200}b, \ref{fig:InertialAvgdLiftkappap05}b and \ref{fig:InertialAvgdLiftkappap01}b. Finally, it is worth mentioning that the divergence of the tumble period might appear to lead to a breakdown of the time-averaged analysis which, as mentioned at the beginning of $\S$\ref{sec:Reclarge}, relies on a separation between the rotation and migration time scales. However, $\langle S_{ij} \rangle$, and thence, $\langle V_p \rangle$, remains finite despite the divergence of the tumble period, ensuring that the time-averaged lift velocity of a tumbling spheroid, for $Re_c < Re_c^\text{arrest}$, transitions in a continuous manner to the steady lift of a stationary spheroid for $Re_c > Re_c^\text{arrest}$.

\begin{figure}
	\centering
	\includegraphics[width=0.9\textwidth]{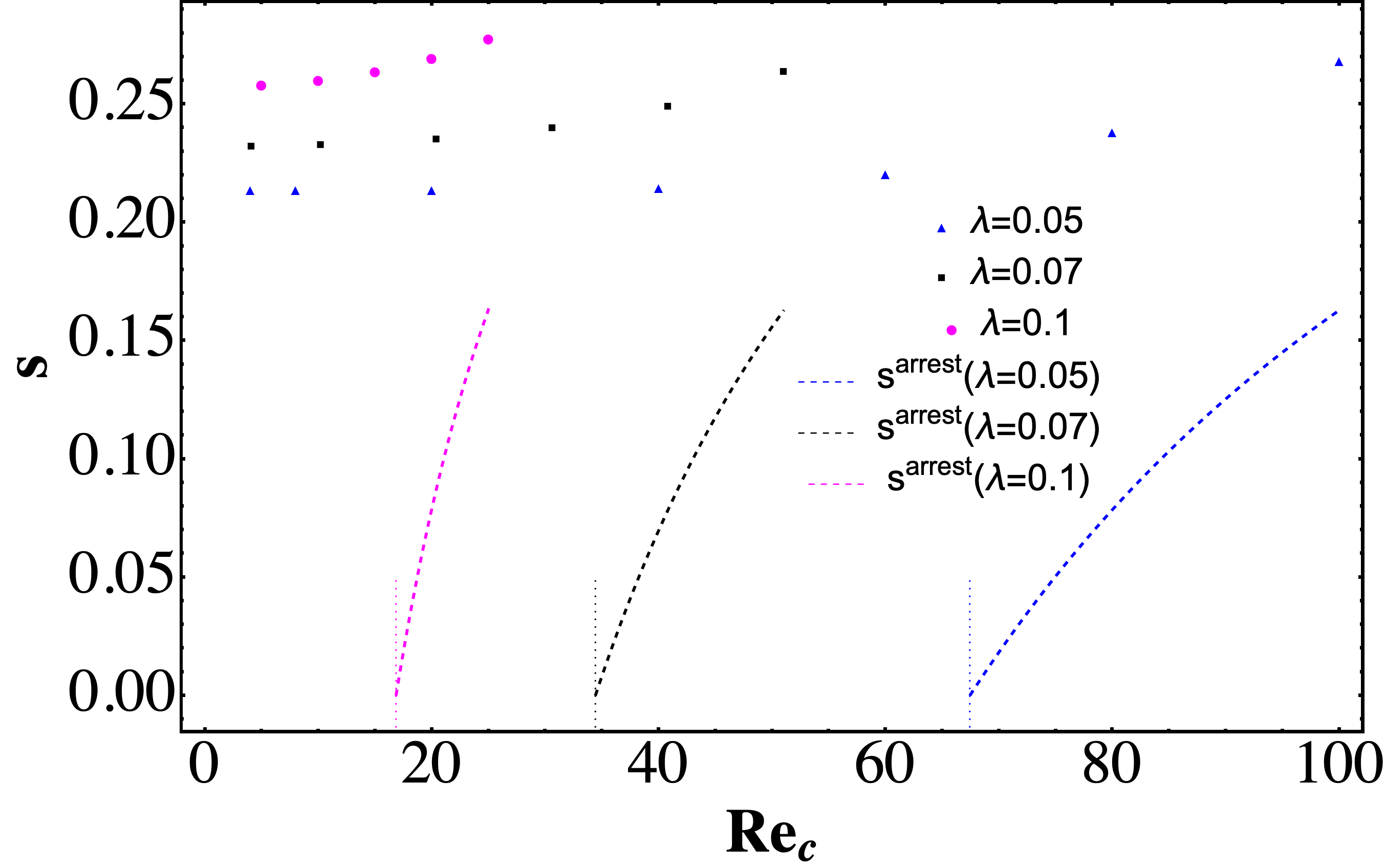}
	\caption{The symbols denote the equilibrium loci plotted as a function of $Re_c$, for different confinement ratios\,($\lambda=0.05$, $0.07$ and $0.1$), for $\kappa=200$; the $Re_c$-intervals under consideration are $[4,Re_c^\text{max}] \equiv [4/4\lambda^2]$. The dashed curves denote the corresponding rotation-arrest loci that begin at $Re_c^\text{arrest}\,(s^\text{arrest}_- = 0)$,with $s^\text{arrest}_- \rightarrow 1/2$ for $Re_c \rightarrow \infty$. $(Re_c^\text{arrest},Re_c^\text{max})$ are given by $(67.4,100)$, $(34.4,51)$, $(16.8,25)$ for $\lambda = 0.05$, $0.07$ and $0.1$, respectively; 
    }
\label{fig:Locikappa200lambdap1}
\end{figure}

All of the plots shown until now pertained to $\lambda = 0.05$, and Figure~\ref{fig:Locikappa200lambdap1} therefore examines the effect of varying $\lambda$ on the equilibrium loci. This is done for a prolate spheroid with $\kappa=200$, for $\lambda = 0.05$, $0.07$ and $0.1$, and with the equilibrium loci having been plotted until the respective $Re_c^\text{max}$ values. The overall behavior of each equilibrium locus remains the same as that seen earlier in Figures~\ref{fig:LociArrestProlate} and \ref{fig:LociArrestoblate}- in that, it starts out on a plateau without any significant dependence on $Re_c$ until arriving close to the rotation-arrest locus, at which point it starts to move upward.  
The rotation-arrest locus moves to smaller $Re_c$ with increasing $\lambda$\,(as $\lambda$ increases from $0.05$ to $0.1$, $Re_c^\text{arrest}$ decreases from $67.4$ to $16.8$), which leads to the equilibrium locus for a larger $\lambda$ moving upward at a smaller $Re_c$. Similar to $\lambda=0.05$, the arrest locus never crosses the equilibrium locus for the $Re_c$'s examined. 
Also evident from the figure is the fact that, at a given $Re_c$, the equilibrium locations for larger spheroids are closer to the centerline.

\begin{figure}
	\centering
    \begin{subfigure}{0.49\textwidth}
		\includegraphics[width=\textwidth]{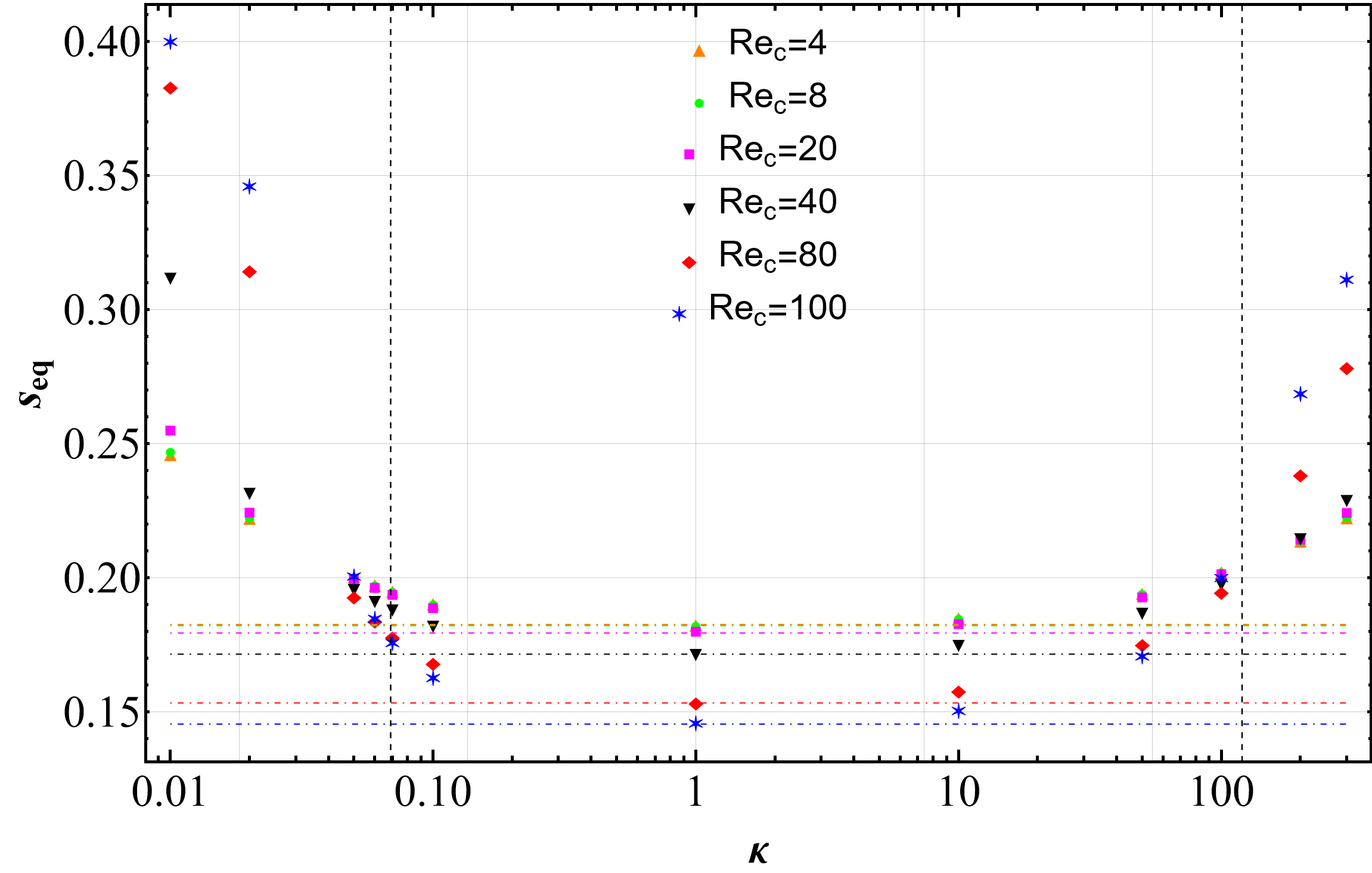}
		\caption{}
	\end{subfigure}
    \begin{subfigure}{0.49\textwidth}
	 	\includegraphics[width=\textwidth]{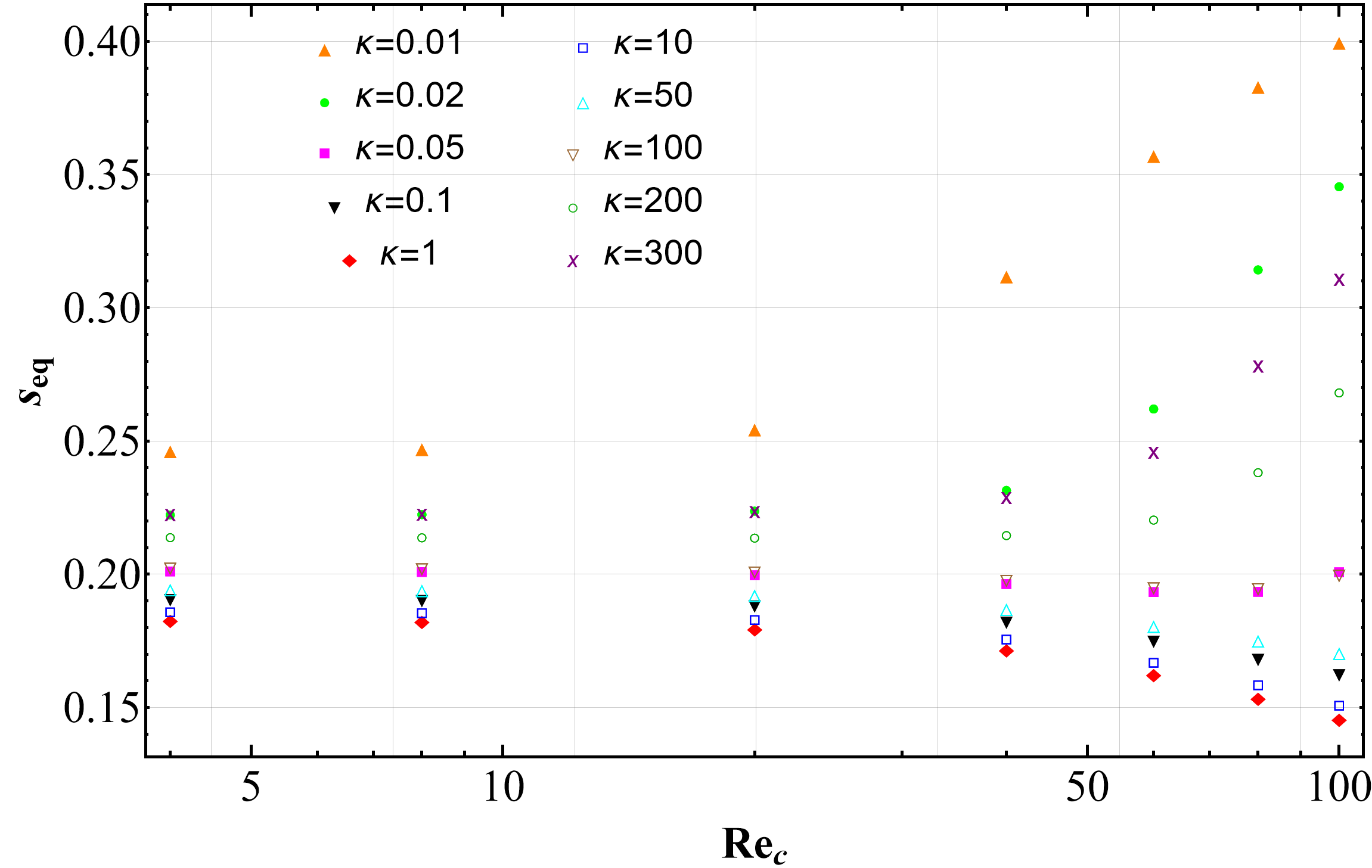}
	 	\caption{}
	\end{subfigure}
	\caption{(a) Loci of equilibrium locations, plotted as a function of $\kappa$, for different $Re_c$'s, for $\lambda=0.05$. The pair of vertical dashed lines correspond to critical aspect ratios on the prolate\,($\kappa\approx120$) and oblate\,($\kappa\approx0.069$) sides, for which the arrest location just enters the channel; at the said aspect ratios, $s^\text{arrest}_-\lesssim0.005$ for $Re_c^\text{arrest}\approx99$\,($Re_c^\text{max}=100$ for the chosen $\lambda$). The region between the dashed lines, $0.069 \lesssim \kappa \lesssim 120$, therefore corresponds to cases where spheroids always tumble regardless of their transverse location within the channel. (b) Loci of equilibrium locations, plotted as a function of $\kappa$, for different $Re_c$'s, for $\lambda=0.05$. The lowest curve is the baseline scenario for spheres\,($\kappa = 1$); for $\kappa$ departing from unity in either direction, the equilibrium locus transitions from a wallward movement, to one towards the centerline, with increasing $Re_c$.}
\label{fig:Locivskappa}
\end{figure}

Figures~\ref{fig:Locivskappa}a and b summarize the $\kappa$-dependence of equilibrium positions, accounting for all of the lift velocity profiles shown thus far. In Figure~\ref{fig:Locivskappa}a, the equilibrium position is plotted as a function of $\kappa$, for different $Re_c$, for $\lambda=0.05$. A pair of vertical lines, corresponding to $\kappa = 0.069$ and $120$, demarcate the range of aspect ratios\,($0.069 \lesssim \kappa \lesssim 120$) for which spheroids tumble at all locations within the channel, from the aspect-ratio intervals\,($\kappa \gtrsim 120$, $\kappa \lesssim 0.069$) where the channel is divided into a near-centerline tumbling region and a pair of near-wall rotation-arrested regions. The $\kappa$-independent Jeffery-averaged equilibria appear on this plot as dashed horizontal lines which move downward with increasing $Re_c$, reflecting the wallward movement of the equilibrium position. All of the $s_{eq}$-loci start from the Jeffery-averaged baseline at $\kappa = 1$, and curve upward\,(indicative of a movement towards the centerline) as $\kappa$ departs from unity. The upward shift becomes significant on the outside of the aforementioned vertical lines, and further, is significantly more pronounced on the oblate side. The upward shift in the $s_{eq}$-loci is also greater for the larger $Re_c$ values. This is because, while the largest $Re_c$'s lead to the smallest $s_{eq}$ for $\kappa = 1$, spheroids at these $Re_c$'s are also the most affected by the inertial slowing down and eventual rotation arrest; the latter leads to $s_{eq}$ increasing more rapidly with changing $\kappa$. The result of the above is that there exist intermediate aspect ratios\,($\kappa \approx 100$ and $\kappa \approx 0.05$ on the prolate and oblate sides, respectively) at which the $s_{eq}$-loci for different $Re_c$ approximately converge, leading to a rather muted dependence on $Re_c$. Figure~\ref{fig:Locivskappa}b plots equilibrium positions as a function of $Re_c$, for different $\kappa$, again for $\lambda = 0.05$. A given $Re_c$ corresponds to a fixed flow rate in an experiment, and the figure thus highlights the possibility of using inertial lift forces towards shape-sorting particles of different aspect ratios, but with the same maximum dimension\,(in this case, the semi-major axis that is used to define $Re_p$, and thence, $Re_c$), by appropriate choice of the flow rate. The lowest curve in this figure is the baseline scenario for  spheres\,(or, equivalently, the Jeffery-averaged approximation for spheroids), and decreases with increasing $Re_c$ owing to wallward migration of the equilibrium location. With increasing $\kappa$ or $\kappa^{-1}$, the $s_{eq}$-loci depart from this curve, the departure being earlier for the more extreme aspect ratios. Similar to Figure~\ref{fig:Locivskappa}a, the departure of thin oblate spheroids is much more pronounced than that of slender prolate spheroids. For instance, $s_{eq} \approx 0.4$ for $\kappa = 0.01$, at $Re_c = 100$; in contrast, $s_{eq} \approx 0.2$ with this value being fairly insensitive to $Re_c$, for the complementary prolate spheroid with $\kappa = 100$.

\section{Conclusions and Discussion}\label{sec:conclusion}

In this paper, we have shown that accounting for the effect of fluid inertia on spheroid rotation leads to lift velocity profiles, in a wall-bounded plane Poiseuille flow, that are qualitatively different from those for a sphere. While the assumption of a spheroid rotating in the inertia-stabilized tumbling mode, but with the Jeffery angular velocity, leads to time-averaged lift velocity profiles that have the same shape as those for a sphere, but with a $\kappa$-dependent amplitude~\citep{anand2023Jeff}, accounting for inertia-induced slowdown of tumbling prolate and oblate spheroids allows one to go beyond the Segre-Silberberg picture for a sphere~\citep{segresilberberg1962a,segresilberberg1962b}.  In the latter scenario, and within a point-particle approximation, spheres migrate to a pair of symmetrically located equilibrium positions intermediate between the channel centerline and walls, for small $Re_c$~\citep{holeal1974,vasseur1976}, with these locations migrating wallward with increasing $Re_c$~\citep{schonberghinch1989,matas2009}. The time-averaged analysis based on the inertial angular velocity, in $\S$\ref{sec:Reclarge} and in Appendix \ref{App:LiftSmallRec}, shows that the equilibrium locations for spheroids depend on $\kappa$ for all $Re_c$. Within the small-$Re_p$ theoretical framework used here, this dependence is very weak for $\kappa \sim O(1)$, with equilibrium positions that are only marginally closer to the channel centerline than the Segre-Silberberg ones for a sphere; the weak dependence being due to inertia having only a perturbative effect on spheroid rotation for $\kappa \sim O(1)$ for small but finite $Re_p$ . 

For increasingly slender prolate spheroids or thin oblate spheroids, the 
inertial correction to the spheroid angular velocity becomes comparable to the leading order Jeffery contribution, eventually equalling it when the local-shear-based Reynolds number, $Re_s$, equals $Re^\text{arrest}$, which leads to spheroid rotation being arrested. $Re^\text{arrest}$ is a function of the spheroid aspect ratio, being $O(\ln \kappa/\kappa)$ and $O(\kappa)$ in the limits of large and small aspect ratios, respectively. 
The pair of arrest locations coincide with the channel walls at $Re_c^\text{arrest} =Re^\text{arrest}/4\lambda^2$, and move in towards the centerline with a further increase in $Re_c$. For any $Re_c > Re_c^\text{arrest}$, the channel is divided into a pair of symmetrically located near-wall regions wherein spheroids adopt a stationary orientation, and a central region where spheroids continue to tumble. The stationary spheroids in the near-wall regions are nearly flow-aligned in the prolate case, and almost gradient-aligned in the oblate case; the spheroids rotate in the central tumbling zone although, unlike the Stokesian case, inertia causes them to spend a longer time in the extensional quadrants. Importantly, for both prolate and oblate spheroids, the appearance of rotation-arrested regions within the channel triggers an accelerated movement of the equilibrium positions towards the centerline.
In the prolate case, we find both the arrest and equilibrium loci to move towards the centerline at almost the same rate with increasing $Re_c$, and to never intersect over the range of $Re_c$'s examined\,(this range being determined by the requirement that $Re_s$ is less than unity). As a result, slender prolate spheroids continue to tumble at their equilibrium locations, regardless of $Re_c$. In sharp contrast, the movement of the rotation-arrest locus is more rapid for the oblate case, leading to sufficiently thin oblate spheroids becoming rotation arrested, at their equilibrium locations, beyond a threshold $Re_c$. Figures \ref{fig:Locivskappa}a and b serve to summarize the overall inertial migration scenario as a function of $\kappa$ and $Re_c$. The vertical lines corresponding to $\kappa \approx 0.069$ and $120$ in Figure~\ref{fig:Locivskappa}a mark the onset of rotation arrest within the channel, and act as rough markers for a transition in the behavior of the equilibrium loci - from being almost $\kappa$-independent for $\kappa\,\in\,(0.069,120)$, to exhibiting a strong dependence for $\kappa$ outside this interval.

One of the central findings in this manuscript is that fluid-inertia-induced slowdown of spheroid rotation retards and even reverses the known wallward movement of the equilibrium position for spheres with increasing $Re_c$. As elaborated upon in the previous paragraph, the rate of slowdown and the reversal depends on $\kappa$, and it is this $\kappa$-dependence that then provides a theoretical basis for shape-sorting using inertial lift forces~\citep{pregibon2007,decuzzi2009intravascular,mitragotri2009,dicarlo2010}. The $\kappa$-dependence is qualitatively consistent with the shape dependence observed in the experiments of \citet{masaeli2012} discussed in the introduction\,(see Figure~\ref{fig:Masaeliequilibria}); although, the restriction to small $Re_p$ leads to the $\kappa$-dependence in theory being postponed until aspect ratios much higher than those in the said experiments. It is also worth noting that the experiments did not observe rotation arrest. This might, on one hand, be due to the arrest location being too close to the wall\,(the spheroid positions were not controlled at the channel inlet), and on the other hand, due to the particle Reynolds numbers being smaller than the arrest values. In principle, spheroids of order unity aspect ratios should also exhibit rotation arrest for sufficiently large $Re_p$. However, as evident from Figure~\ref{fig:Rearrest&stresslet}a, the arrest values appear to have only been determined\,(numerically) for the two-dimensional case. It is worth emphasizing that both the onset of rotation arrest and the associated $\kappa$-dependence occurs at less extreme aspect ratios on the oblate side. This awaits experimental verification, especially the prediction that sufficiently thin oblate spheroids ought to be rotation arrested at their equilibrium locations. Despite disk-shaped particles having been used in the experiments of \citet{hur2011}, the lack of detailed information preclude a detailed comparison. 

\begin{figure}
	\centering
    \begin{subfigure}{0.49\textwidth}
    \includegraphics[width=\textwidth]{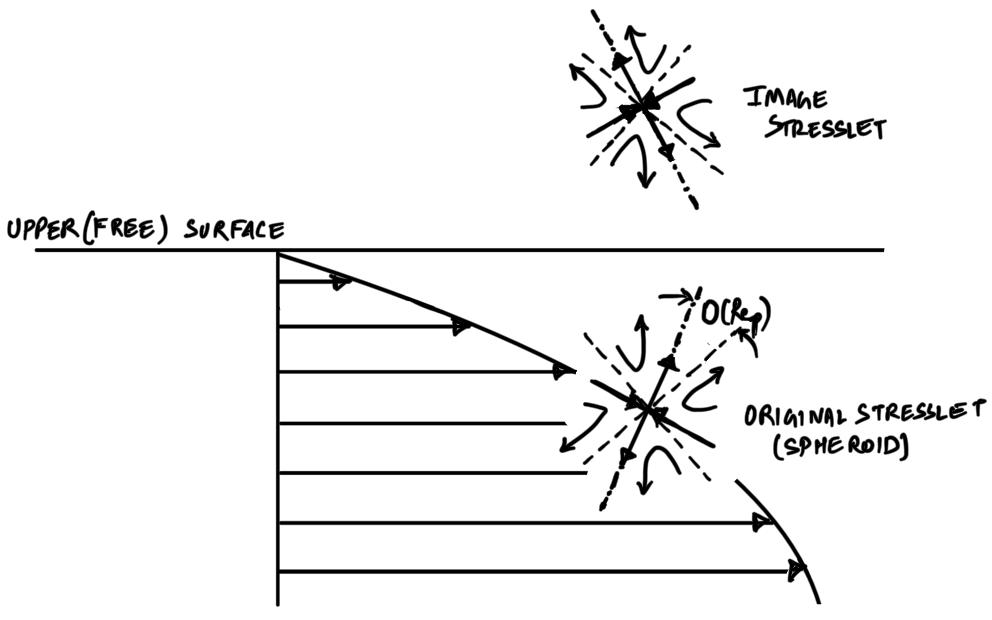}
    \caption{}
    \end{subfigure}
    \begin{subfigure}{0.49\textwidth}
    \includegraphics[width=\textwidth]{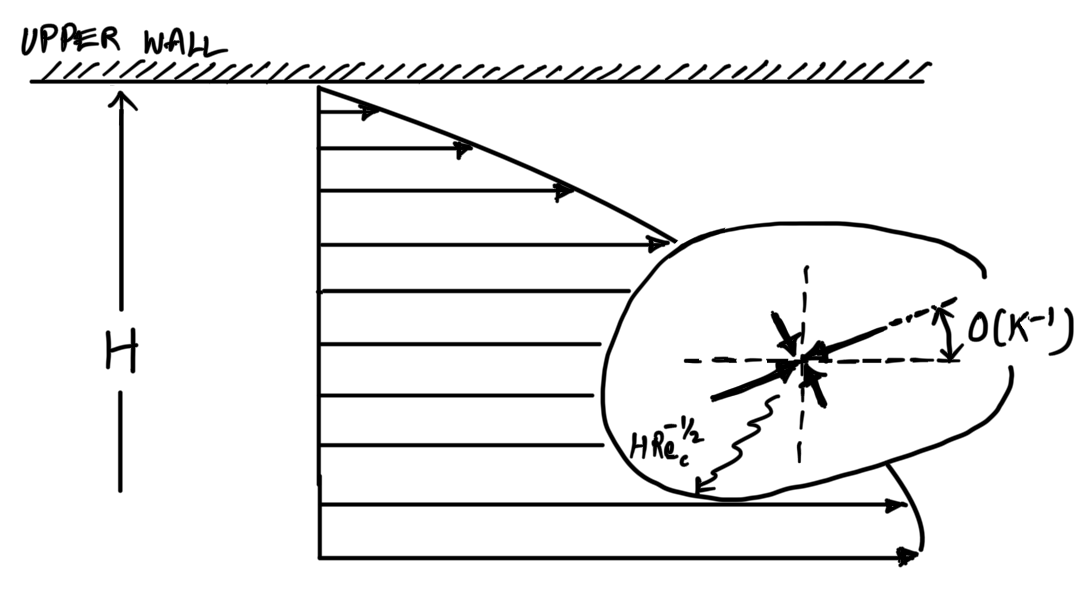}
    \caption{}
    \end{subfigure}
    \caption{(a) Repulsive interaction with a time-averaged image stresslet\,(pertaining to a plane free surface replacing the no-slip upper wall), that drives migration towards the centerline for small $Re_c$, and for sufficiently small or large $\kappa$. The pair of dashed lines along the local extensional and compressional directions characterize the orientation of the Jeffery-averaged stresslet; the arrows denote the inertial stresslet whose orientation is obtained via an $O(Re_p)$ rotation in the anticlockwise direction. (b) Stresslet-wall interaction at $Re_c$ close to rotation arrest for $\kappa \gg 1$. The compressive force-dipole now makes a small angle of $O(\kappa^{-1})$ with the flow direction, although the Stokesian velocity field, depicted in (a), only extends until the inertial screening length of $O(H Re_c^{-\frac{1}{2}})$.}
	\label{fig:imagestress}
\end{figure}

Based on the analysis in Appendix~\ref{App:LiftSmallRec}, for small $Re_c$ and within the point-particle framework used here, the mechanism underlying the movement of the equilibrium location towards the centerline may be interpreted as a time-averaged repulsive interaction of the spheroid\,(regarded as a stresslet) with an image-system that accounts for the wall boundary condition. The eigenvectors of the inertially averaged stresslet, in the flow-gradient plane, are along $\frac{r_2}{r_1}=\left(\pm 1-\beta Re_p \frac{B_{11}(\kappa)-B_{22}(\kappa)}{2\langle S_{12}\rangle_J(\kappa)}\right)$, 
with the associated eigenvalues being  $\lambda_{1,2}=\beta \langle S_{12}\rangle_J \pm \beta Re_p(B_{11}+B_{22})/2$. 
The local shear rate $\beta$ is positive in the lower half and negative in the upper half of the channel, with $B_{11},\langle S_{12}\rangle_J<0$ and $B_{22}>0$ throughout the channel. It is the $O(Re_p)$ modification of the eigendirections that is critical since, in the Jeffery-averaged approximation, these directions make angles of $\pi/4$ and $3\pi/4$ with the flow, and the neutral direction is therefore aligned with the vertical; the image-induced velocity along this direction is zero, leading to no migration. The eigenvectors of the inertially averaged stresslet exhibit a clockwise\,(anticlockwise) rotation in the lower\,(upper) half, relative to the Jeffery-averaged ones. The resulting interaction with the upper wall is pictorially depicted in Figure~\ref{fig:imagestress}a, where the image stresslet shown corresponds to a plane free surface instead. A repulsion arises in this case due to the image-stresslet-induced velocity at the spheroid\,(the original stresslet) location, that owes its origin to the aforementioned inertia-induced rotation of the Jeffery-averaged stresslet. Note that a no-slip condition at a plane wall leads to a more complex image system with a stresslet of the opposite sign, and that is only one of multiple singularities, all of which contribute comparably at distances of order the spheroid separation. Nevertheless, the resulting migration is qualitatively similar to that due to the lone image-stresslet induced by a free surface, as has been shown in earlier efforts\,(see Appendix B2 in \citep{Spagnolie_Lauga_2012}). Image-stresslet-based mechanisms similar to the above have been postulated to explain attractive interactions between microswimmers and plane boundaries~\citep{BerkeLauga_PRL2008}, and are also thought to govern shear-induced migration of polymer molecules and deformable particles in confined shear flows\citep{graham2011}. 

The above stresslet-repulsion picture is restricted to small $Re_c$, however. The inertial stresslet orientation changes with increasing $Re_c$, with its form close to rotation arrest being given in $\S$\ref{sec:results}. While the changed orientation would still appear consistent with a repulsive wall-interaction based on the Stokesian stresslet field\,(see Fig.\ref{fig:imagestress}b), it is difficult to say whether this small-$Re_c$ mechanism is relevant anymore. This is because, for large $Re_c$, the stresslet velocity is nontrivially altered on length scales larger than the inertial screening length\,($H Re_c^{-1/2}$), and it is not at all obvious if the resulting outer-region velocity field is still consistent with a repulsive interaction. Signatures of the outer-region alteration can be seen, for instance, in Figure 9 in \cite{Morris_JFM2004}, which shows the structure of the disturbance field around a torque-free sphere, in plane Couette flow, at finite $Re_p$. The present scenario is further complicated by the role of the quadratic component of the ambient plane Poiseuille profile.

\begin{figure}
	\centering
    \includegraphics[width=0.8\textwidth]{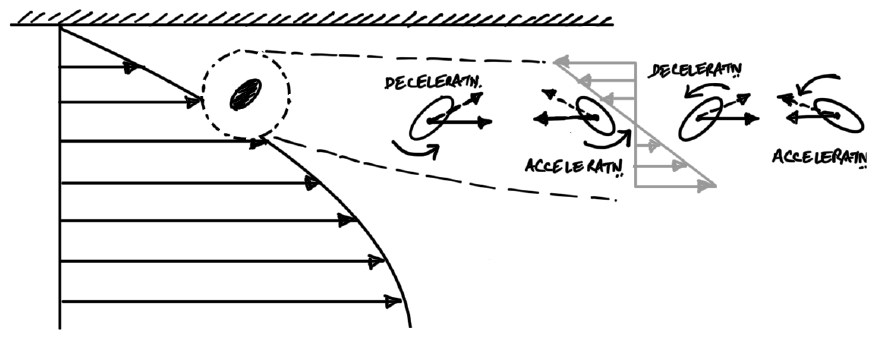}
    \caption{(a) Particle-inertia-driven migration of a freely rotating spheroid, towards the (upper)\,wall, in an ambient plane Poiseuille flow. The local simple shear flow causes the spheroid to tumble, while the quadratic component leads to a time periodic translation velocity in the flow direction. The solid arrows denote the inertial reaction due to the associated acceleration, while the dashed arrows denote the wallward drift in response to this inertial force at different times during a single tumble.}
	\label{fig:liftpartinertia}
\end{figure}

We mention the effort of Tornberg and co-workers~\citep{Tornberg_JFM2021} who have analyzed the role of particle inertia in leading to a cross-stream migration for an unbounded plane Poiseuille profile. This is because the resulting direction of migration is predicted to be opposite to that induced by fluid inertia above. The importance of particle inertia is measured by the Stokes number, defined as $St = \rho_p V_{max}L^2/H\mu$, $\rho_p$ being the particle density. Although the analysis in \citet{Tornberg_JFM2021} was for arbitrary $St$, for the neutrally buoyant particles\,($\rho_p = \rho_f$) of interest here, $St = Re_p \ll 1$. Figure~\ref{fig:liftpartinertia} serves to clarify the physics underlying this `wallward' migration for small $Re_p$. The migration arises as the time-averaged lateral drift of a freely rotating spheroid translating under the action of an imposed force. The instantaneous drift is the result of an orientation-dependent mobility contracted with the force that is minus the translational acceleration\,($-St\frac{d{\bm U}_p}{dt} = -Re_p\frac{d{\bm U}_p}{dt}$ in dimensionless terms), with the time average, for small $Re_p$, being over orientations sampled by the spheroid over a Jeffery period. A tumbling spheroid in Poiseuille flow moves fastest when aligned with the flow direction, and slowest when oriented perpendicular to it, and the acceleration above is associated with this time varying velocity. As the spheroid rotates from the flow-aligned to the vertical orientation, it decelerates, so that $-Re_p\frac{d{\bm U}_p}{dt}$ points to the right; the resulting drift is along the dashed arrow, which points upward. In the next quarter period, the spheroid accelerates as it rotates into flow-alignment, leading to a reversal in the direction $-Re_p\frac{d{\bm U}_p}{dt}$. The changed orientation, however, ensures that the lateral drift continues to point upward. The same sequence repeats over the next half cycle, which allows for a net wallward drift over a full tumble period.

The above contribution arises from the term proportional to $\langle {\bm U}^t \cdot \frac{d{\bm U}_p}{dt}\rangle$ in a reciprocal theorem formulation; see \eqref{eq:RTinstliftvel} in Appendix \ref{App:LiftSmallRec}. This may be used to show that the particle inertial contribution is $O(\lambda Re_p)$, and therefore, asymptotically small compared to the fluid inertial contributions examined here, that are $O(Re_p)$ in the limit $\lambda \rightarrow 0$; we have verified this smallness based on an explicit evaluation. For the extreme aspect ratios of interest, the particle inertial contribution is, in fact, even smaller, on account of the mass of a slender fiber/thin disk being asymptotically small compared to that of a spherical blob of fluid\,(of the same density) circumscribing it and that determines the fluid inertial forces. This contribution could, however, be important for spheroids with order unity aspect ratios. In this case, the particle-inertia-induced drift will compete with an $O(\lambda Re_p)$ fluid inertial contribution that must arise from a Jeffery-averaged evaluation of the nonlinear inertial terms, with the dominant contributions originating in a region of $O(L^3)$ around the rotating spheroid~\citep{anand2023Jeff}. These finite-size contributions are to be contrasted with the contributions evaluated here, that pertain to a point-particle framework, and originate on length scales greater than $O(L)$, regardless of $Re_c$. Despite being $O(\lambda)$ smaller for $Re_c \sim O(1)$, the finite-size contributions are expected to become important for sufficiently large $Re_c$. Indeed, the recent effort of \citet{anand2024FiniteSize} shows that the $O(\lambda Re_p)$ finite-size contribution for spheres becomes comparable to the point-particle contribution for large $Re_c$, leading to the emergence of a new pair of inner equilibrium locations~\citep{anand2024FiniteSize}. The resulting migration diagram in the $(Re_c,\lambda)$-plane has served to rationalize a series of observations dating back to the discovery of the inner equilibrium ring by \citet{matas2004}. The calculation of the $O(\lambda Re_p)$ contribution for spheroids\,(which, unlike spheres, arises due to both fluid and particle inertial effects), and its implication for migration behavior, will be reported in a separate communication.

\begin{bmhead}[Acknowledgements]
The first author\,(Prateek Anand) was partly supported by SERB grant number CRG/2020/004137.
\end{bmhead}
\begin{bmhead}[Declaration of Interests]The authors report no conflict of interest.
\end{bmhead}
\begin{bmhead}[Author ORCID]
Prateek Anand, https://orcid.org/0000-0001-6922-0894; Ganesh Subramanian, https://orcid.org/0000-0003-4314-3602
\end{bmhead}

\begin{appen}
\section{Derivation of jump conditions for the Fourier transformed velocity and pressure fields} \label{App:JumpCondns}
In this Appendix, we derive the jump conditions given in (\ref{eq:FTJumpCondns}a-d). To avoid cumbersome notation, we avoid use of time-averaging brackets for the dependent variables in (\ref{eq:FTODEs}a,b); thus, we use $\hat{\langle P\rangle} \equiv \hat{\text{P}}$ and $\hat{\langle U_2\rangle} \equiv \hat{\text{U}}_2$, which leads to the governing equations in the form:
\begin{subequations}
	\begin{align}
	\frac{d^2 \hat{\text{P}}}{dR_2^2}-k_\perp^2\hat{\text{P}}&=2\iota k_1  \hat{\text{U}}_2(\beta+2\gamma R_2 Re_c^{-1/2})+\beta(k_1^2\langle S_{11}\rangle +k_3^2\langle S_{33}\rangle )\delta(R_2)\nonumber\\
	&+2\iota \beta k_1 \langle S_{12}\rangle \delta'(R_2)-\beta \langle S_{22}\rangle\delta''(R_2),
	\label{eq:PressureFTODE1}\\
	\frac{d^2\hat{\text{U}}_2}{dR_2^2}-k_\perp^2\hat{\text{U}}_2&=\frac{d\hat{\text{P}}}{dR_2}-\iota k_1\hat{\text{U}}_2(\beta R_2+\gamma R_2^2 Re_c^{-1/2})-\iota \beta k_1 \langle S_{12}\rangle\delta(R_2)+\beta \langle S_{22}\rangle\delta'(R_2).
	\label{eq:VelocityFTODE1}
	\end{align} 
\end{subequations}
Here, we have explicitly written down the form of the inertially averaged stresslet forcing, given by (\ref{eq:inertialstressletforcing}) in the main manuscript; note that the different components have been plotted as a function of $Re_c$ in Figure~\ref{fig:Rearrest&stresslet}b. The most singular term in this forcing, the one proportional to $\delta''(R_2)$ in (\ref{eq:PressureFTODE1}), only leads to a localized pressure contribution proportional to $\delta(R_2)$, with no accompanying velocity field. The aforesaid singular forcing can therefore be accommodated by defining a modified pressure field, $\hat{\text{T}}=\hat{\text{P}}+\beta \langle S_{22}\rangle\delta(R_2)$, whence one obtains:
\begin{subequations}
	\begin{align}
	\frac{d^2 \hat{\text{T}}}{dR_2^2}-k_\perp^2\hat{\text{T}}&=2\iota k_1  \hat{\text{U}}_2(\beta+2\gamma R_2 Re_c^{-1/2})+\beta[(k_1^2-k_3^2)\langle S_{11}\rangle-(k_1^2+2 k_3^2)\langle S_{22}\rangle]\delta(R_2) +2\iota \beta k_1 \langle S_{12}\rangle \delta'(R_2), 
	\label{eq:PressureFTODE2}\\
	\frac{d^2\hat{\text{U}}_2}{dR_2^2}-k_\perp^2\hat{\text{U}}_2&=\frac{d\hat{\text{T}}}{dR_2}-\iota k_1\hat{\text{U}}_2(\beta R_2+\gamma R_2^2 Re_c^{-1/2})-\iota \beta k_1 \langle S_{12}\rangle\delta(R_2)
	\label{eq:VelocityFTODE2}.
	\end{align} 
\end{subequations}
Based on the order of the derivatives in (\ref{eq:PressureFTODE2}) and (\ref{eq:VelocityFTODE2}), and the singular forcings on the RHS, one postulates the following forms:
\begin{align}
\hat{\text{T}}&=\hat{\text{T}}^-(R_2)+[\hat{\text{T}}^+(R_2)-\hat{\text{T}}^-(R_2)]\mathcal{H}(R_2), \label{eq:T1}\\
\hat{\text{U}}_2'&=\hat{\text{U}}_2^{-'}(R_2)+[\hat{\text{U}}_2^{+'}(R_2)-\hat{\text{U}}_2^{-'}(R_2)]\mathcal{H}(R_2),
\label{eq:U21}
\end{align}
where the superscripts `+' and `-' denote function definitions for $R_2>0$ and $R_2<0$, respectively, and $\mathcal{H}(R_2)$ is the Heaviside function. Thus, $\hat{T}$ and $\hat{U}_2'$ undergo step jumps across $R_2 = 0$. The latter implies that $\hat{U}_2$ is itself continuous across $R_2 = 0$, that is, $\hat{U}^{+'}_2(0) = \hat{U}^{-'}_2(0)$, which serves as the first of the matching conditions; this corresponds to (\ref{eq:FTJumpCondns}c) in the main manuscript. 

Differentiating \eqref{eq:T1} and \eqref{eq:U21}, one obtains: 
\begin{subequations}
\begin{align}
\hat{\text{T}}'&=\hat{\text{T}}^{-'}(R_2)+[\hat{\text{T}}^{+'}(R_2)-\hat{\text{T}}^{-'}(R_2)]\mathcal{H}(R_2)+[\hat{\text{T}}^{+}(0)-\hat{\text{T}}^{-}(0)]\delta(R_2), \label{eq:T2}\\
\hat{\text{T}}''&=\hat{\text{T}}^{-''}(R_2)+[\hat{\text{T}}^{+''}(R_2)-\hat{\text{T}}^{-''}(R_2)]\mathcal{H}(R_2)+2[\hat{\text{T}}^{+'}(0)-\hat{\text{T}}^{-'}(0)]\delta(R_2), \nonumber\\
&+[\hat{\text{T}}^{+}(R_2)-\hat{\text{T}}^{-}(R_2)]\delta'(R_2), \label{eq:T3}\\
\hat{\text{U}}_2''&=\hat{\text{U}}_2^{-''}(R_2)+[\hat{\text{U}}_2^{+''}(R_2)-\hat{\text{U}}_2^{-''}(R_2)]\mathcal{H}(R_2)+[\hat{\text{U}}_2^{+'}(0)-\hat{\text{U}}_2^{-'}(0)]\delta(R_2).
\label{eq:U22}
\end{align}
\end{subequations}
The piecewise regular functions in \eqref{eq:T1}-\eqref{eq:U22} can be Taylor-expanded about the particle position ($R_2=0$), which, at leading order, yields:
\begin{align}
\hat{\text{T}}&\approx\hat{\text{T}}^-(0)+[\hat{\text{T}}]_0\,\mathcal{H}(R_2), \label{eq:ST1}\\
\hat{\text{T}}'&\approx\hat{\text{T}}^{-'}(0)+[\hat{\text{T}}']_0\,\mathcal{H}(R_2)+[\hat{\text{T}}]_0\,\delta(R_2), \label{eq:ST2}\\
\hat{\text{T}}''&\approx\hat{\text{T}}^{-''}(0)+[\hat{\text{T}}'']_0\,\mathcal{H}(R_2)+[\hat{\text{T}}']_0\,\delta(R_2)+[\hat{\text{T}}]_0\,\delta'(R_2), \label{eq:ST3}\\
\hat{\text{U}}_2&\approx\hat{\text{U}}_2^{-}(0),\label{eq:SU21}\\
\hat{\text{U}}_2'&\approx\hat{\text{U}}_2^{-'}(0)+[\hat{\text{U}}_2']_0\,\mathcal{H}(R_2),\label{eq:SU22}\\
\hat{\text{U}}_2''&\approx\hat{\text{U}}_2^{-''}(0)+[\hat{\text{U}}_2'']_0\,\mathcal{H}(R_2)+[\hat{\text{U}}_2']_0\,\delta(R_2),\label{eq:SU23}
\end{align}
where the symbol $[.]_0$ denotes the jump in function value across $R_2=0$. 
To obtain the three remaining jump conditions\,((\ref{eq:FTJumpCondns}ab) and (\ref{eq:FTJumpCondns}d) in the main manuscript), we substitute the expansions \eqref{eq:ST1}-\eqref{eq:SU23} in \eqref{eq:PressureFTODE2} and \eqref{eq:VelocityFTODE2}, multiply both sides of the equation with $R_2^n$, and integrate from $R_2=0^-$ to $R_2=0^+$. This yields: 
\begin{itemize}
	\item For $n=0$:
\begin{align}
[\hat{\text{T}}']_0&= \beta \left[(k_1^2-k_3^2)\langle S_{11}\rangle-(k_1^2+2 k_3^2)\langle S_{22}\rangle\right], \\
[\hat{\text{U}}_2']_0&=\iota \beta k_1 \langle S_{12}\rangle. \\
\end{align}
	\item and, for $n=1$:
\begin{align}
[\hat{\text{T}}]_0&=2\iota \beta k_1 \langle S_{12}\rangle.
\label{eq:Jump2}
\end{align}
\end{itemize}
The second relation for $n=1$, and all of the relations obtained for $n \geq 2$ are trivial identities.

\section{Inertial lift velocity for $Re_c\ll1$}\label{App:LiftSmallRec}


We start off by rewriting the time-averaged governing equations, given in $\S$\ref{sec:Reclarge}\,(see (\ref{eq:NSOuterEqnTA}a,b) therein), in the following form:
\begin{subequations}
	\begin{align}
	\nabla'^2 \langle \bm{u}' \rangle -\bm{\nabla}' \langle p' \rangle 
    &=Re_c\Big(\langle \bm{u}'\bm{\cdot\nabla}' \bm{u}' \rangle +\langle \bm{u}' \rangle \bm{\cdot\nabla}' \bm{u}^{\infty'}+\bm{u}^{\infty'}\bm{\cdot\nabla}'\langle\bm{u}'\rangle \Big)+\langle\bm{S}\rangle\bm{\cdot\nabla}'\delta(\bm{r}'),\\
	\bm{\nabla}'\bm{\cdot} \langle \bm{u}' \rangle &=0,
	\end{align} \label{eq:NS1_app}
\end{subequations}
where $\langle\bm{u}'\rangle$ satisfies the following boundary conditions:
\begin{subequations}
	\begin{align} 
	\langle\bm{u}'\rangle&\rightarrow 0 \text{ for } r_1',r_3'\rightarrow \infty\,(r_2'\hspace*{0.05in}\text{fixed}),\\
	\langle\bm{u}'\rangle&=0 \text{ at } r_2'=-s, 1-s.
	\end{align} \label{eq:JABC} 
\end{subequations}
Anticipating the dominant scale that contributes to the inertial lift velocity for small $Re_c$~\citep{anand2023Jeff}, we have used the channel width $H$ for purposes of scaling lengths and $V_\text{max}\lambda$ for scaling velocities in (\ref{eq:NS1_app}a,b), as a result of which $Re_c = \lambda^{-2} Re_p$, rather than $Re_p$ appears in front of the inertial terms. Further, as mentioned in $\S$\ref{sec:form}, inertia for small $Re_c$ acts as a regular perturbation, and therefore, the inertial acceleration terms have been shifted to the RHS of (\ref{eq:NS1_app}a), consistent with interpreting them as a forcing function to the Stokes equations; here $\bm{u}^{\infty'}=(\beta r_2'+\gamma r_2'^2)\bm{1}_1$. Also note that the boundary condition on the surface of the spheroid\,(equation (\ref{eq:BC1}a) in $\S$\ref{sec:form}) has now been replaced by the time-averaged stresslet forcing, which is the second term on the RHS of (\ref{eq:NS1_app}). This is consistent with the neutrally buoyant spheroid in an ambient shearing flow acting as a force-dipole singularity on length scales of $O(H)$, at leading order, for small $\lambda$. Next, based on the scaling arguments given in \citet{anand2023Jeff}, the nonlinear inertial terms in (\ref{eq:NS1_app}a) are neglected, whence one has to solve:
\begin{subequations}
	\begin{align}
	\nabla'^2 \langle \bm{u}' \rangle -\bm{\nabla}' \langle p'\rangle 
    &=Re_c\Big(\langle\bm{u}'\rangle\bm{\cdot\nabla}' \bm{u}^{\infty'}+\bm{u}^{\infty'}\bm{\cdot\nabla}'\langle\bm{u}'\rangle \Big)+\langle\bm{S}\rangle\bm{\cdot\nabla}'\delta(\bm{r}'),\\
	\bm{\nabla}'\bm{\cdot} \langle \bm{u}' \rangle &=0,
	\end{align} \label{eq:linearNS1}
\end{subequations}
for small $Re_c$, subject to the conditions (\ref{eq:JABC}a,b).

For sufficiently small $Re_c$, so one is well away from rotation arrest regardless of $\kappa$ or the channel location, one may expand the time averaging operator in the form:
\begin{align}
\langle f(\phi_j)\rangle=\frac{1}{T}\int_{2\pi}^{0}f(\phi_j) \dfrac{d\phi_j}{\dot{\phi}_j} = \frac{1}{T_{\text{jeff}}}\left[ \int_{2\pi}^{0}\frac{f(\phi_j)}{\dot{\phi}^0_j} d\phi_j - Re_c\int_{2\pi}^{0}\frac{f(\phi_j)\dot{\phi}^1_j}{(\dot{\phi}^0_j)^2} d\phi_j\right],
\label{Perturbed_Timeavg}
\end{align}
which is the sum of an $O(1)$ average based on the Jeffery angular velocity, and an $O(Re_c)$ inertial correction; note that the time period $T$ remains unchanged from its Jeffery value to this order. Next, writing $\langle . \rangle = \langle .\rangle_J + Re_c\langle . \rangle_I$,  and expanding the velocity and pressure fields as $\langle\bm{u}' \rangle=\langle\bm{u}' \rangle_0 +Re_c \langle\bm{u}' \rangle_1$ and $\langle p \rangle=\langle p \rangle_0+Re_c \langle p'\rangle_1$, one obtains at $O(1)$:
\begin{subequations}\begin{align}
\nabla^{'2} \langle\bm{u}'\rangle_0-\bm{\nabla}' \langle p'\rangle_0 &=\langle\bm{S}\rangle_J\bm{\cdot\nabla}'\delta(\bm{r}'),\\
\bm{\nabla}'\bm{\cdot} \langle\bm{u}'\rangle_0&=0,
\end{align} \label{eq:O1GovEqn}
\end{subequations}
with
\begin{subequations}
	\begin{align} 
	\langle\bm{u}'\rangle_0&\rightarrow 0 \text{ for } r_1',r_3'\rightarrow \infty\,(r_2'\hspace*{0.05in}\text{fixed}),\\
	\langle\bm{u}'\rangle_0&=0 \text{ at } r_2'=-s, 1-s,
	\end{align} \label{eq:O1BC} 
\end{subequations}
and at $O(Re_c)$:
\begin{subequations} \begin{align}
\nabla^{'2} \langle\bm{u}'\rangle_1-\bm{\nabla}' \langle p'\rangle_1 &= \langle\bm{u} \rangle_0\cdot\bm{\nabla' u}^{\infty'}+\bm{u}^{\infty'}\cdot\bm{\nabla'} \langle \bm{u}\rangle_0 +\langle\bm{S}\rangle_I\bm{\cdot\nabla}'\delta(\bm{r}'),\\
\bm{\nabla'\cdot} \langle\bm{u}'\rangle_1 &=0,
\end{align} \label{eq:JAGovEqn}
\end{subequations}
with
\begin{subequations}
	\begin{align} 
	\langle\bm{u}'\rangle_1&\rightarrow 0 \text{ for } r_1',r_3'\rightarrow \infty\,(r_2'\hspace*{0.05in}\text{fixed}),\\
	\langle\bm{u}'\rangle_1&=0 \text{ at } r_2'=-s, 1-s.
	\end{align} \label{eq:ORecBC} 
\end{subequations}
The $Re_c$-scalings for the various components of $\langle\bm{S}\rangle_I$, in (\ref{eq:JAGovEqn}a), away from rotation arrest, are indicated in Figure~\ref{fig:Rearrest&stresslet}b. Since $\langle S_{ii} \rangle_I \sim O(Re_c)$, but $\langle S_{12}\rangle_I \sim O(Re_c^2)$, the inertial correction to the stresslet, $\langle \bm S \rangle_I$, may be approximated as a diagonal tensor at the above order.

The $O(1)$ problem defined by (\ref{eq:O1GovEqn}a-\ref{eq:O1BC}b) yields the Jeffery-averaged stresslet velocity field, $\langle \bm{u}' \rangle_0$, that satisfies no-slip boundary conditions on a pair of plane parallel boundaries~\citep{anand2023Jeff}. Since $\langle \bm S\rangle_J \propto {\bm E}$ with $\langle S_{12} \rangle_J$ being the only nontrivial component\,(see discussion in the  paragraph following (\ref{eq:NSOuterBCTA}c)), $\langle \bm{u}'\rangle_0$ is, to within a proportionality factor, identical to the velocity field induced by a neutrally buoyant sphere\,(treated as a point force-dipole) confined between plane parallel walls. There is, of course, no lift at this order, on account of Stokesian reversibility constraints.

In light of the homogeneous boundary conditions, the $O(Re_c)$ velocity field, defined by (\ref{eq:JAGovEqn}a-\ref{eq:ORecBC}b), may evidently be written as the sum of two contributions: $\langle \bm u' \rangle_{11}$ that is driven by the linearized inertial terms evaluated using the aforementioned Jeffery-averaged stresslet field; and  $\langle \bm u' \rangle_{12}$ that is driven by the inertial correction to the Jeffery-averaged stresslet forcing. The first contribution, $\langle \bm u \rangle_{11}$, was determined in \cite{anand2023Jeff} using a reciprocal theorem formulation; see section 3 therein which pertained to the small-$Re_c$ calculational framework. The solution to (\ref{eq:JAGovEqn}a-\ref{eq:ORecBC}b) may be formally written in terms of a convolution integral involving a bounded-domain Green's function, $\bm G(\bm r'|\bm r'')$, that satisfies no-slip boundary conditions at the two walls. Thus,
\begin{align}
\langle \bm u' \rangle_1(\bm r'|\bm r'') =& \langle \bm u' \rangle_{11}(\bm r'|\bm r'') + \langle \bm u' \rangle_{12}(\bm r'|\bm r'')\, \nonumber\\
=& \displaystyle\int d\bm r'' \bm G(\bm r'|\bm r'') \bm\cdot[\langle\bm u'\rangle_0(\bm r'|\bm r'') \bm{\cdot\nabla}''\bm u^{\infty'}(\bm r'')+\bm u^{\infty'}(\bm r'')\bm{\cdot\nabla}'' \langle \bm u'\rangle_0(\bm r'|\bm r'')] \nonumber\\
&+ \displaystyle\int d\bm r'' \bm G(\bm r'|\bm r'') \bm\cdot[\langle\bm S \rangle_I\bm{\cdot\nabla}''\delta(\bm r'')].
\end{align}
Since the inertial lift velocity is essentially $\langle  u_2\rangle_1$ at $\bm r' = 0$, corresponding to the location of the spheroid\,(interpreted as a time-averaged point singularity at the origin), one may write the following formal expression for the inertial lift velocity:
\begin{align}
\langle V_p\rangle =-Re_p\int_{V^F+V^P}d\bm r'' &(\bm{1}_2 \bm{\cdot G}(\bm 0|\bm r'')) \bm\cdot [\langle\bm u' \rangle_0(\bm 0|\bm r'')\bm{\cdot\nabla}'' \bm u^{\infty'}(\bm r'')+\bm u^{\infty'}(\bm r'')\bm{\cdot\nabla}'' \langle \bm u\rangle_0(\bm 0|\bm r'')] \nonumber\\
&-Re_p\displaystyle\int_{V^F+V^P} d\bm r'' (\bm 1_2\bm{\cdot G}(\bm 0|\bm r'')) \bm\cdot [\langle\bm{S}\rangle_I\bm{\cdot\nabla}'\delta''(\bm r'')],
\label{eq:Full_RecContribution}
\end{align}
where, on account of the point-particle\,(stresslet/Stokeslet) approximation, the domain of integration in (\ref{eq:Full_RecContribution}) includes the entire volume between the plane boundaries\,($V^F + V^P$), including the volume inside the original finite-sized particle\,($V^P$). Noting that the $2$-component of the Greens function is essentially the wall-bounded velocity field\,(${\bm u}_{St}$) driven by a gradient-aligned Stokeslet at the origin, and further, using $\langle \bm u \rangle_0 \equiv \langle {\bm u}_{str} \rangle_J$ for the wall-bounded Jeffery-averaged stresslet velocity field, the first term in (\ref{eq:Full_RecContribution}) may be rewritten in the form:
\begin{align}
\langle V_p\rangle_1 =&-Re_p\int_{V^F+V^P}\bm u'_{St}\bm\cdot\left(\langle\bm u'_{str}\rangle_J \bm{\cdot\nabla}' \bm u^{\infty'}+\bm u^{\infty'}\bm{\cdot\nabla}'\langle\bm u'_{str}\rangle_J \right) dV'.
\label{eq:RTVolIntegral}
\end{align}
 The above is the expression for the Jeffery-averaged lift velocity obtained in \citet{anand2023Jeff}. Note that $\langle\bm{u}_{str}'\rangle_J$ only involves $\langle \bm{S}_{12} \rangle_J$ which may be pulled out of the volume integral, as a result of which (\ref{eq:RTVolIntegral}) may be written as the product of a $\kappa$-dependent factor and an integral that gives the inertial lift velocity for a sphere. It is this multiplicative decomposition that ensures that, within a Jeffery-averaged framework, the equilibrium locations for a spheroid are identical to those for a sphere.  
The actual calculation of the integral in (\ref{eq:RTVolIntegral}) is more easily done in Fourier space, via use of the convolution theorem, which leads to the following expression~\citep{anand2023Jeff}:
\begin{align}
\langle V_p\rangle_1=\langle V_p\rangle_J = Re_p\langle S_{12}\rangle_J(\kappa)\Big(\beta^2 F(s)+\beta\gamma G(s)\Big).
\label{eq:VpJeffAvgd}
\end{align}
The expression for $\langle S_{12} \rangle_J$ is given by \eqref{eq:S12J} in the main manuscript. The functions $F(s)$ and $G(s)$ are given by:
\begin{align}
F(s)&=\int_0^\infty dk_\perp' \dfrac{k_\perp'\,\,e^{-k_\perp' (27 s+16)} I(k_\perp',s)}{48 \pi \left(e^{2 k_\perp'}-1\right) \left[-2 e^{2 k_\perp'} \left(2 k_\perp'^2+1\right)+e^{4 k_\perp'}+1\right]^2}\label{eq:Fs},\\ 
G(s)&=\int_0^\infty dk_\perp' \dfrac{e^{-k_\perp' (27 s+16)} J(k_\perp',s)}{192 \pi\, k_\perp'^2 \left(e^{2 k_\perp'}-1\right) \left[-2 e^{2 k_\perp'} \left(2 k_\perp'^2+1\right)+e^{4 k_\perp'}+1\right]^2}\label{eq:Gs},
\end{align}
where $k_\perp'^2=k_1'^2+k_3'^2$, and $I(k_\perp',s)$ and $J(k_\perp',s)$ are defined as:
\allowdisplaybreaks
\begin{align}
I(k_\perp',s)&=-e^{k_\perp' (25 s+18)}(s-1)^2\left[3 k_\perp'^2 (s-1)^2-2 k_\perp' (s-1)+3\right]+e^{k_\perp' (29 s+24)}(s-1)^2\nonumber\\
&\big[3 k_\perp'^2 (s-1)^2+2 k_\perp'(s-1)+3\big]-2 (2 s-1) e^{3 k_\perp' (9 s+8)} \big[6 k_\perp'^3 (s-1) s\nonumber\\
&-4 k_\perp'^2 (s-1) s-3\big]-2 (2 s-1) e^{9 k_\perp' (3 s+2)} \left[6 k_\perp'^3 (s-1) s+4 k_\perp'^2 (s-1) s+3\right]\nonumber\\
&-s^2 e^{k_\perp' (25 s+26)} \left(3 k_\perp'^2 s^2-2 k_\perp' s+3\right)+s^2 e^{k_\perp' (29 s+16)} \left(3 k_\perp'^2 s^2+2 k_\perp's+3\right)\nonumber\\
&-2 e^{k_\perp' (27 s+20)} \big[8 k_\perp'^4 s \big(2 s^2-3 s+1\big)-6 k_\perp'^3 s \left(2 s^2-3 s+1\right)\nonumber\\
&-12 k_\perp'^2 \left(2 s^3-3 s^2+3 s-1\right)-18 s+9\big]+2 e^{k_\perp' (27 s+22)} \big[8 k_\perp'^4 s \left(2 s^2-3 s+1\right)\nonumber\\
&+6 k_\perp'^3 s \left(2 s^2-3 s+1\right)-12 k_\perp'^2 \left(2 s^3-3 s^2+3 s-1\right)-18 s+9\big]\nonumber\\
&+e^{5 k_\perp' (5 s+4)} \big[12 k_\perp'^4 (s-1)^2 s^2+4 k_\perp'^3 (s-1)^2 (4 s-1)+3 k_\perp'^2 \big(4 s^4-12 s^3+14 s^2\nonumber\\
&-12 s+5\big)-2 k_\perp' \left(4 s^3-9 s^2+9 s-3\right)+3 \left(4 s^2-6 s+3\right)\big]\nonumber\\
&-e^{k_\perp' (29 s+22)} \big[12 k_\perp'^4 (s-1)^2 s^2-4 k_\perp'^3 (s-1)^2 (4 s-1)+3 k_\perp'^2 \big(4 s^4-12 s^3+14 s^2\nonumber\\
&-12 s+5\big)+2 k_\perp' \left(4 s^3-9 s^2+9 s-3\right)+3 \left(4 s^2-6 s+3\right)\big]\nonumber\\
&+e^{k_\perp' (25 s+24)} \big[12 k_\perp'^4 (s-1)^2 s^2+4 k_\perp'^3 s^2 (4 s-3)+3 k_\perp'^2 \left(4 s^4-4 s^3+2 s^2+4 s-1\right)\nonumber\\
&+k_\perp' \left(-8 s^3+6 s^2-6 s+2\right)+12 s^2-6 s+3\big]-e^{k_\perp' (29 s+18)} \big[12 k_\perp'^4 (s-1)^2 s^2\nonumber\\
&-4 k_\perp'^3 s^2 (4 s-3)+3 k_\perp'^2 \left(4 s^4-4 s^3+2 s^2+4 s-1\right)+k_\perp' \left(8 s^3-6 s^2+6 s-2\right)\nonumber\\
&+12 s^2-6 s+3\big]-e^{k_\perp' (25 s+22)} \big[24 k_\perp'^4 (s-1)^2 s^2+4 k_\perp'^3 (2 s-1)^3\nonumber\\
&+3 k_\perp'^2 \big(6 s^4-12 s^3+10 s^2-4 s+3\big)-6 k_\perp' \left(2 s^3-3 s^2+3 s-1\right)+9 \left(2 s^2-2 s+1\right)\big]\nonumber\\
&+e^{k_\perp' (29 s+20)} \big[24 k_\perp'^4 (s-1)^2 s^2-4 k_\perp'^3 (2 s-1)^3+3 k_\perp'^2 \left(6 s^4-12 s^3+10 s^2-4 s+3\right)\nonumber\\
&+6 k_\perp' \left(2 s^3-3 s^2+3 s-1\right)+9 \left(2 s^2-2 s+1\right)\big], \label{eq:IJeff}\\
J(k_\perp',s)&=e^{k_\perp' (25 s+18)} \big[8 k_\perp'^5 (s-1)^5-6 k_\perp'^4 (s-1)^4+60 k_\perp'^3 (s-1)^3\nonumber\\
&+24 k_\perp'^2 (s-1)^2+54 k_\perp' (s-1)+27\big]-27 e^{k_\perp' (27 s+16)}-27 e^{k_\perp' (27 s+26)}\nonumber\\
&+e^{k_\perp' (29 s+24)} \big[-8 k_\perp'^5 (s-1)^5-6 k_\perp'^4 (s-1)^4-60 k_\perp'^3 (s-1)^3+24 k_\perp'^2 (s-1)^2\nonumber\\
&-54 k_\perp' (s-1)+27\big]-e^{k_\perp' (29 s+16)}\big[8 k_\perp'^5 s^5+6 k_\perp'^4 s^4+60 k_\perp'^3 s^3-24 k_\perp'^2 s^2+54 k_\perp' s-27\big]\nonumber\\
&+e^{k_\perp' (25 s+26)} \big[8 k_\perp'^5 s^5-6 k_\perp'^4 s^4+60 k_\perp'^3 s^3+24 k_\perp'^2 s^2+54 k_\perp' s+27\big]\nonumber\\
&+e^{9 k_\perp' (3 s+2)} \big[32 s (3 s^3-6 s^2+4 s-1) k_\perp'^6+8 s (7 s^3-14 s^2+10 s-3) k_\perp'^5\nonumber\\
&+240 (s-1) s k_\perp'^4+24 (s-1) s k_\perp'^3+108 k_\perp'^2+108 k_\perp'+81\big]\nonumber\\
&+e^{3 k_\perp' (9 s+8)} \big[32 s (3 s^3-6 s^2+4 s-1) k_\perp'^6-8 s (7 s^3-14 s^2+10 s-3) k_\perp'^5\nonumber\\
&+240 (s-1) s k_\perp'^4-24 (s-1) s k_\perp'^3+108 k_\perp'^2-108 k_\perp'+81\big]\nonumber\\
&-2 e^{k_\perp' (27 s+22)} \big[16 s (5 s^3-10 s^2+6 s-1) k_\perp'^7+16 s (3 s^3-6 s^2+4 s-1) k_\perp'^6\nonumber\\
&-12 s (7 s^3-14 s^2-26 s+33) k_\perp'^5+120 (s-1) s k_\perp'^4-36 (s^2-s+6) k_\perp'^3+54 k_\perp'^2\nonumber\\
&-162 k_\perp'+27\big]+2 e^{k_\perp' (27 s+20)} \big[16 s \left(5 s^3-10 s^2+6 s-1\right) k_\perp'^7\nonumber\\
&-16 s (3 s^3-6 s^2+4 s-1) k_\perp'^6-12 s (7 s^3-14 s^2-26 s+33) k_\perp'^5\nonumber\\
&-120 (s-1) s k_\perp'^4-36 (s^2-s+6) k_\perp'^3-54 k_\perp'^2-162 k_\perp'-27\big]\nonumber\\
&+2 e^{k_\perp' (29 s+18)} \big[16 (s-1)^3 s^2 k_\perp'^7-4 (2-3 s)^2 s^2 k_\perp'^6+4 (4 s^5-5 s^4+28 s^3\nonumber\\
&-22 s^2-5 s+1) k_\perp'^5+3 (4 s^4-4 s^3-50 s^2-4 s+1) k_\perp'^4+6 (20 s^3-15 s^2+13 s+5) k_\perp'^3\nonumber\\
&-6 (8 s^2-4 s+11) k_\perp'^2+27 (4 s-1) k_\perp'-54\big]-2 e^{k_\perp' (25 s+24)} \big[16 (s-1)^3 s^2 k_\perp'^7\nonumber\\
&+4 s^2 (2-3 s)^2 k_\perp'^6+4 (4 s^5-5 s^4+28 s^3-22 s^2-5 s+1) k_\perp'^5-3 (4 s^4-4 s^3-50 s^2\nonumber\\
&-4 s+1) k_\perp'^4+6 (20 s^3-15 s^2+13 s+5) k_\perp'^3+6 (8 s^2-4 s+11) k_\perp'^2+27 (4 s-1) k_\perp'\nonumber\\
&+54\big]+2 e^{k_\perp' (29 s+22)} \big[16 (s-1)^2 s^3 k_\perp'^7-4 (3 s^2-4 s+1)^2 k_\perp'^6+4 (4 s^5-15 s^4\nonumber\\
&+48 s^3-72 s^2+35 s-1) k_\perp'^5+3 (4 s^4-12 s^3-38 s^2+100 s-53) k_\perp'^4\nonumber\\
&+6 (20 s^3-45 s^2+43 s-23) k_\perp'^3-6 (8 s^2-12 s+15) k_\perp'^2+27 (4 s-3) k_\perp'-54\big]\nonumber\\
&-2 e^{5 k_\perp' (5 s+4)} \big[16 (s-1)^2 s^3 k_\perp'^7+4 (3 s^2-4 s+1)^2 k_\perp'^6+4 (4 s^5-15 s^4+48 s^3-72 s^2\nonumber\\
&+35 s-1) k_\perp'^5-3 (4 s^4-12 s^3-38 s^2+100 s-53) k_\perp'^4+6 (20 s^3-45 s^2+43 s-23) k_\perp'^3\nonumber\\
&+6 (8 s^2-12 s+15) k_\perp'^2+27 (4 s-3) k_\perp'+54\big]+2 e^{k_\perp' (25 s+22)} \big[16 (s-1)^2 s^2 (2 s-1) k_\perp'^7\nonumber\\
&+4 (18 s^4-36 s^3+26 s^2-8 s+1) k_\perp'^6+4 (6 s^5-15 s^4+66 s^3-84 s^2+25 s+1) k_\perp'^5\nonumber\\
&-3 (6 s^4-12 s^3-94 s^2+100 s-53) k_\perp'^4+6 (30 s^3-45 s^2+41 s-13) k_\perp'^3\nonumber\\
&+72 (s^2-s+2) k_\perp'^2+81 (2 s-1) k_\perp'+81\big]-2 e^{k_\perp' (29 s+20)} \big[16 (s-1)^2 s^2 (2 s-1) k_\perp'^7\nonumber\\
&-4 (18 s^4-36 s^3+26 s^2-8 s+1) k_\perp'^6+4 (6 s^5-15 s^4+66 s^3-84 s^2+25 s+1) k_\perp'^5\nonumber\\
&+3 (6 s^4-12 s^3-94 s^2+100 s-53) k_\perp'^4+6 (30 s^3-45 s^2+41 s-13) k_\perp'^3\nonumber\\
&-72 (s^2-s+2) k_\perp'^2+81 (2 s-1) k_\perp'-81\big].\label{eq:JJeff}
\end{align}
For purposes of the figures that follow, the $k_\perp'$-integrals above are evaluated numerically for various $s$, using Gauss-Legendre quadrature, after replacing the infinite interval with a finite one - $(0,K_\text{max})$; while the value of $K_\text{max}$ needed diverges with decreasing wall separation, $K_\text{max}\approx10000$ suffices for obtaining lift profiles down to $s\approx\ 0.001$.

From (\ref{eq:Full_RecContribution}), one notes that the second lift velocity contribution, arising from the stresslet forcing, may be written in the form:
\begin{align}
\langle V_p\rangle_2=&-Re_p \displaystyle\int_{V^F+V^P} d\bm r' \{ \bm 1_2 \cdot \left[{\bm G}(\bm{0}|{\bm r}') - \frac{1}{8\pi}(\frac{\bm I}{r'} + \frac{{\bm r}'\bm{r}'}{r'^3})\right] \}{\cdot}[\langle \bm{S} \rangle_I \cdot {\bm \nabla}\delta(\bm{r}')] dV',
\label{eq:RTSIVolIntegral}
\end{align}
where we have now subtracted the singular unbounded-domain contribution\,(the Oseen-Burgers tensor), so as to isolate the finite contribution due to the image-singularities that lead to the lift. The generalized function in (\ref{eq:RTSIVolIntegral}) implies that this second contribution is essentially the gradient of the regularized Green's function, at $\bm r' = 0$, contracted with the inertial stresslet tensor.
While the above approach clarifies the physical origin of the second $O(Re_c)$ contribution to the inertial lift\,(see Figure~\ref{fig:imagestress} in $\S$\ref{sec:conclusion}), and the gradient of the bounded-domain Green's function in (\ref{eq:RTSIVolIntegral}) may be evaluated in principle, there exists a simpler alternative. The latter is especially in light of the analysis already carried out in $\S$\ref{sec:Reclarge}, to derive the partially Fourier transformed equations governing the pressure and gradient component of the velocity fields; see (\ref{eq:FTODEs}a,b). Rather than solve these ODEs numerically, as is done in implementing the shooting method protocol in $\S$\ref{sec:Reclarge}, we now solve them analytically via a regular perturbation expansion for $Re_c\ll1$, using both the boundary conditions (\ref{eq:FTBC}a,b) and the jump conditions (\ref{eq:FTJumpCondns}a-d). Again, keeping in mind that inertia is a regular perturbation for small $Re_c$, we rescale the governing equations, the boundary conditions, and jump conditions, using: $(k_1,k_3)=Re_c^{-1/2}(k_1',k_3')$, $R_2=Re_c^{1/2} R_2'$, $\langle\hat{U}_2\rangle=\langle\hat{U}_2'\rangle$ and $\langle\hat{T}\rangle=Re_c^{-1/2}\langle\hat{T}'\rangle$, to obtain:
\begin{subequations}
	\begin{align}
	\frac{d^2 \langle \hat{T}'\rangle}{dR_2'^2}-k_\perp'^2\langle \hat{T}'\rangle&=2\iota k_1' \langle \hat{U}'_2\rangle Re_c(\beta+2\gamma R_2'),\\
	\frac{d^2\langle\hat{U}_2'\rangle}{dR_2'^2}-k_\perp'^2\langle\hat{U}_2'\rangle&=\frac{d\langle \hat{T}'\rangle}{dR_2'}-\iota k_1'\langle\hat{U}_2'\rangle Re_c(\beta R_2'+\gamma R_2'^2),
	\end{align} \label{eq:SmallRecODEs}
\end{subequations}
with the boundary conditions:
\begin{align} 
\langle\hat{U}_2'\rangle&=\frac{d\langle\hat{U}_2'\rangle}{dR_2'}= 0 \text{ at } R_2'=-s,1-s,\label{eq:SmallRecBC}
\end{align} 
and (modified)\,jump conditions:
\begin{subequations}
	\begin{align}
	\langle \hat{T}'\rangle^+ (k_1',0,k_3') -\langle \hat{T}'\rangle^-(k_1',0,k_3') &=2\iota k_1' \beta\langle S_{12}\rangle,\\
	\frac{d\langle \hat{T}'\rangle^+}{dR_2'}(k_1',0,k_3')-\frac{d\langle \hat{T}'\rangle^-}{dR_2'}(k_1',0,k_3') &= \beta \big[(k_1'^2-k_3'^2)\langle S_{11}\rangle-(k_1'^2+2 k_3'^2)\langle S_{22}\rangle\big]\\
	\langle\hat{U}_2'\rangle^+(k_1',0,k_3')&=\langle\hat{U}_2'\rangle^- (k_1',0,k_3') \\
	\frac{d\langle\hat{U}_2'\rangle^+}{dR_2'}(k_1',0,k_3')-\frac{d\langle\hat{U}_2'\rangle^-}{dR_2'}(k_1',0,k_3') &=\iota k_1' \beta \langle S_{12}\rangle.
	\end{align} \label{eq:SmallRecJumpCondns}
\end{subequations}

Next, we define $\langle S_{11}\rangle=Re_c F_{11}(\kappa)$, $\langle S_{22}\rangle=Re_c F_{22}(\kappa)$\,(with $\langle S_{33} \rangle = -(\langle S_{11} \rangle + \langle S_{22} \rangle)$) and $\langle S_{12}\rangle=\langle S_{12}\rangle_J+O(Re_c^2)$, where $F_{11}$ and $F_{22}$ may be obtained numerically as in Fig.\ref{fig:Rearrest&stresslet}a. Using the expansions $\langle\hat{T}'\rangle=\langle\hat{T}'\rangle_0+Re_c(\langle\hat{T}'\rangle_{11}+\langle\hat{T}'\rangle_{12})+O(Re_c^2)$ and $\langle\hat{U}_2'\rangle=\langle\hat{U}_2'\rangle_0+Re_c(\langle\hat{U}_2'\rangle_{11}+\langle\hat{U}_2'\rangle_{12})+O(Re_c^2)$ in (\ref{eq:SmallRecODEs}a,b), (\ref{eq:SmallRecBC}) and (\ref{eq:SmallRecJumpCondns}a-d), one obtains the following systems of equations at $O(1)$ and $O(Re_c)$:
\\\\
{O(1)}:
\begin{subequations}
	\begin{align}
	\frac{d^2 \langle \hat{T}'\rangle_0}{dR_2'^2}-k_\perp'^2\langle \hat{T}'\rangle_0&=0,\\
	\frac{d^2\langle\hat{U}_2'\rangle_0}{dR_2'^2}-k_\perp'^2\langle\hat{U}_2'\rangle_0&=\frac{d\langle \hat{T}'\rangle_0}{dR_2'},
	\end{align} \label{eq:SmallRecODE1}
\end{subequations}
with the following jump conditions:
\begin{subequations}
	\begin{align}
	\langle \hat{T}'\rangle_0^+ (k_1',0,k_3') -\langle \hat{T}'\rangle_0^- (k_1',0,k_3') &=2\iota k_1' \beta\langle S_{12}\rangle_J,\\
	\frac{d\langle \hat{T}'\rangle_0^+}{dR_2'}(k_1',0,k_3')-\frac{d\langle \hat{T}'\rangle_0^-}{dR_2'}(k_1',0,k_3') &= 0\\
	\langle\hat{U}_2'\rangle_0^+(k_1',0,k_3')&=\langle\hat{U}_2'\rangle_0^- (k_1',0,k_3') \\
	\frac{d\langle\hat{U}_2'\rangle_0^+}{dR_2'}(k_1',0,k_3')-\frac{d\langle\hat{U}_2'\rangle_0^-}{dR_2'}(k_1',0,k_3') &=\iota k_1' \beta \langle S_{12}\rangle_J.
	\end{align} \label{eq:SmallRecJumpCondns1}
\end{subequations}
\\
\textbf{O($Re_c$)}: First Inertial contribution
\begin{subequations}
	\begin{align}
	\frac{d^2 \langle \hat{T}'\rangle_{11}}{dR_2'^2}-k_\perp'^2\langle \hat{T}'\rangle_{11}&=2\iota k_1' \langle\hat{U}_2'\rangle_0(\beta+2\gamma R_2'),\\
	\frac{d^2\langle\hat{U}_2'\rangle_{11}}{dR_2'^2}-k_\perp'^2\langle\hat{U}_2'\rangle_{11}&=\frac{d\langle \hat{T}'\rangle_{11}}{dR_2'}-\iota k_1'\langle\hat{U}_2'\rangle_0(\beta R_2'+\gamma R_2'^2),
	\end{align} \label{eq:SmallRecODE2}
\end{subequations}
with there being no kink across $R_2' = 0$.\\
\textbf{O($Re_c$)}: Second Inertial contribution
\begin{subequations}
	\begin{align}
	\frac{d^2 \langle \hat{T}'\rangle_{12}}{dR_2'^2}-k_\perp'^2\langle \hat{T}'\rangle_{12}&=0,\\
	\frac{d^2\langle\hat{U}_2'\rangle_{12}}{dR_2'^2}-k_\perp'^2\langle\hat{U}_2'\rangle_{12}&=\frac{d\langle \hat{T}'\rangle_{12}}{dR_2'},
	\end{align} \label{eq:SmallRecODE3}
\end{subequations}
with the following jump conditions:
\begin{subequations}
	\begin{align}
	\langle \hat{T}'\rangle^+_{12} (k_1',0,k_3') -\langle \hat{T}'\rangle^-_{12} (k_1',0,k_3') &=0,\\
	\frac{d\langle \hat{T}'\rangle^+_{12}}{dR_2'}(k_1',0,k_3')-\frac{d\langle \hat{T}'\rangle^-_{12}}{dR_2'}(k_1',0,k_3') &= \beta \big[F_{11}(k_1'^2-k_3'^2)- F_{22}(k_1'^2+2 k_3'^2)\big]\\
	\langle\hat{U}_2'\rangle^+_{12}(k_1',0,k_3')&=\langle\hat{U}_2'\rangle^-_{12} (k_1',0,k_3') \\
	\frac{d\langle\hat{U}_2'\rangle^+_{12}}{dR_2'}(k_1',0,k_3')-\frac{d\langle\hat{U}_2'\rangle^-_{12}}{dR_2'}(k_1',0,k_3') &=0.
	\end{align} \label{eq:SmallRecJumpCondns3}
\end{subequations}
All of the aforementioned fields are subject to the no-slip conditions \eqref{eq:SmallRecBC}, since these are independent of $Re_c$. Note that the first inertial contribution is driven by the leading order Jeffery-averaged field acting as a forcing function. As a result, $\langle \hat{U}_2' \rangle_{11}$ and $\langle \hat{T}'\rangle_{11}$, defined by (\ref{eq:SmallRecODE1}a,b), are twice differentiable at $R_2' = 0$, owing to $\langle U_2' \rangle_0$ being continuous at this location- this then leads to the absence of any jump relations for these fields. Solving the combined system of equations (\ref{eq:SmallRecODE1}a,b) and (\ref{eq:SmallRecJumpCondns1}a-d) alongwith (\ref{eq:SmallRecODE2}a,b), and evaluating the inverse transform at the origin\,(see \eqref{eq:VpHinch} in $\S$\ref{sec:Reclarge}), leads to the Jeffery-averaged lift, $\langle V_p\rangle_1$, derived in \citet{anand2023Jeff}, and the expression for which was given above; see \eqref{eq:VpJeffAvgd}-\eqref{eq:JJeff}. 

Of interest is the solution of (\ref{eq:SmallRecODE3}a,b) and (\ref{eq:SmallRecJumpCondns3}a-d) which corresponds to the second inertial contribution, and is driven by to the inertial correction to the stresslet. Note that in this case, $\langle \hat{U}_2' \rangle_{12}$ and $\langle \hat{T}'\rangle_{12}$ are solely driven by the jump conditions which contain the $O(Re_c)$ stresslet corrections\,($F_{11}$ and $F_{22}$ in (\ref{eq:SmallRecJumpCondns3}b)). 
The above system of equations can be solved analytically using variation of parameters~\citep{arfkenweber} to obtain $\langle\hat{U}_2'\rangle_{12}$. The outer-region rescalings for $(k_1,k_3)$ alongwith the regular expansion  $Re_c\langle\hat{U}_2'\rangle_{12}$ for the second inertial contribution mentioned earlier can be substituted in \eqref{eq:VpHinch}, thereby yielding the following lift velocity:
\begin{align}
\langle V_p\rangle_2&=\frac{Re_p}{4\pi^2}\Re\left\{\int_{-\infty}^\infty \int_{-\infty}^\infty \langle\hat{U}_2'\rangle^\pm_{12}(k_1',0,k_3')dk_1' dk_3'\right\}\nonumber\\ 
&= Re_p F_{22}(\kappa) \int_0^\infty dk_\perp'\frac{N}{2\pi\left(-2 k_\perp'^2 +\cosh \left(2 k_\perp' \right)-1\right)},
\label{eq:Vp2ndInert}
\end{align}
where 
\begin{align}
N=&\,\,3 k_\perp'^3 (2s-1) \left(-s^2 \cosh \left(2 k_\perp' (s-1)\right)+(s-1)^2 \cosh \left(2 k_\perp' s\right)+2s-1\right), \label{eq:N_defn}
\end{align}
with the $\kappa$-dependence being entirely contained within $F_{22}(\kappa)$. The integral in \eqref{eq:Vp2ndInert} is again evaluated numerically using Gauss-Legendre quadrature, with the upper limit taken to be a suitably large number~($400$ in the present case) to ensure accuracy. The total time-averaged lift velocity  for small $Re_c$ is:
\begin{align}
\langle V_p\rangle(s;\kappa)=\langle V_p\rangle_{1}(s;\kappa)+\langle V_p\rangle_{2}(s;\kappa). \label{eq:VpsmallRec}
\end{align}

We will now analyze the lift close to the channel walls. Sufficiently close to the lower wall, for instance, the lift should only depend on the distance from the wall\,(that is, $s$), and not the channel width\,(via $Re_c$). Therefore, one can define a rescaled wavenumber $k'_\perp=k_w/s$ in \eqref{eq:VpJeffAvgd} and \eqref{eq:Vp2ndInert}. Next, expanding the respective integrands about $s=0$ with $k_w$ fixed, the near-wall limits of the Jeffery-averaged lift and second inertial contribution take the form:
\begin{align}
\lim_{s\to 0}\,\,\langle V_p\rangle_1=-\frac{Re_p\langle S_{12}\rangle_J(\kappa)}{3\pi} \int_0^\infty dk_w\,\, e^{-2 k_w} k_w (3 k_w^2-2 k_w+3) = -\dfrac{11 Re_p\langle S_{12}\rangle(\kappa)}{24\pi},\label{eq:VpJeffnearWall} \\
\lim_{s\to 0}\,\,\langle V_p\rangle_2= \frac{3 Re_p F_{22}(\kappa)}{2\pi s^2} \int_0^\infty dk_w\,\, e^{-2 k_w} k_w^3 = \dfrac{9 Re_p F_{22}(\kappa)}{16\pi s^2}.
\label{eq:Vp2ndInertnearWall}
\end{align}
Thus, the limiting values of the inertial lift velocity at the two walls are given by:
\begin{align}
\langle V_p\rangle^\text{lower wall}=  Re_p \left[-\dfrac{11 \langle S_{12}\rangle(\kappa)}{24\pi} + \dfrac{9  F_{22}(\kappa)}{16\pi s^2} \right], \label{eq:VpnearWall1} \\
\langle V_p\rangle^\text{upper wall}=  Re_p \left[\dfrac{11 \langle S_{12}\rangle(\kappa)}{24\pi} - \dfrac{9  F_{22}(\kappa)}{16\pi (1-s)^2} \right]. \label{eq:VpnearWall2}
\end{align}
It was shown in \cite{anand2023Jeff} that the first terms in (\ref{eq:VpnearWall1}) and (\ref{eq:VpnearWall2}) are, in fact, independent of $Re_c$. For small $Re_p$, these $Re_c$-independent plateau values pertain to near-wall regions, defined by $\lambda\ll s,(1-s) \ll Re_c^{-1/2}$, where inertia acts as a regular perturbation. These plateau values must simultaneously correspond to the far-field limit of the lift on a finite-size spheroid translating in a linear shear flow, parallel to, and at a distance of $O(L)$ from a single plane boundary. This connection has been established for a sphere\citep{anand2023Jeff}, although the corresponding calculation for a spheroid, translating near a single plane wall, has not yet been done; the added difficulty comes from one not being able to use a bispherical coordinate system for purposes of this calculation, as has been done for a sphere\,\citep{cherukat1994}. Note that the lift velocity must eventually tend to zero with approach towards either wall\,(that is, for $s, (1-s) \ll \lambda$), on account of lubrication forces. 

The second terms in (\ref{eq:VpnearWall1}) and (\ref{eq:VpnearWall2}) exhibit an inverse-squared divergence with decreasing wall separation. This scaling is characteristic of a stresslet-wall interaction - the lift is essentially due to the disturbance velocity field of the image system induced by the spheroid\,(approximated as a stresslet) that, to leading order, enforces the wall boundary condition. These terms reveal the physical mechanism underlying slowdown\,(and reversal) of the wallward movement of the equilibrium location, with increasing $Re_c$, relative to a sphere; see $\S$\ref{sec:conclusion} for a more detailed discussion. The aforesaid divergent behavior only prevails for $s,(1-s) \gg \lambda$, and will be regularized for $s,(1-s) \sim O(\lambda)$ when the finite size of the spheroid becomes important at leading order, and the stresslet approximation breaks down. It will be seen below\,(in Figures~\ref{fig:JeffvsNonJeffK2}a-d and \ref{fig:JeffvsNonJeffK50}a-d) that the divergent asymptote above 
is again independent of $Re_c$, and agrees with the numerical solution, $\langle V_p\rangle-\langle V_p\rangle_1$, for sufficiently small $s$, for all $Re_c$ examined\,($\langle V_p \rangle$ being determined numerically, as in $\S$\ref{sec:Reclarge}). Now, the small-$Re_c$ expansion can break down even before the finite size of the spheroid becomes important, owing to the divergent second inertial contribution becoming comparable to the finite Jeffery-averaged plateau. For $\kappa \sim O(1)$, this happens for an $s\sim O(F_{22}(\kappa)/\langle S_{12}\rangle_J(\kappa))^{1/2}$ which turns out to be very small numerically. 

In Figures~\ref{fig:1stvs2ndInertContr}a-d, we plot the Jeffery-averaged lift profile $\langle V_p\rangle_{1}(s)$ given by (\ref{eq:VpJeffAvgd}), the second small-$Re_c$ contribution $\langle V_p\rangle_{2}(s)$ given by (\ref{eq:Vp2ndInert}), and the sum of the two, (\ref{eq:VpsmallRec}), on both linear and logarithmic ordinate scales, for $\kappa = 2$ and $50$ and with $Re_c = 1$, $\lambda = 0.05$\,(the lift magnitude is plotted on the logarithmic scale). For $\kappa=2$, Figure~\ref{fig:1stvs2ndInertContr}a shows the Jeffery-averaged contribution to be orders of magnitude larger than the second inertial contribution, down to $s = 0.01$. Accordingly, the linear plot in Figure~\ref{fig:1stvs2ndInertContr}b shows almost exact agreement of the total lift profile with the Jeffery-averaged one. This confirms the accuracy of the Jeffery-averaged description~\citep{anand2023Jeff} for neutrally buoyant spheroids with order unity aspect ratios. On the other hand, Figure~\ref{fig:1stvs2ndInertContr}c for $\kappa=50$ shows that the second inertial contribution starts to become comparable in the bulk\,(for $s \lesssim 0.1$), with both Figures~\ref{fig:1stvs2ndInertContr}c and d showing a very slight shift in the equilibrium location towards the centerline. This shifted locus, obtained with the addition of the second small-$Re_c$ contribution, has been plotted in the insets in Figures~\ref{fig:InertialAvgdLiftkappagt1}, \ref{fig:InertialAvgdLiftkappalt1}, \ref{fig:Liftkappa200}a, \ref{fig:InertialAvgdLiftkappap05}a and \ref{fig:InertialAvgdLiftkappap01}a in the main manuscript.

\begin{figure}
	\centering
    \begin{subfigure}{0.49\textwidth}
    	\includegraphics[width=\textwidth]{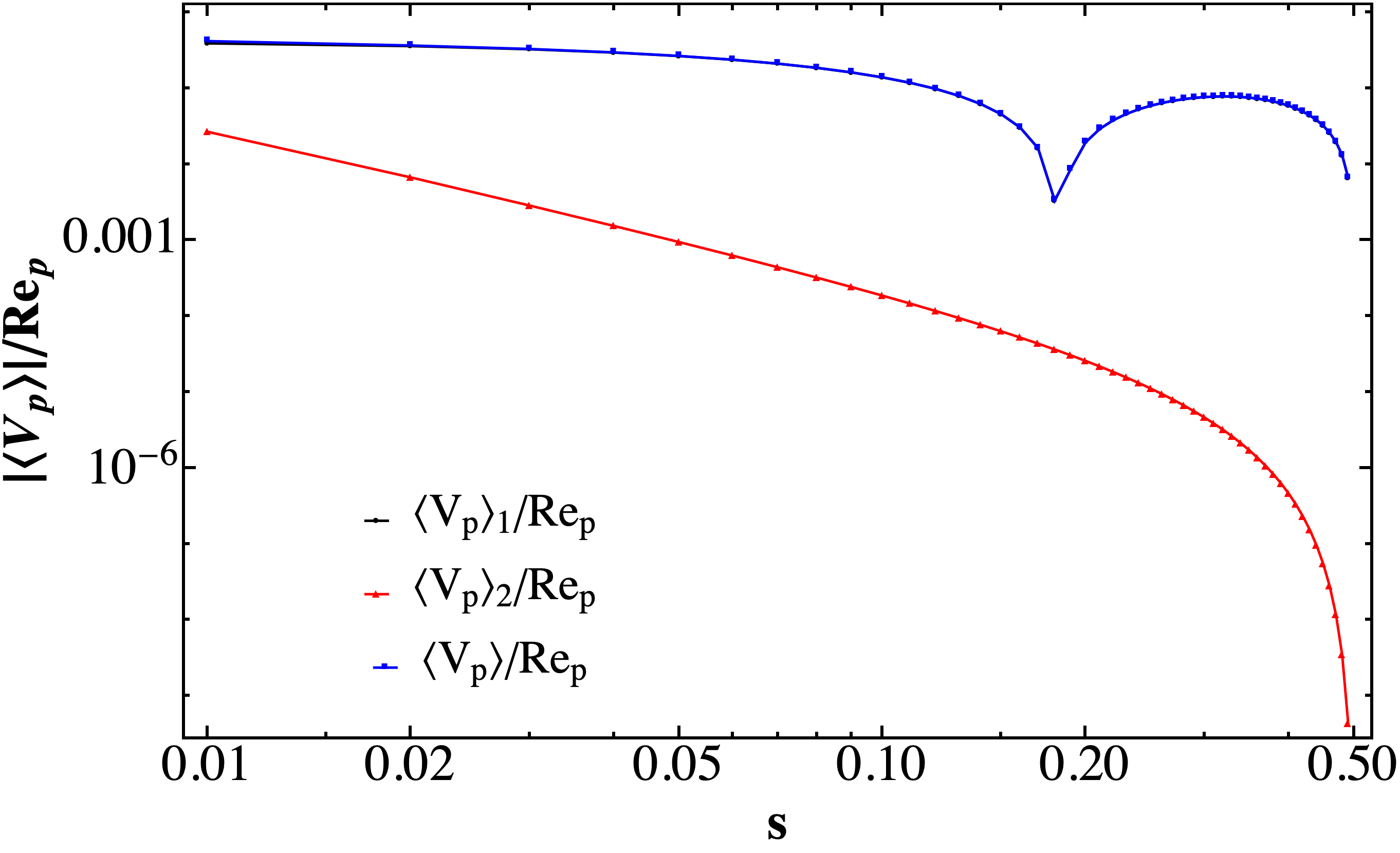}
    	\caption{$\kappa=2$}
    \end{subfigure}
    \begin{subfigure}{0.49\textwidth}
	\includegraphics[width=\textwidth]{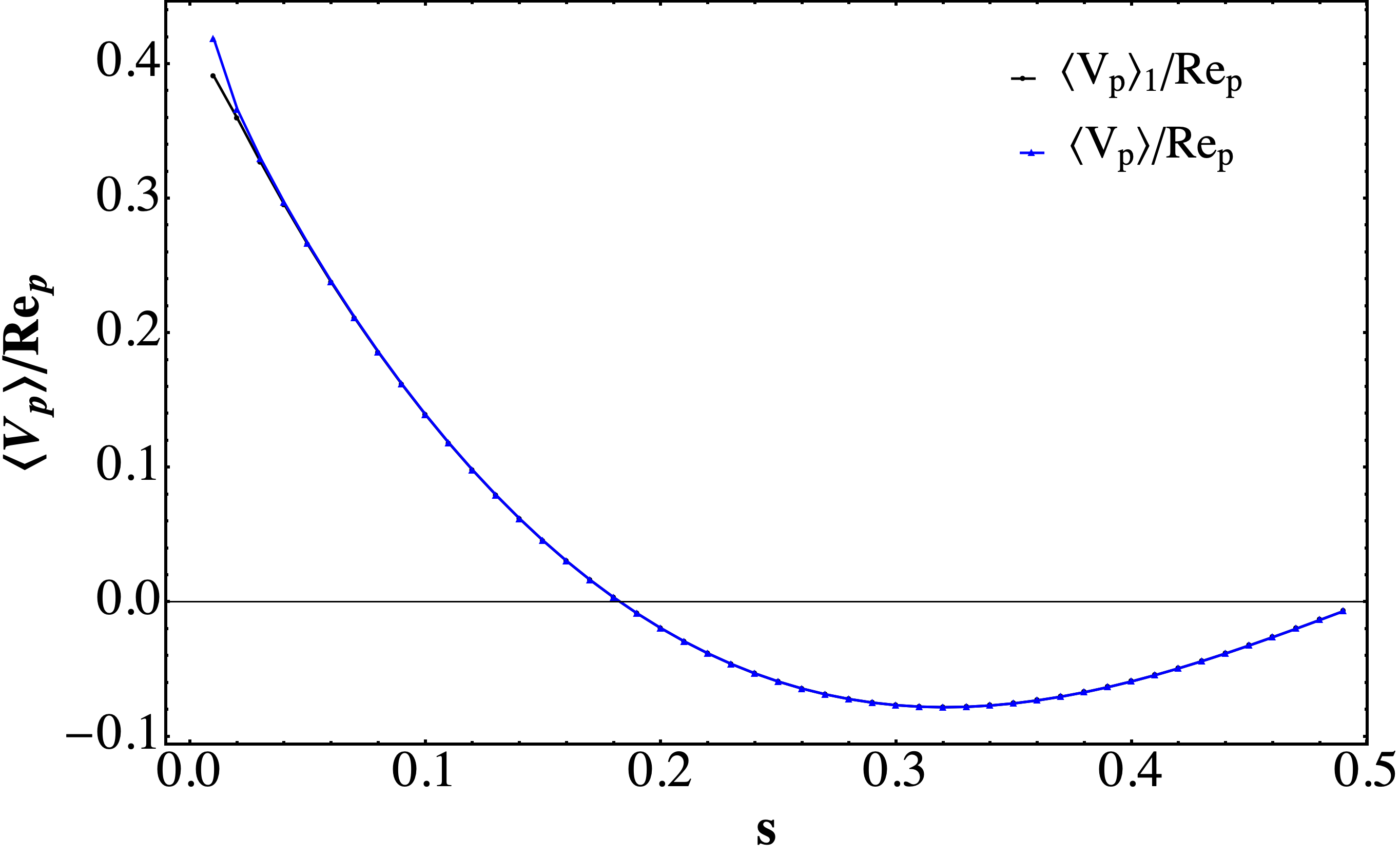}
	\caption{$\kappa=2$}
    \end{subfigure}
    \begin{subfigure}{0.49\textwidth}
	\includegraphics[width=\textwidth]{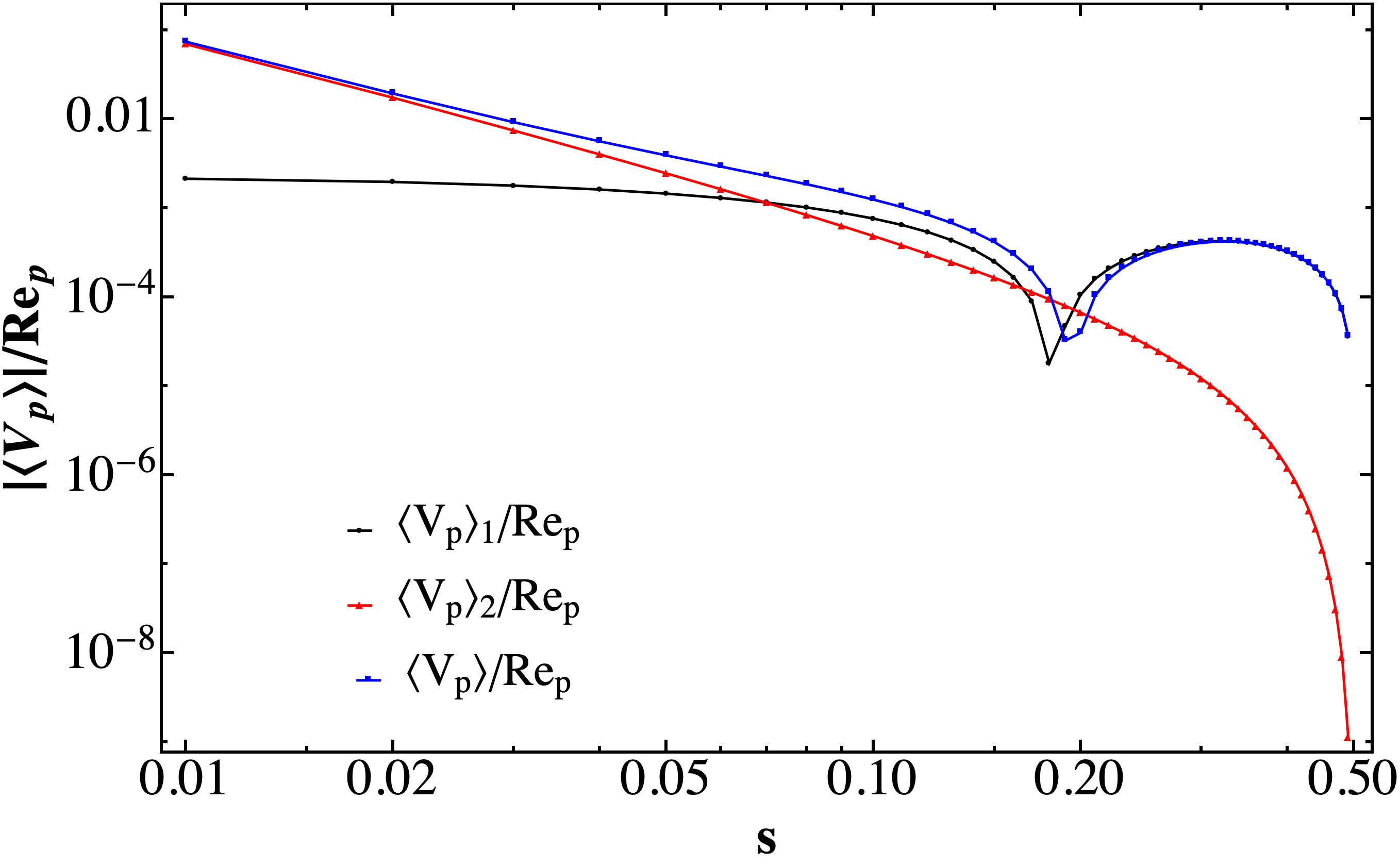}
	\caption{$\kappa=50$}
    \end{subfigure}
    \begin{subfigure}{0.49\textwidth}
	\includegraphics[width=\textwidth]{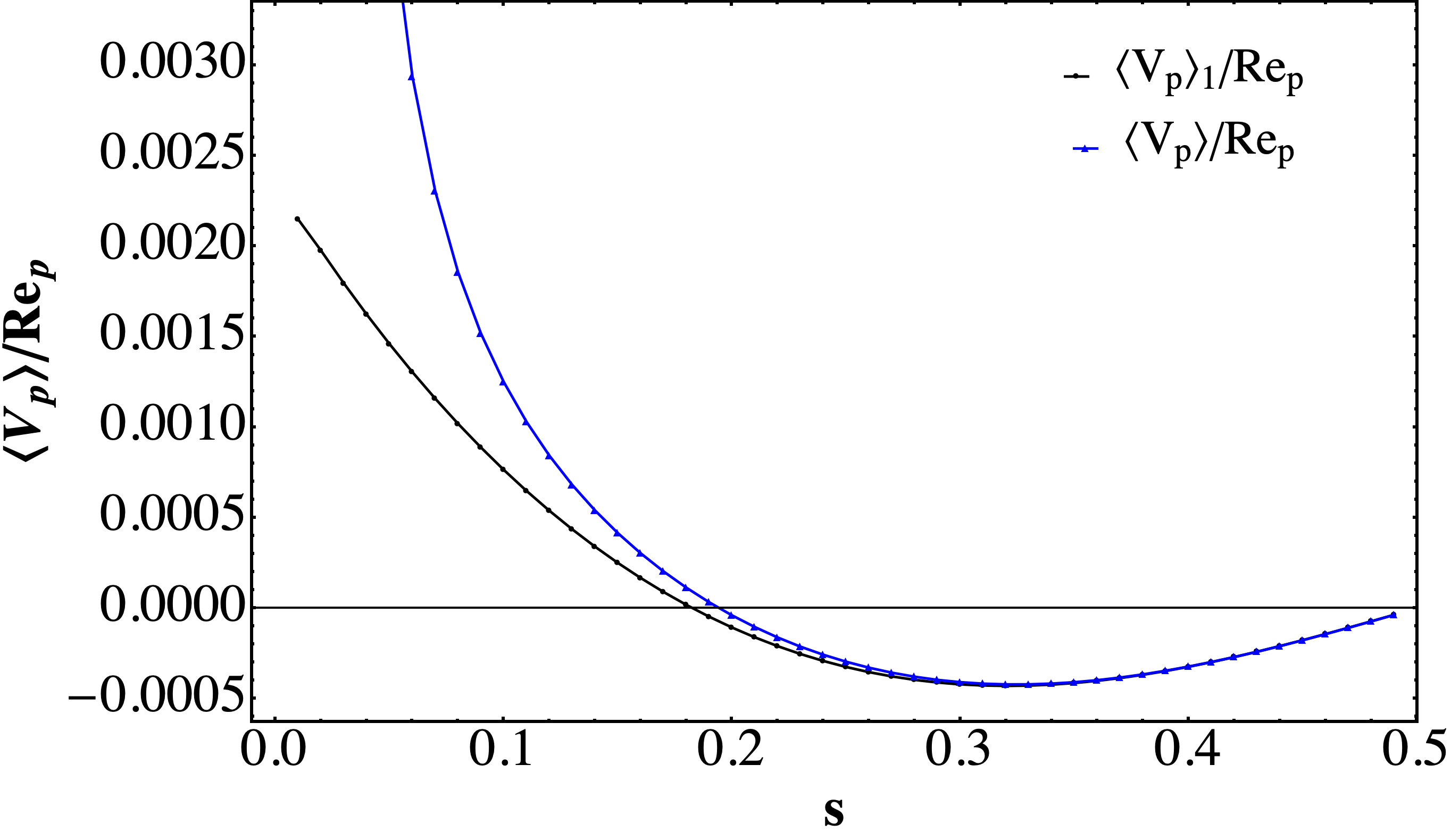}
	\caption{$\kappa=50$}
    \end{subfigure}
	\caption{Comparison of the magnitudes for the Jeffery-averaged lift~($\langle V_p\rangle_1/Re_p$), the second inertial inertial contribution~($\langle V_p\rangle_2/Re_p$) and the inertially-averaged lift~($\langle V_p\rangle/Re_p$) for $Re_c=1$ and $\lambda=0.05$, for tumbling prolate spheroids  of the indicated aspect ratios; (a) and (c) present the comparison on a log-log scale, and (b) and (d) on a linear scale.}
	\label{fig:1stvs2ndInertContr}
\end{figure}

Next, we compare the second inertial contribution obtained semi-analytically\,(that is, \eqref{eq:Vp2ndInert}) with the difference $(\langle V_p\rangle-\langle V_p\rangle_1)/Re_p$, calculated numerically using the shooting method and (\ref{eq:VpJeffAvgd}), for $\kappa=2$ and $50$, for various $Re_c$. While one expects a good comparison between the two for small $Re_c$, Figures~\ref{fig:JeffvsNonJeffK2}a-d show that, for $\kappa=2$, the small-$Re_c$ asymptote remains valid at least upto $Re_c\approx8$. While the magnitude of the inertial correction is seen to be larger in figures~\ref{fig:JeffvsNonJeffK50}a-d, rather surprisingly, agreement with the small-$Re_c$ asymptote is achieved until a larger $Re_c$. 
While the main plots in the figures above use a logarithmic scale to focus on the inertial correction, the insets in each of these plots show the total lift profiles based on the Jeffery-averaged and inertially-averaged approximations on a linear scale. The insets in figures~\ref{fig:JeffvsNonJeffK2}a-d show that the Jeffery-averaged approximation remains accurate, for $\kappa=2$, for all the $Re_c$'s. However, for $\kappa=50$, the larger magnitude of the second inertial contribution leads to deviations from the Jeffery-averaged approximation, with the deviation growing with increasing $Re_c$. 

\begin{figure}
	\centering
    \begin{subfigure}{0.49\textwidth}
    	\includegraphics[width=\textwidth]{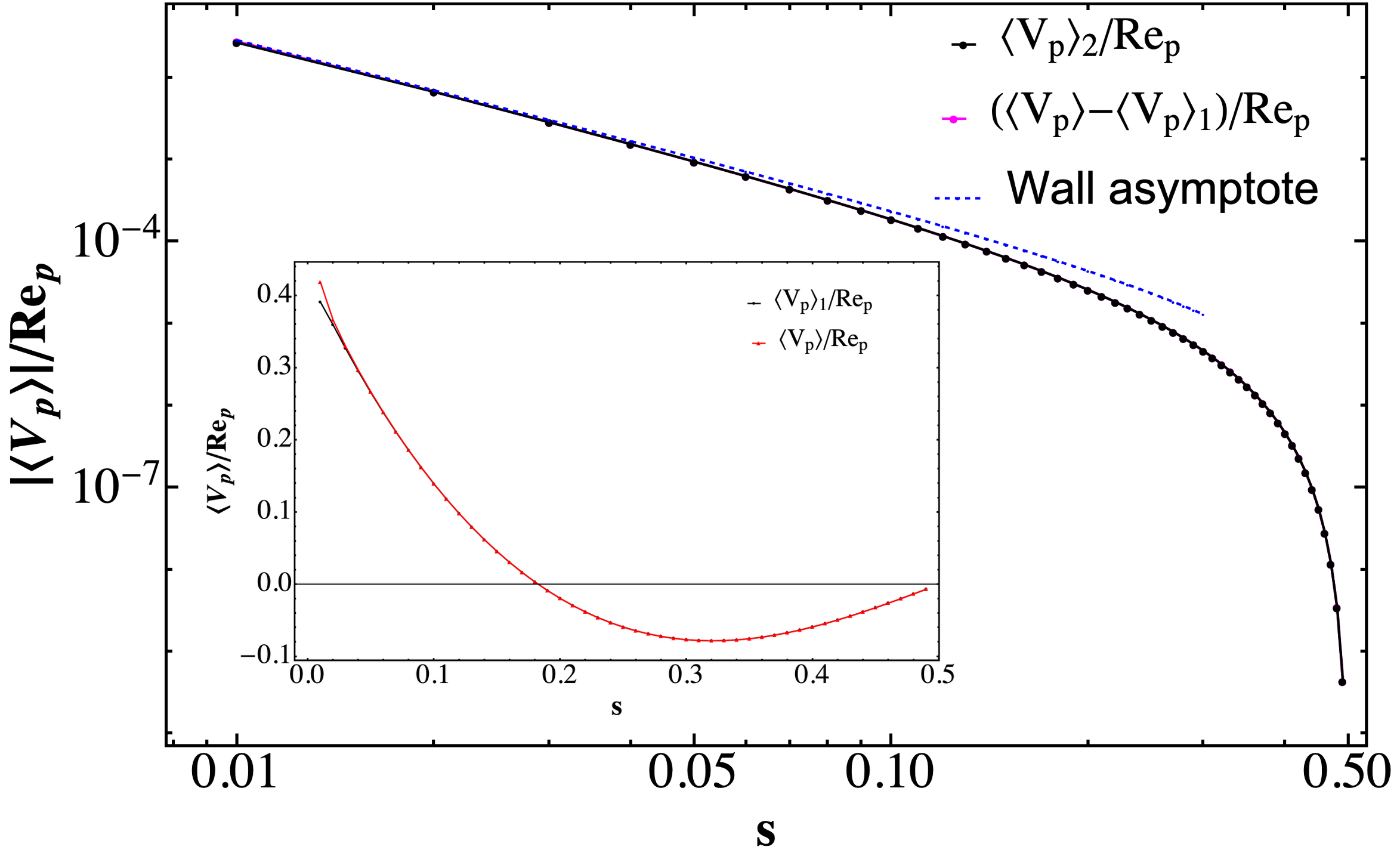}
    	\caption{$Re_c=1$}
    \end{subfigure}
    \begin{subfigure}{0.49\textwidth}
	\includegraphics[width=\textwidth]{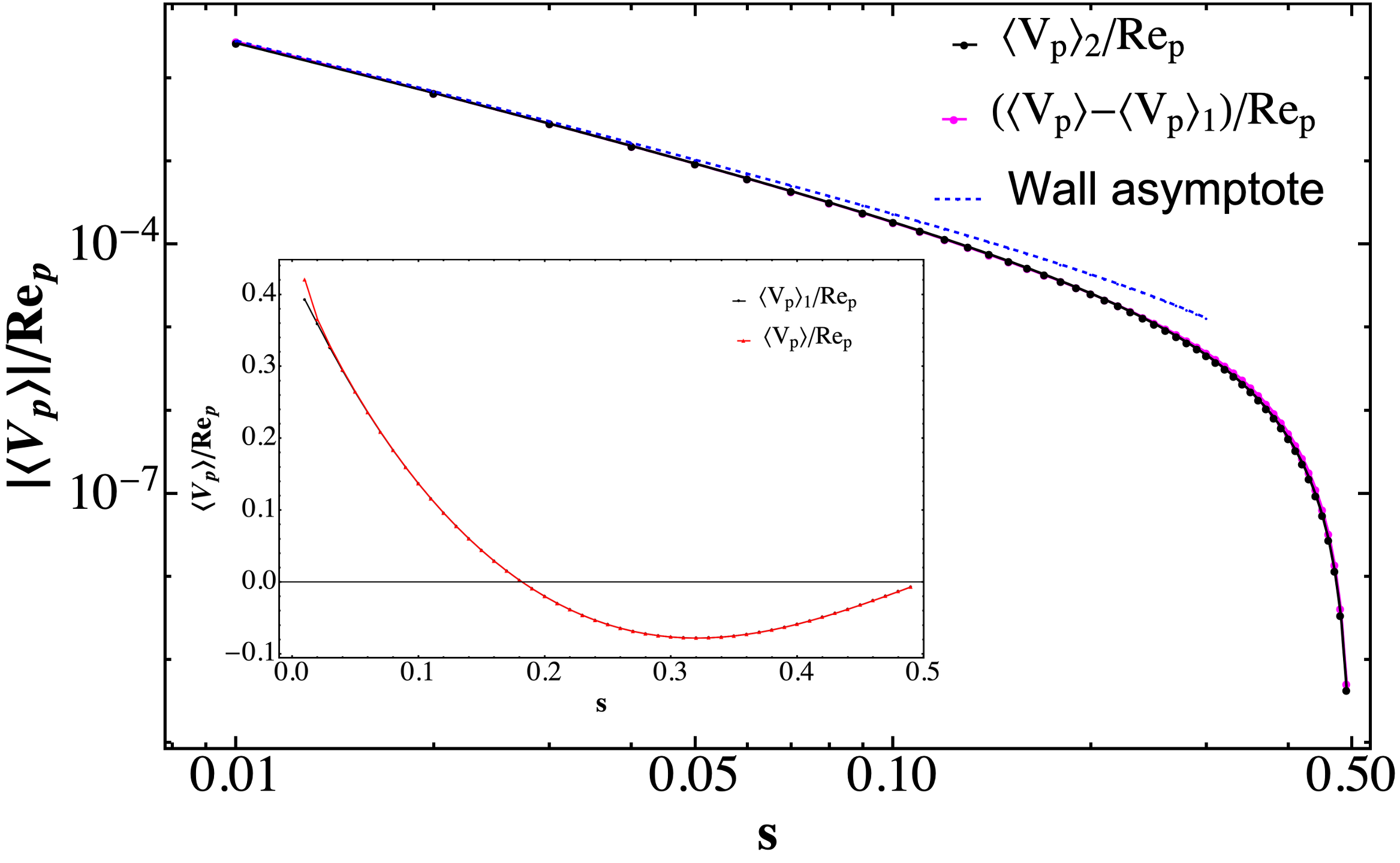}
	\caption{$Re_c=8$}
    \end{subfigure}
    \begin{subfigure}{0.49\textwidth}
	\includegraphics[width=\textwidth]{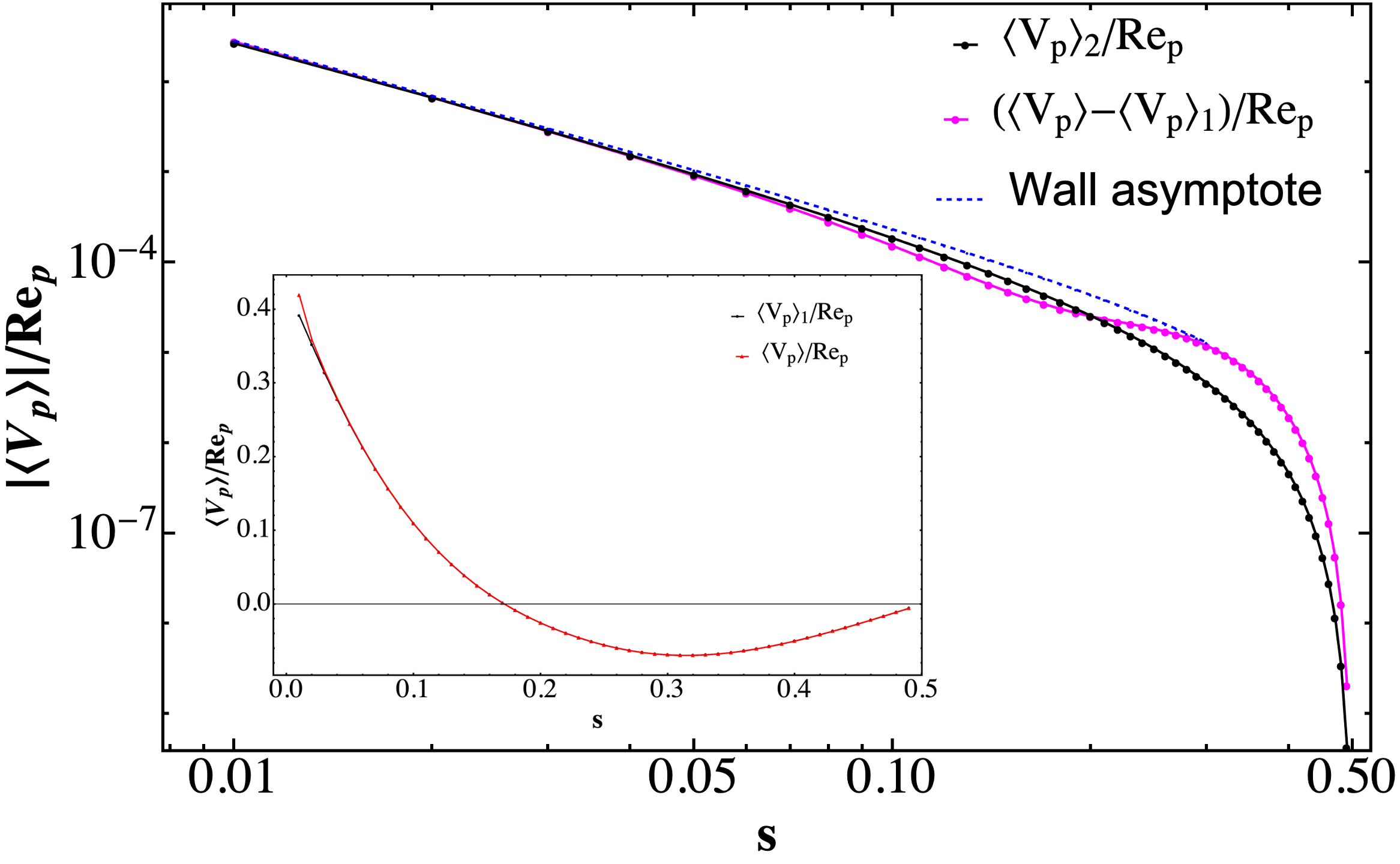}
	\caption{$Re_c=40$}
    \end{subfigure}
    \begin{subfigure}{0.49\textwidth}
	\includegraphics[width=\textwidth]{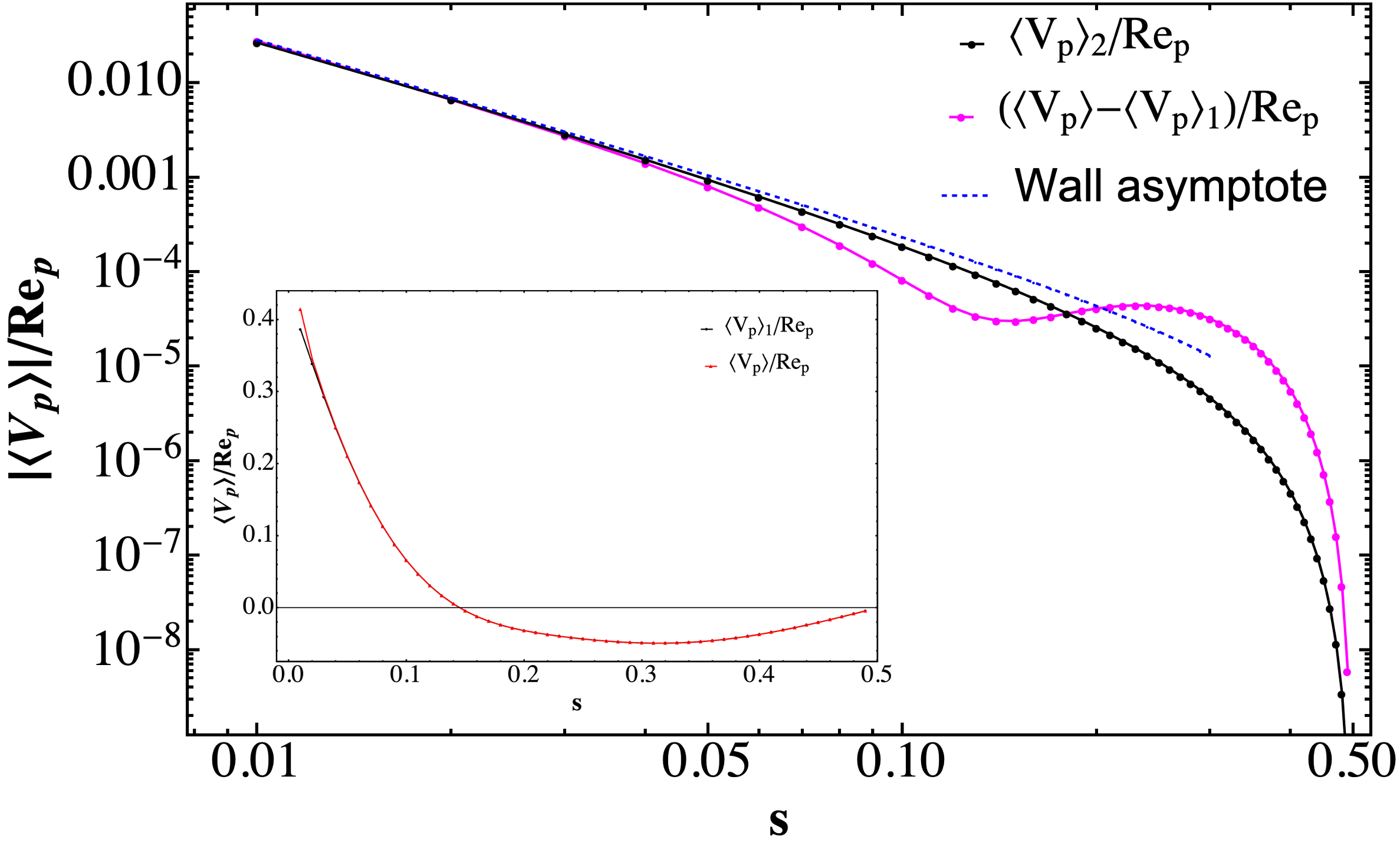}
	\caption{$Re_c=100$}
    \end{subfigure}
	\caption{Comparison of the second inertial contribution, $\langle V_p\rangle_2/Re_p$, with the difference, $(\langle V_p\rangle-\langle V_p \rangle_1)/Re_p$, for a tumbling prolate spheroid with $\kappa=2$, for $\lambda=0.05$. The two curves show good agreement upto $Re_c\sim O(10)$. The blue dashed line is the $O(1/s^2)$ wall asymptote, given by the second term in \eqref{eq:VpnearWall1}, and agrees well with the full profile upto $s\approx0.05$ for all four values of $Re_c$ shown. The insets show a comparison between the inertially averaged\,($\langle V_p\rangle/Re_p$) and Jeffery-averaged\,($\langle V_p \rangle_1/Re_p$) lift profiles for the respective $Re_c$'s.}
	\label{fig:JeffvsNonJeffK2}
\end{figure}

\begin{figure}
	\centering
    \begin{subfigure}{0.49\textwidth}
    	\includegraphics[width=\textwidth]{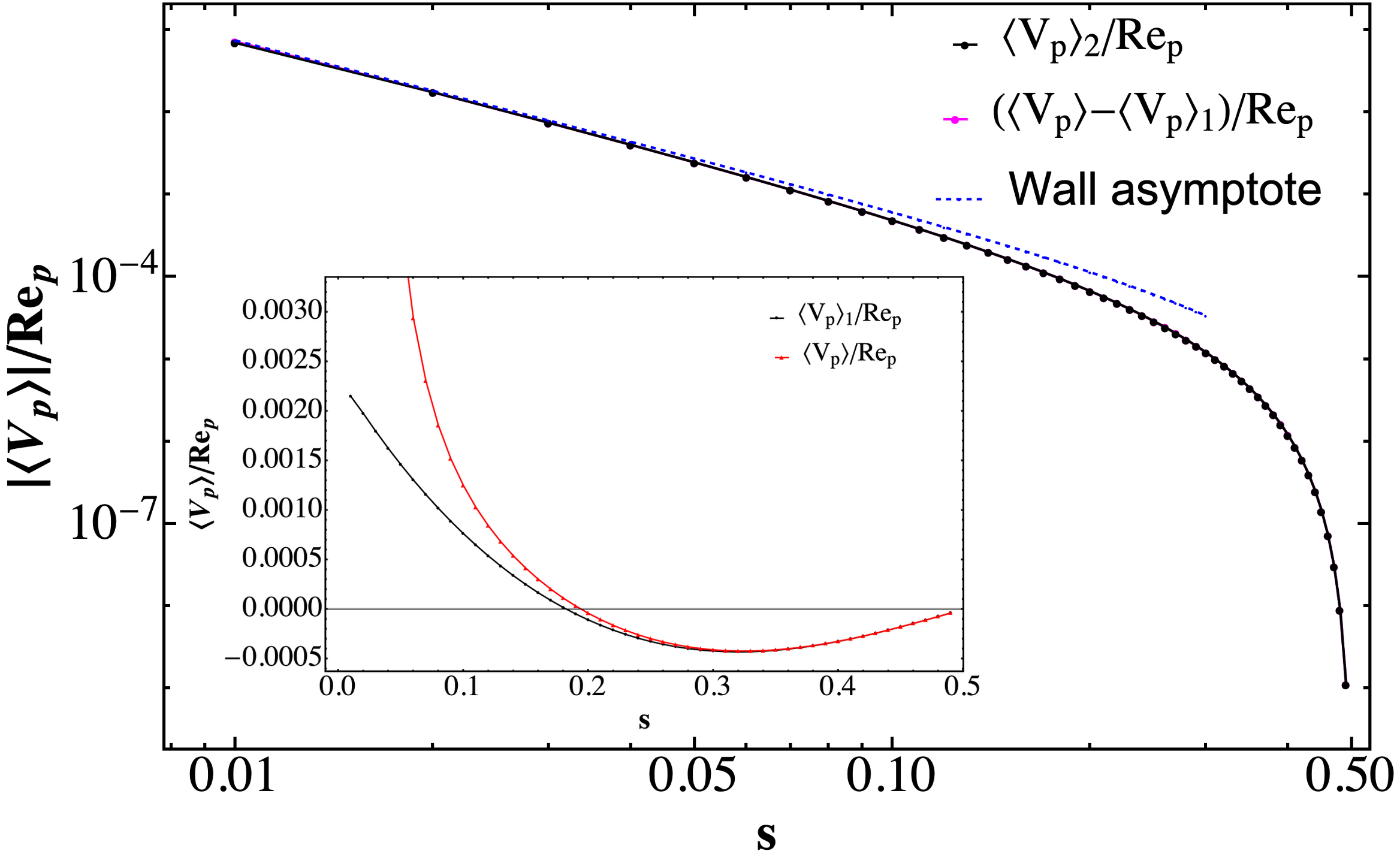}
    	\caption{$Re_c=1$}
    \end{subfigure}
    \begin{subfigure}{0.49\textwidth}
	\includegraphics[width=\textwidth]{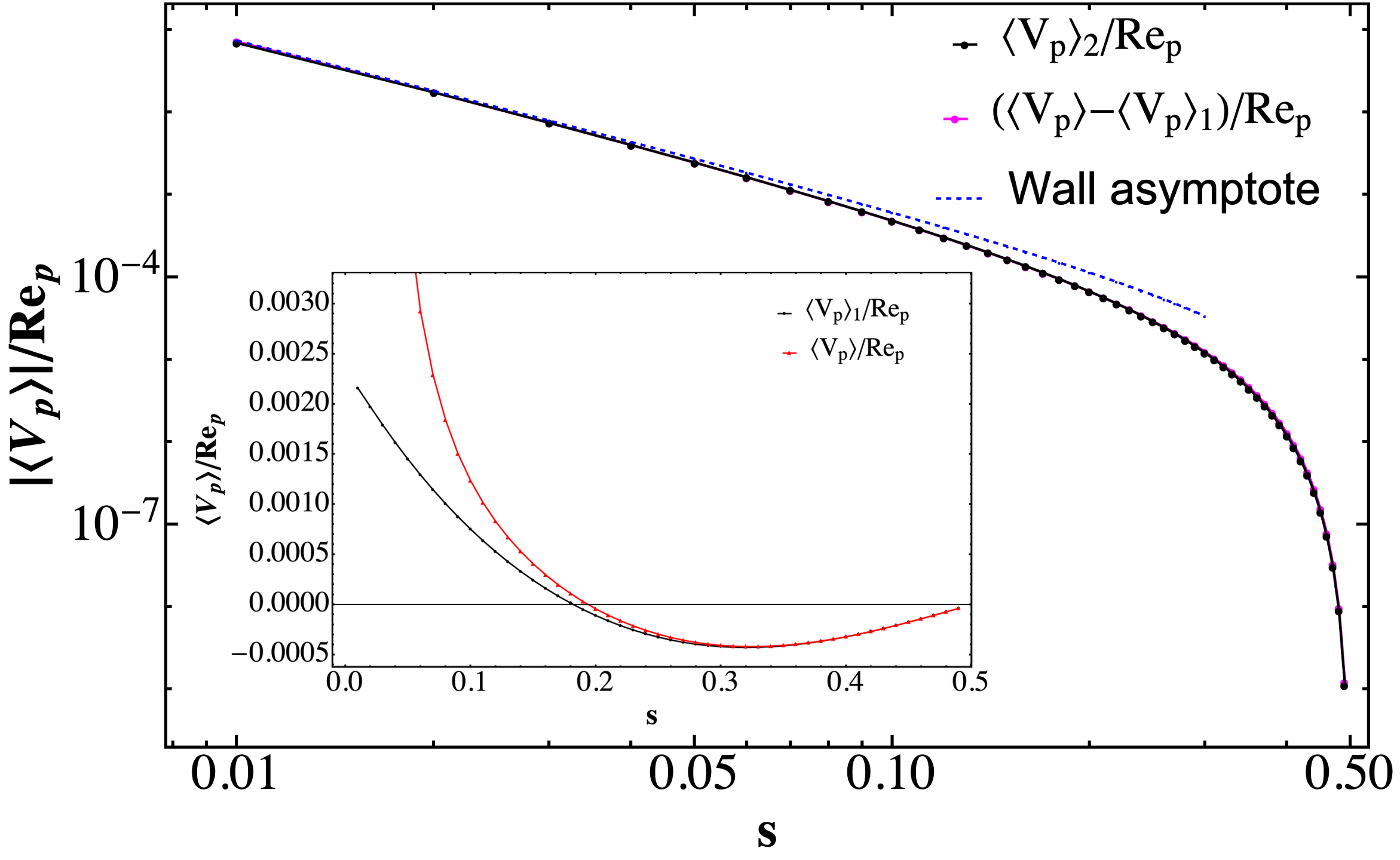}
	\caption{$Re_c=8$}
    \end{subfigure}
    \begin{subfigure}{0.49\textwidth}
	\includegraphics[width=\textwidth]{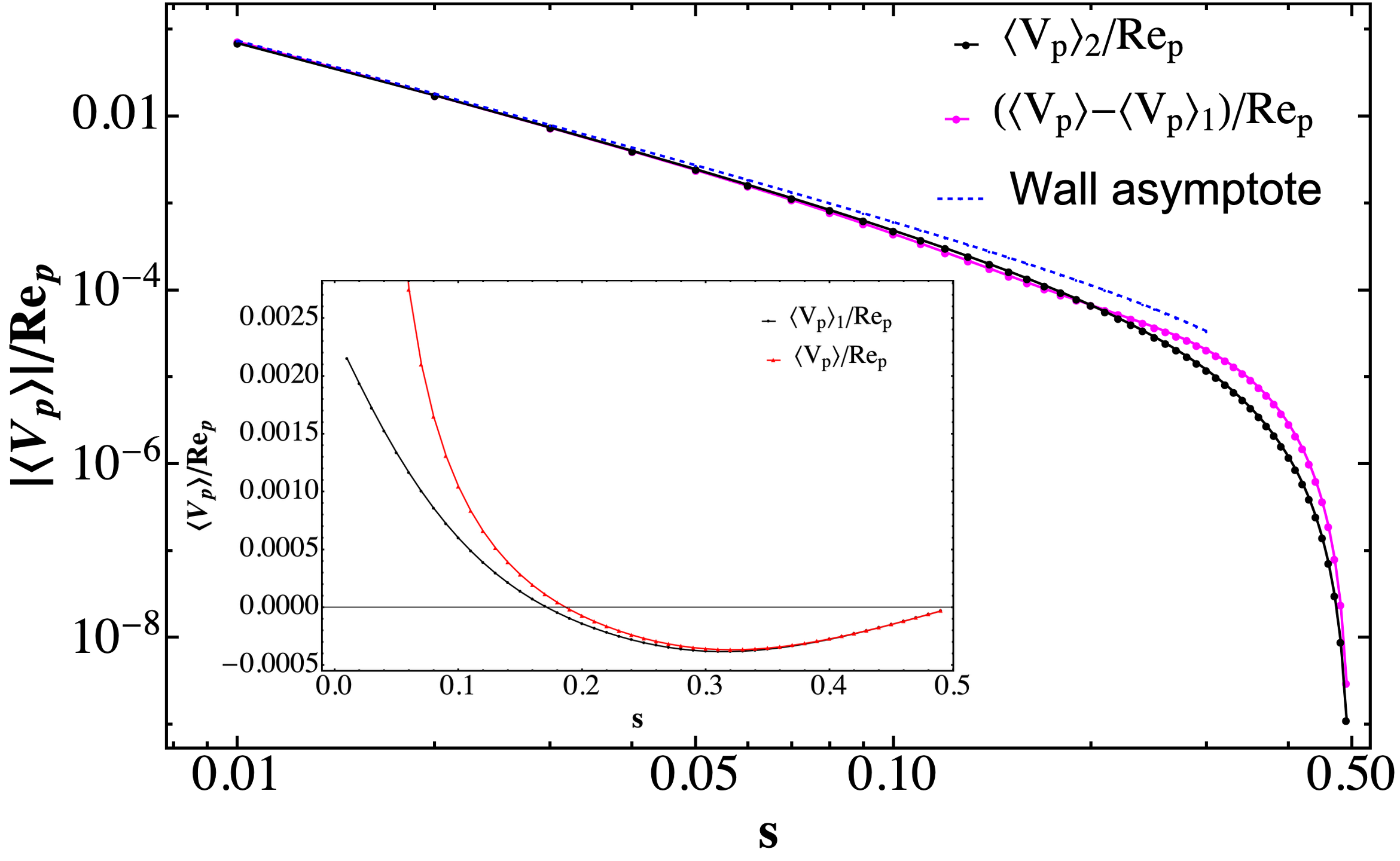}
	\caption{$Re_c=40$}
    \end{subfigure}
    \begin{subfigure}{0.49\textwidth}
	\includegraphics[width=\textwidth]{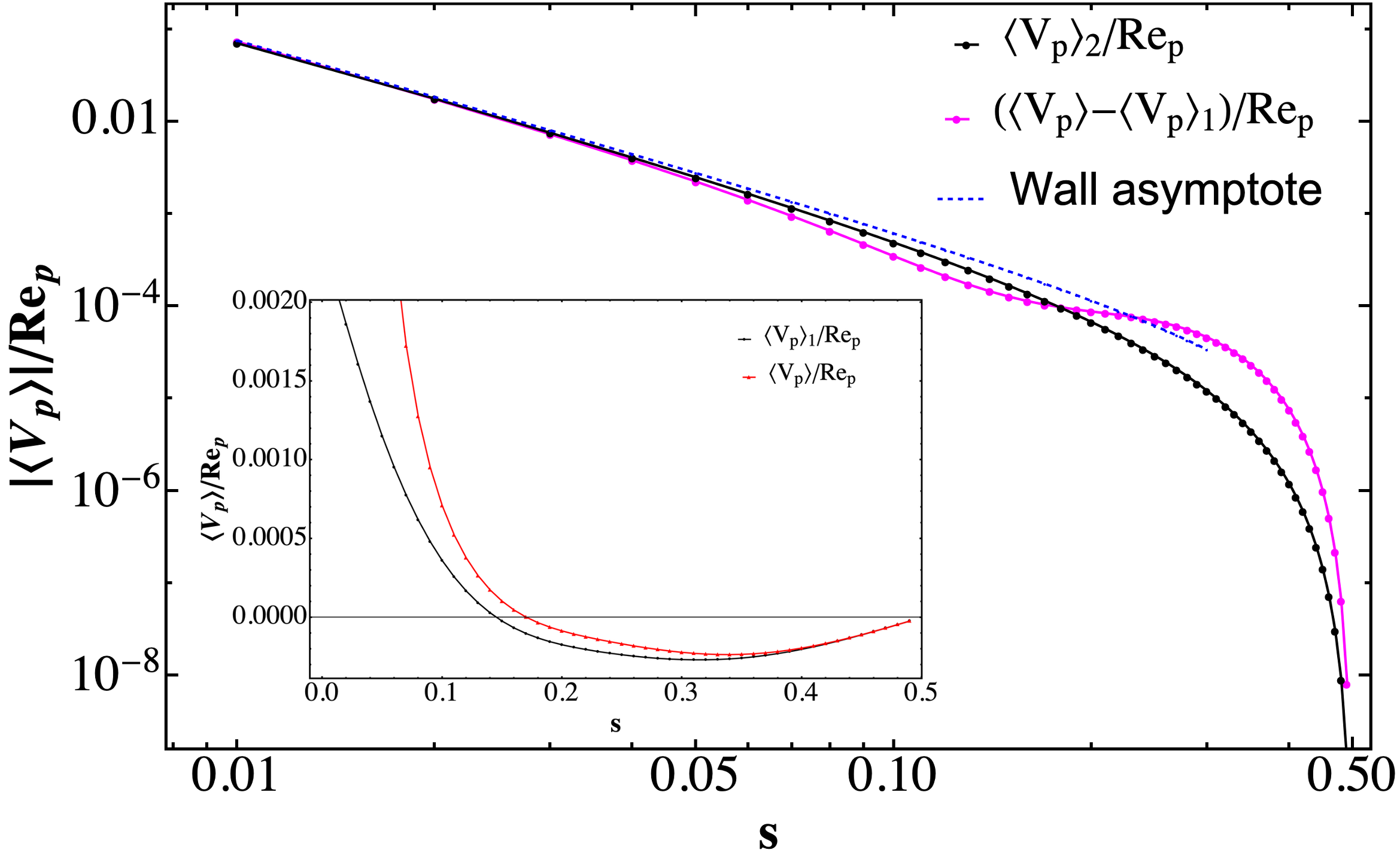}
	\caption{$Re_c=100$}
    \end{subfigure}
	\caption{Comparison of the second inertial contribution, $\langle V_p\rangle_2/Re_p$, against the difference, $(\langle V_p\rangle-\langle V_p \rangle_1)/Re_p$, for a tumbling prolate spheroid with $\kappa=50$, for $\lambda=0.05$. The two curves show good agreement upto $Re_c\sim O(40)$. The blue dashed line is the $O(1/s^2)$ wall asymptote, given by the second term in \eqref{eq:VpnearWall1}, and agrees well with the full profile $s\approx0.05$ in all cases. The insets show a comparison between the inertially averaged\,($\langle V_p\rangle/Re_p$) and the Jeffery-averaged\,($\langle V_p \rangle_1/Re_p$) lift profiles for the respective $Re_c$'s.}
	\label{fig:JeffvsNonJeffK50}
\end{figure}

Finally, it is worth mentioning that the generalized reciprocal theorem used in \citet{anand2023Jeff} leads to the following exact expression for the instantaneous lift velocity:
\begin{align}
 V_p=&-Re_p\int_{V^F}\bm{u}^{t}\cdot\left(\frac{\partial\bm{u}_s}{\partial t}+\bm{u}_s\cdot\bm{\nabla u}_s+\bm{u}_s\cdot\bm{\nabla u}^\infty+\bm{u}^\infty\cdot\bm{\nabla u}_s\right) dV\nonumber\\
 &-Re_p\,\,\bm{U}_p^{t}\cdot \frac{d\bm{U}_p}{dt}-Re_p\,\,\bm{\Omega}_p^{t}\cdot\frac{d(\bm{I}_p\cdot\bm{\Omega}_p)}{dt}.
\label{eq:RTinstliftvel}
\end{align}
for a spheroid in wall-bounded plane Poiseuille flow. 
The integral above is over the fluid domain\,($V^F$) outside the spheroid, but within the plane parallel boundaries constituting the channel, and involves a dot product between the inertial terms on the RHS of (\ref{eq:NS1}a), and a test velocity field $\bm{u}^t$. Note that, to leading order in $Re_p$, the actual velocity field $\bm{u}$ in (\ref{eq:NS1}a) has now been replaced by its Stokesian approximation $\bm{u}_s$. The velocity field $\bm{u}^t$ is the Stokes disturbance due to a spheroid, with the same\,(instantaneous) orientation as the original neutrally buoyant spheroid, and translating under the action of a unit force oriented normal to the walls; the term $d\bm{U}_p/dt$ present in (\ref{eq:NS1}a) does not contribute to (\ref{eq:RTinstliftvel}) since $\bm{u}_t$ does has net zero momentum. The second and third terms in \eqref{eq:RTinstliftvel} involve the translational and angular accelerations, respectively, of the spheroid in the actual problem. Here, $\bm{I}_p$ is the spheroid moment of inertia tensor, with $\bm{U}_p$ and $\bm{\Omega}_p$ being the translational and angular velocities that, to leading order in $Re_p$, may be obtained using the respective Faxen's laws\,(\eqref{eq:UpFaxen}  and \eqref{eq:OmegapFaxen}); $\bm{U}_p^t$ and $\bm{\Omega}_p^t$ are the translational and angular velocities of the spheroid in the test problem. 


Time averaging (\ref{eq:RTinstliftvel}) leads to:
\begin{align}
\langle V_p\rangle=&-Re_p\int_{V^F+V^P}\bm{u}_{St}\cdot\left(\langle\bm{u}_{str}\rangle\cdot\bm{\nabla u}^\infty+\bm{u}^\infty\cdot\bm{\nabla \langle u}_{str}\rangle\right) dV-Re_p\,\,\left\langle\bm{U}_p^t\cdot \frac{d\bm{U}_p}{dt}\right\rangle\nonumber\\
&-Re_p\,\,\left\langle\bm{\Omega}_p^t\cdot\frac{d(\bm{I}_p\cdot\bm{\Omega}_p)}{dt}\right\rangle. \label{eq:RTtimeavgdliftvel}
\end{align}
For small $Re_c$, the point-particle approximation of the volume integral in (\ref{eq:RTinstliftvel}), along with an approximate time averaging based on the Jeffery angular velocity, leads to (\ref{eq:RTVolIntegral}) for the Jeffery-averaged lift $\langle V_p \rangle_1$. However, the exact relation, (\ref{eq:RTtimeavgdliftvel}), also identifies additional contributions due to the second and third terms, that are of a smaller order in $\lambda$, and that are associated with the translational and rotational accelerations of an inertial spheroid in plane Poiseuille flow. The second term in \eqref{eq:RTtimeavgdliftvel} involves a correlation between the time-periodic variations of $d\bm{U}_p/dt$ along the flow direction, and the analogous variation of the flow-directed component of $\bm{U}_p^t$, the latter arising due to the periodically varying test spheroid orientation acted on by the gradient-directed unit force. The third term involves a correlation between the angular acceleration of the actual spheroid, and the angular velocity of the test spheroid. These finite-size contributions due to the translational and angular acceleration terms may be shown to be $O(\lambda Re_p)$ and $O(\lambda^2 Re_p)$, respectively; the translational acceleration term was erroneously estimated as being $O(\lambda^2 Re_p)$ in \cite{anand2023Jeff}. The lift contribution due to the time-averaged translational acceleration contribution has been calculated by \citet{Tornberg_JFM2021}, and the underlying mechanism is discussed in $\S$\ref{sec:conclusion}.

\section{Angular velocity coefficients, $G_i$'s}\label{App:AngVel}
The detailed expressions for the aspect-ratio-dependent functions $G_1(\kappa)-G_4(\kappa)$, that appear in (\ref{inertcorr_Re}), were derived in \citet{navaneeth2016}, and are reproduced below for convenience:
\allowdisplaybreaks
\begin{align}
G_1(\kappa)&=\Bigg(9 \kappa ^2 (4 \kappa ^6+8 \kappa ^4-\kappa ^2-2) \sqrt{\kappa ^2-1} \coth ^{-1}\left(\frac{\kappa }{\sqrt{\kappa ^2-1}}\right)^4+\kappa ^2 (4 \kappa ^{10}+106 \kappa ^8\nonumber\\
&-78 \kappa ^6-214 \kappa ^4+209 \kappa ^2+54) (\kappa ^2-1)^{5/2}-\kappa(76 \kappa ^{10}+298 \kappa ^8-1029 \kappa ^6\nonumber\\
&+878 \kappa ^4+137 \kappa ^2-36) (\kappa ^2-1)^2 \coth ^{-1}\left(\frac{\kappa }{\sqrt{\kappa ^2-1}}\right)+3 \kappa  (8 \kappa ^{12}-16 \kappa ^{10}\nonumber\\
&-18 \kappa ^8-93 \kappa ^6+133 \kappa ^4+9 \kappa ^2-23) \coth ^{-1}\left(\frac{\kappa }{\sqrt{\kappa ^2-1}}\right)^3-\sqrt{\kappa ^2-1} (8 \kappa ^{14}\nonumber\\
&-204 \kappa ^{12}+630 \kappa ^{10}-697 \kappa ^8-453 \kappa ^6+945 \kappa ^4-211 \kappa ^2-18)\coth ^{-1}\left(\frac{\kappa }{\sqrt{\kappa ^2-1}}\right)^2\Bigg)\nonumber\\
&\times \Bigg[40 (\kappa ^3+\kappa)^2 \Big(\sqrt{\kappa ^2-1} (\kappa ^2+2)-3 \kappa\coth^{-1}\left(\frac{\kappa }{\sqrt{\kappa ^2-1}}\right)\Big) \Big(2 \kappa ^5-7 \kappa ^3\nonumber\\
&+3 \sqrt{\kappa ^2-1} \coth ^{-1}\left(\frac{\kappa }{\sqrt{\kappa ^2-1}}\right)+5 \kappa \Big) \Big(3 \kappa (\kappa ^2-1)\nonumber\\
&-\sqrt{\kappa ^2-1} (2 \kappa ^2+1) \coth ^{-1}\left(\frac{\kappa }{\sqrt{\kappa ^2-1}}\right)\Big)\Bigg]^{-1},\\
G_2(\kappa)&=\Big(-9 \kappa ^4 \sqrt{\kappa^2-1} (4 \kappa ^4-1) \coth ^{-1}\left(\frac{\kappa }{\sqrt{\kappa ^2-1}}\right)^4-\kappa ^2 (4 \kappa ^{10}+10 \kappa ^8-210 \kappa ^6\nonumber\\
&+284 \kappa ^4-43 \kappa ^2-18) (\kappa ^2-1)^{5/2}+\kappa  (12 \kappa ^{10}-230 \kappa ^8+207 \kappa ^6+150 \kappa ^4-43 \kappa ^2\nonumber\\
&+12) (\kappa^2-1)^2 \coth^{-1}\left(\frac{\kappa}{\sqrt{\kappa^2-1}}\right)-3 \kappa(8 \kappa^{12}-32 \kappa^{10}-2 \kappa^8+47 \kappa^6-31 \kappa^4\nonumber\\
&+5 \kappa ^2+5) \coth ^{-1}\left(\frac{\kappa }{\sqrt{\kappa ^2-1}}\right)^3+(8 \kappa ^{14}+68 \kappa ^{12}-210 \kappa ^{10}+75 \kappa ^8+67 \kappa ^6-63 \kappa ^4\nonumber\\
&+49 \kappa ^2+6) \sqrt{\kappa ^2-1} \coth ^{-1}\left(\frac{\kappa }{\sqrt{\kappa ^2-1}}\right)^2\Big)\times\Big[40 (\kappa ^3+\kappa )^2 \Big(\sqrt{\kappa ^2-1} (\kappa ^2+2)\nonumber\\
&-3 \kappa  \coth ^{-1}\left(\frac{\kappa }{\sqrt{\kappa ^2-1}}\right)\Big) \Big(2 \kappa ^5-7 \kappa ^3+3 \sqrt{\kappa ^2-1} \coth ^{-1}\left(\frac{\kappa }{\sqrt{\kappa ^2-1}}\right)+5 \kappa \Big)\nonumber\\
&\Big(3 \kappa (\kappa ^2-1)-\sqrt{\kappa^2-1} (2 \kappa ^2+1) \coth ^{-1}\left(\frac{\kappa }{\sqrt{\kappa ^2-1}}\right)\Big)\Big]^{-1},\\
G_3(\kappa)&=-(\kappa ^2-1) \Big(27 \kappa ^2 (4 \kappa ^6+4 \kappa ^4-\kappa ^2-1) \sqrt{\kappa ^2-1} \coth ^{-1}\left(\frac{\kappa }{\sqrt{\kappa ^2-1}}\right)^4+\kappa ^2 (28 \kappa ^{10}\nonumber\\
&+226 \kappa ^8-1128 \kappa ^6-505 \kappa ^4+659 \kappa ^2+342) (\kappa ^2-1)^{5/2}-\kappa (220 \kappa ^{10}-814 \kappa ^8\nonumber\\
&-2573 \kappa ^6+1568 \kappa ^4+583 \kappa ^2+44) (\kappa ^2-1)^2 \coth ^{-1} \left(\frac{\kappa }{\sqrt{\kappa ^2-1}}\right)+9 \kappa  (8 \kappa ^{12}\nonumber\\
&-16 \kappa ^{10}+2 \kappa ^8-39 \kappa ^6+57 \kappa ^4+15 \kappa ^2-27) \coth ^{-1}\left(\frac{\kappa }{\sqrt{\kappa ^2-1}}\right)^3-3 \sqrt{\kappa ^2-1} (8 \kappa ^{14}\nonumber\\
&+12 \kappa ^{12}+406 \kappa ^{10}-467 \kappa ^8-335 \kappa ^6+533 \kappa ^4-135 \kappa ^2-22) \coth ^{-1}\left(\frac{\kappa }{\sqrt{\kappa ^2-1}}\right)^2\Big)\nonumber\\
&\times\Big[120\kappa ^2 (\kappa ^2+1)^3 \Big(\sqrt{\kappa ^2-1} (\kappa ^2+2)-3 \kappa  \coth ^{-1}\left(\frac{\kappa }{\sqrt{\kappa ^2-1}}\right)\Big) \Big(2 \kappa ^5-7 \kappa ^3\nonumber\\
&+3 \sqrt{\kappa ^2-1} \coth ^{-1}\left(\frac{\kappa }{\sqrt{\kappa ^2-1}}\right)+5 \kappa \Big) \Big(3 \kappa  (\kappa ^2-1)\nonumber\\
&-\sqrt{\kappa ^2-1} (2 \kappa ^2+1) \coth ^{-1}\left(\frac{\kappa }{\sqrt{\kappa ^2-1}}\right)\Big)\Big]^{-1},\\
G_4(\kappa)&=-G_3(\kappa).
\end{align}
The above expressions pertain to a prolate spheroids\,($\kappa > 1$). The corresponding expressions for $G_1-G_4$ for oblate ($\kappa<1$) spheroids can be obtained from \cite{navaneeth2016} by using the transformation $\kappa/(\kappa^2-1)^{1/2}=\iota\kappa/(1-\kappa^2)^{1/2}$ and $d=-\iota d$ ($d$ is half the interfocal distance) in $Re_s\dot{\phi}_j^{(1)}$.

\end{appen}
\bibliographystyle{jfm}
\bibliography{references}

\end{document}